\documentclass[sigconf]{acmart}
\usepackage{booktabs}
\usepackage{multirow}
\usepackage{graphicx}
\usepackage{relsize}
\usepackage{subcaption}
\usepackage{makecell}

\AtBeginDocument{%
  }

\copyrightyear{2026}
\acmYear{2026}
\setcopyright{cc}
\setcctype{by}
\acmConference[CHI '26]{Proceedings of the 2026 CHI Conference on Human Factors in Computing Systems}{April 13--17, 2026}{Barcelona, Spain}
\acmBooktitle{Proceedings of the 2026 CHI Conference on Human Factors in Computing Systems (CHI '26), April 13--17, 2026, Barcelona, Spain}
\acmPrice{}
\acmDOI{10.1145/3772318.3790697}
\acmISBN{979-8-4007-2278-3/2026/04}

\begin{document}

\title[Scaffolding Metacognition with GenAI]{Scaffolding Metacognition with GenAI: Exploring Design Opportunities to Support Task Management for University Students with ADHD}

% GenAI as a Metacognitive Partner: Exploring Design Opportunities for Task Management with University Students with ADHD

% Scaffolding Metacognition:  Exploring Design Opportunities for GenAI to Support University Students with ADHD
%Task Management as a Metacognitive Practice: Co-Designing GenAI Solutions with University Students Experiencing ADHD
%Exploring GenAI Support for Task Management via a Metacognitive Lens: A Case Study on University Students With ADHD

%Metacognitive Time Management with GenAI: Exploring Design Opportunities for Supporting University Students with ADHD

%%GenAI as Productivity Partners: Exploring Design Opportunities for Supporting ADHD University Students

\author{Zihao Zhu}
\affiliation{%
  \department{Department of Computer Science}
  \institution{City University of Hong Kong}
  \city{Hong Kong SAR}
  \country{China}}
\email{zihaozhu9-c@my.cityu.edu.hk}

\author{Junnan Yu}
\affiliation{%
  \department{School of Design}
  \institution{The Hong Kong Polytechnic University}
  \city{Hong Kong SAR}
  \country{China}}
\email{junnan.yu@polyu.edu.hk}

\author{Yuhan Luo}
\affiliation{%
  \department{Department of Computer Science}
  \institution{City University of Hong Kong}
  \city{Hong Kong SAR}
  \country{China}}
\email{yuhanluo@cityu.edu.hk}
\authornote{Corresponding author.}

\begin{abstract}

For university students transitioning to an independent and flexible lifestyle, having ADHD poses multiple challenges to their academic task management, which are closely tied to their metacognitive struggles---difficulties in awareness and regulation of one's own thinking processes. The recently surged Generative AI shows promise to mitigate these gaps with its advanced information understanding and generation capabilities. 
As an exploratory step, we conducted co-design sessions with 20 university students diagnosed with ADHD, followed by interviews with five experts specialized in ADHD intervention. 
Adopting a metacognitive lens, we examined participants' ideas on GenAI-based task management support and experts' assessments, which led to three design directions: providing cognitive scaffolding to enhance task and self-awareness, promoting reflective task execution for building metacognitive abilities, and facilitating emotional regulation to sustain task engagement. 
Drawing on these findings, we discuss opportunities for GenAI to support the metacognitive needs of neurodivergent populations, offering future directions for both research and practice.
\end{abstract}




\maketitle

\section{Introduction}
Attention-Deficit/Hyperactivity Disorder (ADHD) is a neurodevelopmental disorder characterized by inattention, hyperactivity, and impulsivity~\cite{american2013diagnostic, nih_adhd}. Recent studies estimate that at least 3--7\% of adults worldwide are affected by ADHD~\cite{song2021prevalence}, and this number could be largely underestimated due to variations in symptom presentation, underdiagnosis, and stigma~\cite{ginsberg2014underdiagnosis, adamis2022adhd, quinn2004perceptions}. In particular, for university students who are transitioning to adulthood and independent living, having ADHD can impose unique challenges in managing multiple academic responsibilities~\cite{weyandt2006adhd, weyandt2008adhd}. They often struggle to organize tasks, stay focused, and form consistent daily routines, which can lead to incomplete or rushed assignments~\cite{kofler2018working, mirsky1999model,roberts2012constraints}, and in worse cases, result in suspension or dropout~\cite{dupaul2009college,weyandt2006adhd,blase2009self,wolf2001college}.
% can negatively affect their academic performance and mental well-being~\cite{henning2022adhd,weyandt2013performance},

%While there have been a wide range of technologies aimed to support task management, such as to-do lists (e.g., Todoist ~\cite{todoist}), focus reminders (e.g., Forest~\cite{forest_app}), and time monitoring apps (e.g., RescueTime~\cite{rescuetime}), they are not specifically designed for neurodivergent individuals, such as those with ADHD. As shown in previous studies, existing task management tools were largely inadequate in managing the tendency of inattention and hyperactivity of individuals with ADHD~\cite{desrochers2019evaluation, hollis2017annual, hernandez2025decade}. 
%As suggested by Desrocher et al., although individuals with ADHD use digital task management tools in their daily lives, they generally found the tools not effective~\cite{desrochers2019evaluation}. In their review of ADHD apps over the past decade, Hernández et al. argue that apps focused on time management training are underrepresented in the current landscape of ADHD applications \cite{hernandez2025decade}. 
%Partly, these generic reminders and planners are unable to adapt to individuals' attention status and task progress~\cite{nordby2022effect}; also, they are not attractive enough for this special group who tend to be easily distracted~\cite{dovis2012can, powell2017attention}.

Research has suggested that the struggles faced by individuals with ADHD, such as frequent distraction, forgetfulness, and procrastination, stem in part from challenges of \textbf{\textit{metacognition}}, which involve reduced awareness and poor control of one's own cognitive processes~\cite{butzbach2021metacognition, fernie2016contribution, de2017decisional}. %\modify{Beyond cognitive processes, metacognition also relates to awareness of bodily states and emotions, previous research reporting challenges in interoception and alexithymia among individuals with ADHD~\cite{kutscheidt2019interoceptive, roshani2017comparison}}. 
This line of research has shown that enhancing metacognition---the knowledge of one's cognitive and learning process---has therapeutic effects on ADHD populations, as it targets cultivating awareness of one's own attentions and thought patterns~\cite{lenartowicz2024training, kajka2023assessment}.
As such, researchers have developed metacognitive interventions or therapy designed to improve self-awareness and self-monitoring abilities of individuals with ADHD~\cite{re2015effect, mohammadi2022effectiveness, solanto2010efficacy, kajka2023assessment}. These interventions are not merely simple reminders or prompts, but carefully-designed strategies that promote deep reflection and self-assessment, such as mind maps and sketch-noting, which have proven to be effective in helping individuals with ADHD synthesize information and make plans~\cite{kajka2023assessment, solanto2010efficacy}.
%However, within the computer engineering and HCI communities, research exploring how technology can actively support metacognitive processes for task management--especially for ADHD populations---remains limited.

At the same time, advances in technology open up new possibilities for scaffolding the metacognitive process. In particular, the recent surge of Generative Artificial Intelligence (GenAI) technologies (e.g., GPT-5 and Gemini 2.5 Pro) has been rapidly integrated into everyday productivity tools~\cite{chatgpt2024, deepmind2025gemini2.5pro}. With the advanced capabilities in information understanding and generation, these technologies have been applied from work scheduling~\cite{Grover2020VirtualAgent} to focus companion~\cite{li2024stayfocused, de2024dialogues}, and from creative writing assistance~\cite{qin2024charactermeet} to programming toolkits~\cite{imai2022github}. 
Despite the prevalence, most of these tools were designed for neurotypical populations, and little is known about how individuals with ADHD can benefit from them and what aspects of their needs are yet to be addressed from a metacognitive perspective. 
In this light, our research question is: \textbf{\textit{how GenAI can be utilized to support the metacognitive process of individuals with ADHD in task management}}. University students, who face both academic and developmental transitions, represent a particularly important group for such inquiry. 

To answer the research question, we conducted a series of individual co-design workshops with 20 university students diagnosed with ADHD and interviewed five ADHD experts (coaches or intervention specialists) to evaluate and provide feedback on participants' design ideas. 
%As highlighted by Spiel et al., there is limited research that directly engages individuals with ADHD in co-creating technology solutions tailored to their unique needs and experiences~\cite{spiel2022adhd}. Thus, we adopted a co-design approach by actively engaging the target users as co-creators, aiming to gain an in-depth understanding of their task management challenges and desires for support. 
Drawing from both the co-design sessions and expert interviews, we identified the key challenges that participants encountered in task management in relation to their metacognitive process that spans metacognitive \textit{knowledge} (judging task demands and one’s capacity), \textit{abilities} (monitoring and control), and \textit{emotion} (awareness and regulation).  
%lacking awareness of task sources, time needed, and priority; task initiation barriers in front of information overload, limited motivation, and perfectionism; attention control difficulty with distractions, hyperfocus, and unexpected changes; and emotional barriers that often led to task avoidance. 
To address these challenges, participants created designs that leverage GenAI to integrate fragmented task information, calibrate time allocation, collaboratively decompose tasks, and serve as study companions, etc. While acknowledging the values of these designs, experts pointed their potential risks regarding cognition outsourcing and unhealthy dependence. Synthesizing both perspectives, we distilled three design directions: providing cognitive scaffolding to enhance self and task awareness, promoting reflective task execution for building metacognitive abilities, and facilitating emotional regulation for sustaining task engagement. 

Building on the findings, we discuss how GenAI can be best utilized to enhance metacognition in task management for individuals with ADHD. Specifically, we call for promoting reflection rather than fully automating tasks, balancing intrinsic interests with external constraints, and encouraging emotional growth rather than fostering reliance. Thus, the contributions of this work include: (1) an empirical understanding of the task management challenges faced by university students with ADHD from a metacognitive perspective; (2) design implications for GenAI to scaffold metacognition in supporting task management, grounded in students' lived experience and expert insights; and (3) a practical research agenda for leveraging GenAI to support metacognition in the task management for neurodivergent populations.

\label{intro}

\section{Related Work}
In this section, we first cover related work on ADHD among university students, the unique productivity challenges they face, and existing approaches to support their productivity. We then introduce the concept of metacognition and research on metacognitive developments in ADHD populations, along with related interventions. Next, we discuss emerging opportunities for GenAI to support productivity by enhancing metacognitive knowledge and skills.
%both the lens of broader cognitive perspectives and metacognition.

% merge some content from 2.2 here
% ADHD Among University Students and Existing Support (non-technology and technology)
\subsection{ADHD Among University Students}~\label{adhd}
Attention-deficit/hyperactivity disorder (ADHD) is a neurodevelopmental disorder marked by persistent difficulties with attention, excessive activity, and impulsive behavior \cite{nhsadhd, american2013diagnostic, rosqvist2020neurodiversity, CDC_ADHD_Symptoms}. These symptoms pose barriers for individuals to effectively manage their attention and emotions, 
%and their characteristically reduced working memory capacity further contributes to difficulties in organizing and executing 
often making them struggle when handling complex, multi-step tasks~\cite{niermann2014relation,palmini2008professionally, kofler2018working}. For university students transitioning from adolescence to adulthood, the need to balance independent living with academic workloads can be particularly challenging for those with ADHD~\cite{meaux2009adhd, kwon2018difficulties}. In particular, this group usually faces pressure to meet competing deadlines and manage multiple schedules without the structured support systems they had in their earlier life \cite{meaux2009adhd, kwon2018difficulties}. Research indicates that these students face a heightened risk of academic underperformance, poor time management, and emotional distress \cite{sedgwick2018university, weyandt2006adhd}. Furthermore, ADHD often coexists with other neurodivergences that also affect learning (e.g., dyslexia and dysgraphia), as well as mental health conditions (e.g., anxiety and depression), compounding its impact on students' productivity and well-being \cite{kirino2015sociodemographics, kessler2006prevalence,spencer2006adhd}.

To accommodate the unique learning challenges of students with ADHD, many institutions have implemented accommodations (e.g., extra exam time), learning assistance (e.g., note-taking services)~\cite{wolf2001college, meaux2009adhd,sedgwick2018university}, and assistive technologies (e.g., colour-coding for watched and unwatched videos and checklists for progress tracking)~\cite{tcherdakoff2025burnout}. These actions were designed to alleviate students' anxiety and perceived academic pressure while reducing cognitive load. At the same time, the institutions may provide counseling support or behavioral interventions (e.g., Cognitive Behavioral Therapy, ADHD coaching) to those with special learning needs~\cite{antshel2014cognitive, boland2020literature, shrestha2020non, spiel2022adhd}. For students with more severe conditions, they often need to rely on medications~\cite{boland2020literature}.
\subsection{Metacognition and Metacognitive Developments in ADHD Populations}

% What is metacognition, and how does it relate to ADHD
Metacognition was coined by John Flavell in the late 70s while understanding children's cognitive processes~\cite {flavell1976metacognitive}, referring to ``\textit{one’s knowledge concerning their own cognitive processes and products or anything related}''~\cite{flavell1976metacognitive, wellman1990child}. In later research, metacognition also appears to play an important role in several problem-solving and learning contexts among adults~\cite{swanson1990influence, ertmer1996expert, schraw1998promoting}. For instance, Swanson characterized metacognition as individuals' awareness of their capacity to monitor, regulate, and control their own activities~\cite{swanson1990influence}. While the operation of metacognition may vary by context, researchers are in consensus that it typically has four components: \textit{metacognitive knowledge}, \textit{metacognitive experience}, \textit{goals} (or tasks), and \textit{actions} (or strategies)~\cite{ertmer1996expert, schraw1998promoting, flavell1979metacognition}. Specifically, metacognitive knowledge is about one's awareness of their abilities and surroundings; metacognitive experience refers to one's subjective feelings and implicit cues that inform their cognitive process; goals are the objectives that one aims to complete; and actions involve their plans and behaviours to achieve the goal~\cite{flavell1979metacognition}. Nelson and Narens subsequently elaborated and classified the metacognitive model into meta-level and object-level, explaining the interaction between these two levels through the introduction of monitoring and control~\cite{nelson1990metamemory}. Monitoring involves the passive reception and evaluation of the object-level by the meta-level, allowing individuals to update their understanding of the current context. Control involves active regulation and modification of the object-level by the meta-level, guiding subsequent cognitive actions~\cite{nelson1990metamemory}. Prior work often collectively refers to metacognitive monitoring and metacognitive control as metacognitive abilities \cite{tankelevitch2024metacognitive, tarricone2011taxonomy, tobias1996assessing}.

%a set of high-order cognitive abilities (e.g., working memory, inhibitory control, and cognitive flexibility) that enable people to achieve goals and adapt to everyday contexts \cite{cristofori2019executive}. HCI researchers have extensively researched individuals with ADHD through the lens of EF, including tangible tools that scaffold children’s planning and organization in daily routines~\cite{weisberg2014tangiplan}, mobile apps that promote independence at home by supporting transitional moments (e.g., morning and bedtime)~\cite{sonne2016changing}, and wearable devices that support self-regulation through timely prompts while balancing autonomy and caregiver oversight~\cite{cibrian2020supporting}. Prior work has also highlighted a close relationship between executive functioning and metacognition. For example, Fernandez-Duque et al. characterize metacognition through executive control processes, including selective attention, conflict monitoring, error detection, and inhibitory control~\cite{fernandez2000executive}. However, a key distinction is that EF describes the capacities that carry out goal-directed actions, whereas metacognition encompasses the reflective awareness and monitoring of these processes~\cite{cristofori2019executive, swanson1990influence}. In this study, we focused on the latter to complement the existing EF-centric research, aiming to explore how individuals with ADHD can be empowered to strategically allocate and dynamically adjust their limited cognitive resources in complex academic contexts.}

A related term, executive functioning (EF), is often mixed with metacognition~\cite{diamond2013executive}. %EF refers to direct management and control of one's cognitive process or behaviors~\cite{diamond2013executive}. 
Although similar to metacognition in its monitoring of cognitive activity, EF differs by involving direct control and inhibition; its theoretical frameworks emphasize the ability to manage impulses and attention to override internal predispositions or external distractions~\cite{diamond2013executive, roebers2017executive}. 
For instance, in the context of monitoring excessive smartphone use, metacognition might involve understanding one's own smartphone use habits (awareness)~\cite{casale2021systematic}, while EF may focus on the skill to suppress the urge to use smartphones (control/inhibition)~\cite{rosen2017role}. HCI researchers have extensively researched individuals with ADHD through the lens of EF, such as supporting attention control and self-regulated learning~\cite{cibrian2020supporting, sonne2016changing}. While this line of research has contributed to the underlying control mechanisms that individuals with ADHD struggle with, we adopted the metacognitive lens because it allows us to understand this group from an underexplored perspective, such as the awareness, monitoring, and evaluative processes that drive their thoughts and behaviors.

%of a cognitive enterprise; and actions are the cognitions or other behaviors employed to achieve the goal~\cite{flavell1979metacognition}.
%Metacognitive knowledge includes people's awareness about their own abilities (self-awareness), the available information for the task (task awareness), and the strategies to achieve the goals (strategy awareness) \cite{flavell1976metacognitive}. Metacognitive regulation encompasses three processes: \textit{planning}, which refers to the selection of proper strategies and the provision of resources effective for reaching goals, \textit{monitoring} that refers to the critical analysis of the effectiveness of the strategies being implemented and \textit{evaluation} refers to the examination of process being made towards goals which can trigger further planning, monitoring, and evaluation~\cite{jacobs1987children, schraw1998promoting, schraw1995metacognitive}.

%Metacognition plays a crucial role in an individual's learning and cognitive development processes \cite{vos2004developing}. 
On the one hand, research has shown individuals with ADHD experience challenges in metacognition, characterized by limited awareness and poor control of their own cognitive process~\cite{butzbach2021metacognition}, including both bodily (interoception) and emotional  (alexithymia) awareness~\cite{kutscheidt2019interoceptive, roshani2017comparison}, which manifest as productivity challenges such as procrastination, difficulties in task organization, time management, and attention monitoring~\cite{alderson2013attention, sluiter2020exploring, solanto2010efficacy}. As a result, this group often finds it challenging to stay focused and struggles to develop effective plans to achieve their goals~\cite{butzbach2021metacognition}. In response, researchers have integrated metacognitive interventions into ADHD treatment. For example, Re et al. conducted a group training program, where children with ADHD were taught about strategies to maintain attention and self-control, and to understand the time required to complete an activity, which effectively improved their attention control and working memory, and reduced impulsive behaviors~\cite{re2015effect}. Darehshoori Mohammadi et al. demonstrated that metacognitive therapy, including mindfulness, attention training techniques, and regulatory skills training, successfully alleviated behavioral problems arising from emotional dysregulation in children with ADHD and improved their ability to regulate and manage emotions through cognitive processes \cite{mohammadi2022effectiveness}. There is also metacognitive therapy specifically designed for adults with ADHD to improve their time management ability, which involves contingent self-rewards for completing aversive tasks, decomposing complex tasks into manageable steps, and sustaining motivation by visualizing long-term rewards~\cite{solanto2010efficacy}. 
On the other hand, despite the effectiveness of these metacognitive interventions, there has been limited research examining how these strategies can be operationalized in technology design, particularly with the recent surge of GenAI as part of our everyday productivity life.

\subsection{Productivity Support Technologies for ADHD and Opportunities for GenAI}
% Recent advancements in GenAI, such as BERT and GPT, have expanded the horizons of mental health research \cite{devlin2018bert, radford2018improving}. GenAI have robust natural language understanding capabilities and are particularly adept at processing unstructured text data, such as descriptions and dialogues. GenAI acquires knowledge through extensive pre-training on large-scale data, in contrast to traditional machine learning models trained on specific datasets, providing greater flexibility and generalization capabilities. Moreover, GenAIs can potentially reduce biases arising from human-defined rules compared to rule-based applications.

% Recent studies have shown that GenAIs perform well in mental health-related tasks, such as emotional recognition, emotional reasoning, and the detection of mental health conditions (including stress, depression, and suicidal tendencies), while also exhibiting robustness against noisy data \cite{amin2023will,lamichhane2023evaluation,yang2023evaluations}. Furthermore, their effectiveness in specific tasks can be substantially improved through instruction fine-tuning and training customized for particular downstream applications \cite{xu2024mental, yang2023evaluations}.

Researchers have developed assistive technologies to enhance productivity for individuals with ADHD, ranging from supporting routine formation, behavior monitoring, and emotion regulation. For example, Barriga used mobile phones to help individuals establish structured routines and manage physical activities \cite{barriga2023design}, and Moroyoqui et al. leverage gesture and voice input on smartwatches to simplify task creation and daily activities tracking~\cite{moroyoqui2022smartasko}. In another case, DePrenger et al. designed smart pens with accelerometers to track individuals' attention during reading, sending alerts when prolonged pauses to prompt users to refocus \cite{deprenger2010feasibility}. Similarly, Wills and Mason designed a self-monitoring application and sent regular prompts (e.g., ``\textit{Are you on task?}'') to encourage reflection and improve task engagement in children with ADHD~\cite{wills2014implementation}. Some studies use biofeedback technologies to manage ADHD symptoms, such as DEEP, which employs a stretch sensor belt to help children relax during class, while Jiang and Johnstone used electroencephalogram (EEG) devices to enhance individuals' behavioral control through interactive computer games \cite{van2016deep, jiang2015preliminary}.

%At the same time, Generative AI (GenAI) refers to a series of artificial intelligence models trained on large datasets to generate new content, such as text, images, audio, or videos, by learning patterns and structures from existing data~\cite{feuerriegel2024generative}. Different from traditional AI models, which focus on tasks such as classification, regression, and clustering, GenAI leverages large-scale transformer architectures and unsupervised pretraining on vast corpora and task-specific fine-tuning, enabling the generation of creative, contextually appropriate, and human-like outputs \cite{vaswani2017attention,feuerriegel2024generative}. Recent advancements in GenAI, such as GPT-4 and Claude Sonnet 4, have significantly enhanced individual productivity by transforming workflows and unlocking new possibilities across various domains \cite{openai2025gpt5,anthropic2025claude4}. 

In recent years, the emergence of Generative AI (GenAI) has enabled new forms of productivity support. %such as handling repetitive tasks, processing and summarizing information, assisting in generating ideas, and supporting sustained attention. 
One of the most prominent strengths of GenAI is its ability to handle repetitive tasks, such as composing emails, developing and debugging code, and managing routine customer inquiries \cite{thiergart2021understanding, ulfsnes2024transforming, larsenllm}. In addition, GenAI has proven effective in processing large datasets and summarizing complex information. In patent applications, it can reduce human error and enhance efficiency by analyzing patent claims and abstracts \cite{harjamaki2024report,jiang2024artificial}. Furthermore, GenAI has been shown to foster creativity---studies have shown that ChatGPT can help UX designers discover and define stages by simulating stakeholders and building user profiles, thereby inspiring and shaping innovative design ideas \cite{zhou2024exploring}. Researchers also explored the use of GenAI to help people resist distractions during work hours. For instance, Li and Liang et al. built StayFocused powered by GPT-3, which prompts university students to reflect on their compulsive smartphone use in natural conversations, and was shown to be promising in reducing screen time over a long period of time~\cite{li2024stayfocused}. %While these techniques enhance task-level efficiency by supporting cognitive execution, they are not specifically designed to address the challenges faced by the ADHD populations.
%often overlook the unique difficulties faced by individuals with cognitive impairments, including those with ADHD.

%As aforementioned, individuals with ADHD face unique metacognitive challenges that can significantly impact their daily productivity, especially in activities such as planning (e.g., managing multiple deadlines), self-monitoring (e.g., tracking task progress), and self-regulation (e.g., regulating emotions during stressful situations)~\cite{roberts2012constraints, bunford2015adhd, mette2023time}. These challenges are particularly pronounced in academic environments, where students are required to manage a high volume of tasks and deadlines. 
%While recent research has explored the use of GenAI for the ADHD community, the focus has primarily been on diagnosis and therapeutic support, rather than enhancing their overall productivity~\cite{berrezueta2024future, berrezueta2024exploring, gargari2024diagnostic}. Together with recent studies highlighting the potential of GenAI to assist individuals' metacognition in learning, 

At the same time, researchers have been studying how to enhance individuals' metacognitive skills in GenAI environments~\cite {xu2025enhancing, xu2025enhancing, zhang2025impact}. For example, Xu et al. developed a metacognitive support framework based on the self-regulated learning (SRL) model, which prompted college students to make plans for using GenAI, clarify their prior knowledge, and summarize the concept learned from GenAI, which effectively improved their task strategies and self-evaluation abilities~\cite{xu2025enhancing}. Lin et al. identified several AI-assisted metacognitive strategies in academic reading, including goal setting, planning, and critical thinking~\cite{lin2024genai}. 
%can foster readers’ positive emotions \cite{lin2024genai}. 
In addition, Zhang and Wang found that GenAI can help learners develop stronger analytical thinking in high-level knowledge building, such as proposing research designs based on what they have learned~\cite{zhang2025impact}. 
Meanwhile, researchers recognized the risks of leveraging GenAI for work and study, which may inhibit the metacognitive process as individuals outsource their problems~\cite{li2025generative, ligenai2025, zhang2025impact}. Over time, individuals may develop reliance on GenAI, which could erode their ability to assess progress, detect errors, or adapt strategies~\cite{li2025generative}.
%While these studies highlight GenAI’s potential to scaffold metacognition in general productivity and learning contexts, extending such support to neurodivergent populations remains underexplored, despite they often experience persistent metacognitive challenges in monitoring and regulating their productivity processes. 

These tensions highlight the double-edged nature of GenAI: \textit{while it can scaffold metacognition, it can also undermine the metacognitive process if not carefully designed}.
In light of both opportunities and challenges, our study aims to investigate how GenAI can be best leveraged to help university students with ADHD overcome metacognitive challenges in academic task management. As the first step, we set out to understand the perspectives of both the students with ADHD and experts who are experienced in supporting this group.

\section{Method}
%To explore opportunities for GenAI to support the productivity of university students with ADHD, 

We conducted individual co-design sessions with 20 university students diagnosed with ADHD to examine their unique needs for academic task management support. To further understand the feasibility and potential usage of their design ideas, we conducted in-depth interviews with five experts (intervention specialists and coaches) who have extensive experience treating individuals with ADHD. Their feedback helped us understand participants' design ideas from feasibility and clinical relevance perspectives, as well as identify potential gaps or misconceptions in participants' task management approaches.
%Additionally, given that individuals with ADHD can be easily distracted~\cite{american2013diagnostic}, we aimed to create an environment with minimized distraction. Thus, individual sessions would allow participants to concentrate on their challenges and preferences. 
All the co-design and interview sessions were led by the first author. The study was approved by the author's institutional ethics review committee.

\subsection{Participants}

\setlength{\tabcolsep}{1.5pt}
{\sffamily
\begin{table*}[tbp]
\small
\centering
\caption{Participants' demographic information.}
\label{participants_demographics}

\begin{minipage}{0.95\textwidth}
\centering

\begin{tabular}{l@{\hspace{4pt}} l l l p{3cm} p{4cm} p{4.5cm}}
\toprule
\textbf{ID} &
\makecell[l]{\textbf{Year of}\\\textbf{Study}} &
\makecell[l]{\textbf{Gender/}\\\textbf{Age}} &
\makecell[l]{\textbf{Time Since}\\\textbf{Diagnosis}} &
\makecell[l]{\textbf{GenAI}\\\textbf{Experience}$^{\dagger}$} &
\textbf{Major} &
\makecell[l]{\textbf{Co-occurring Conditions} \\ \textbf{\& Medication Status}} \\
\midrule
P1 & 2nd & M/19 & 1-2 yrs & Daily use & Public Administration & Depression, autism spectrum disorder \\
P2 & 2nd & F/20 & 1-2 yrs & Occasional use & French & — \\
P3 & 4th & F/22 & 1-2 yrs & Regular use & Industrial Engineering & Medicating \\
P4 & 3rd & F/21 & < 1 yr & No experience & Food and Nutrition & Medicating, with depression \\
P5 & 2nd & F/20 & 1-2 yrs & Regular use & Chinese Language and Literature & Bipolar disorder \\
P6 & 4th & F/22 & 1-2 yrs & Regular use & Medical Information Engineering & Bipolar disorder \\
P7 & 4th & F/22 & < 1 yr & Regular use & History & — \\
P8 & 4th & M/22 & < 1 yr & Regular use & Molecular Biology & Medicating \\
P9 & 4th & M/24 & 6-10 yrs & \makecell[l]{Daily use with \\development experience} & Business English & Medicating \\
P10 & 3rd & F/21 & 1-2 yrs & Daily use & International Chinese Education & — \\
P11 & 4th & M/22 & < 1 yr & Daily use & Electronic Information Engineering & — \\
P12 & 4th & M/22 & 6-10 yrs & Regular use & Digital Media Technology & Medicating \\
P13 & 4th & M/23 & < 1 yr & Regular use & Industrial Design & Medicating \\
P14 & 4th & M/23 & 3-5 yrs & Regular use & Physics & \makecell[l]{Medicating, with anxiety and \\depression} \\
P15 & 3rd & F/21 & 3-5 yrs & Regular use & Digital Media Technology & \makecell[l]{Medicating, with obsessive-\\compulsive disorder} \\
P16 & 3rd & F/23 & 3-5 yrs & Daily use & Computer Science & \makecell[l]{Medicating, with anxiety and \\ bipolar disorder} \\
P17 & 2nd & M/20 & > 10 yrs & Regular use & International Business & — \\
P18 & 2nd & M/19 & 1-2 yrs & Daily use & Internet of Things Engineering & Medicating \\
P19 & 3rd & F/21 & < 1 yr & Regular use & Pharmaceutical Engineering & Autism spectrum disorder \\
P20 & 4th & F/22 & 1-2 yrs & Occasional use & Philosophy & — \\
\bottomrule
\end{tabular}

\vspace{3pt}
\parbox{0.95\textwidth}{
\footnotesize
\raggedright
\textit{Note.} $^{\dagger}$ The GenAI Experience (shared by
Tables~\ref{participants_demographics} and~\ref{expert_demographics}) describe participants’ frequency and manner of using GenAI tools. \textit{No experience} indicates that the participant has never used GenAI tools. \textit{Occasional use} refers to infrequent, task-specific engagement without a
consistent usage pattern. \textit{Regular use} describes interaction approximately one to three times per week. \textit{Daily use} denotes near-daily engagement, whereas \textit{Daily use with development experience} additionally indicates experience in developing, fine-tuning, or integrating GenAI applications.
}

\end{minipage}
\end{table*}
}

\setlength{\tabcolsep}{2pt} 
{\sffamily
\begin{table*}[tbp]
\small
\centering
\caption{Expert participants' demographic information.}
\label{expert_demographics}

\begin{minipage}{0.95\textwidth}
\centering

\begin{tabular*}{0.95\textwidth}{@{\extracolsep{\fill}} l l p{2.6cm} p{3.5cm} l p{5.5cm}}
\toprule
\textbf{ID} & 
\makecell[l]{\textbf{Gender/}\\\textbf{Age}} & 
\makecell[l]{\textbf{ADHD Intervention}\\\textbf{Experience}} & 
\textbf{Professional Role} & 
\makecell[l]{\textbf{GenAI}\\\textbf{Experience}$^{\dagger}$} & 
\textbf{Target treatment group} \\ 
\midrule
E1 & F/24 & Three months & ADHD Intervention Specialist & Regular use & Adolescents and young adults in early careers \\
E2 & F/47 & Over a year & ADHD Coach & Regular use & University students and working adults \\
E3 & F/33 & Over three years & ADHD Intervention Specialist & Regular use & Children, adolescents, and university students \\
E4 & F/34 & Over four years & ADHD Intervention Specialist & Regular use & Children and adolescents \\
E5 & F/40 & Over a year & ADHD Coach & Daily use & Children and parents \\
\bottomrule
\end{tabular*}

\end{minipage}
\end{table*}
}

% \setlength{\tabcolsep}{2pt} 
% {\sffamily
% \begin{table*}[tbp]
% \small
% \centering
% \caption{Expert participants' demographic information.}
% \label{expert_demographics}

% \begin{tabular}{l l p{2.6cm} p{3.5cm} l p{5.5cm}}
% \toprule
% \textbf{ID} & 
% \makecell[l]{\textbf{Gender/}\\\textbf{Age}} & 
% \makecell[l]{\textbf{ADHD Intervention}\\\textbf{Experience}} & 
% \textbf{Professional Role} & 
% \makecell[l]{\textbf{GenAI}\\\textbf{Experience}$^{\dagger}$} & 
% \textbf{Target treatment group} \\ 
% \midrule
% E1 & F/24 & Three months & ADHD Intervention Specialist & \makecell[l]{Regular use} & Adolescents and young adults in early careers \\
% E2 & F/47 & Over a year & ADHD Coach & \makecell[l]{Regular use} & University students and working adults \\
% E3 & F/33 & Over three years & ADHD Intervention Specialist & \makecell[l]{Regular use} & Children, adolescents, and university students \\
% E4 & F/34 & Over four years & ADHD Intervention Specialist & \makecell[l]{Regular use} & Children and adolescents \\
% E5 & F/40 & Over a year & ADHD Coach & \makecell[l]{Daily use} & Children and parents home-based intervention training \\
% \bottomrule
% \end{tabular}

% \end{table*}
% }

\subsubsection{University Students with ADHD}
We recruited participants through social media platforms, including RedNote, Douban, and QQ groups, which are popular social media platforms in China. Among the 52 responses we received, 46 people met the following inclusion criteria: individuals who (1) were at least 18 years old and currently enrolled as undergraduate students at a university; (2) had been diagnosed with ADHD and could provide medical evidence (e.g., doctor's note and/or prescription) before participating in the study; (3) self-reported experiencing task management challenges in academia and were interested in improving their academic task management with the help of technology; and (4) had access to computer or tablet for participating in a remote co-design study. 
From the pool of eligible participants, we ensured a balanced gender representation when selecting individuals for the co-design sessions. We continued recruiting participants alongside data analysis until data saturation was reached, meaning no new information or insights emerged regarding the challenges participants mentioned or their proposed new design ideas \cite{fusch2015we}, at which point the final number of participants was determined. Eventually, as shown in Table \ref{participants_demographics}, 20 university students participated in and completed the co-design sessions, including 11 females and 9 males, with ages ranging from 19 to 24 (\textit{M} = 21.4, \textit{SD} = 1.31). All participants were Chinese but residing in different countries and regions: 17 participants were located in Mainland China, and the other three participants resided in South Korea (P4), Australia (P16), and Canada (P8), respectively. Most participants used ChatGPT, while some also used ERNIE Bot (P5, P6, P14), Claude (P16), or other tools like Replika (P5), Midjourney (P5), and Pi (P5) \footnote{Table~\ref{participants_genai_usage} in the Appendix provides detailed information on the GenAI tools participants used and the purposes they used these tools.}. Each participant received 100 RMB as a token of appreciation for their participation. 

\subsubsection{ADHD Experts}
Following the completion of all co‑design sessions and the analysis of participants’ design ideas, we invited experts in ADHD to evaluate the proposed designs. We recruited experts through RedNote based on the following inclusion criteria: (1) currently working as coaches or intervention specialists for individuals with ADHD, (2) at least three months of experience in providing support to individuals with ADHD, and (3) having access to a computer or tablet to participate in a remote interview. Ultimately, five ADHD experts participated in our individual, interview-based evaluation sessions (as shown in Table \ref{expert_demographics}), including three ADHD Intervention Specialists (practitioners who deliver structured, skills-oriented interventions for people with ADHD) and two ADHD Coaches (professionals who provide practical, day‑to‑day support for people with ADHD and are certified in ADHD coaching), aged between 24 and 47 (\textit{M} = 35.6, \textit{SD} = 7.66). All ADHD experts were located in Mainland China, and each was compensated with 250 RMB as a token of appreciation for their participation.

\begin{figure*}[t]
    \centering
    \includegraphics[width=\linewidth]{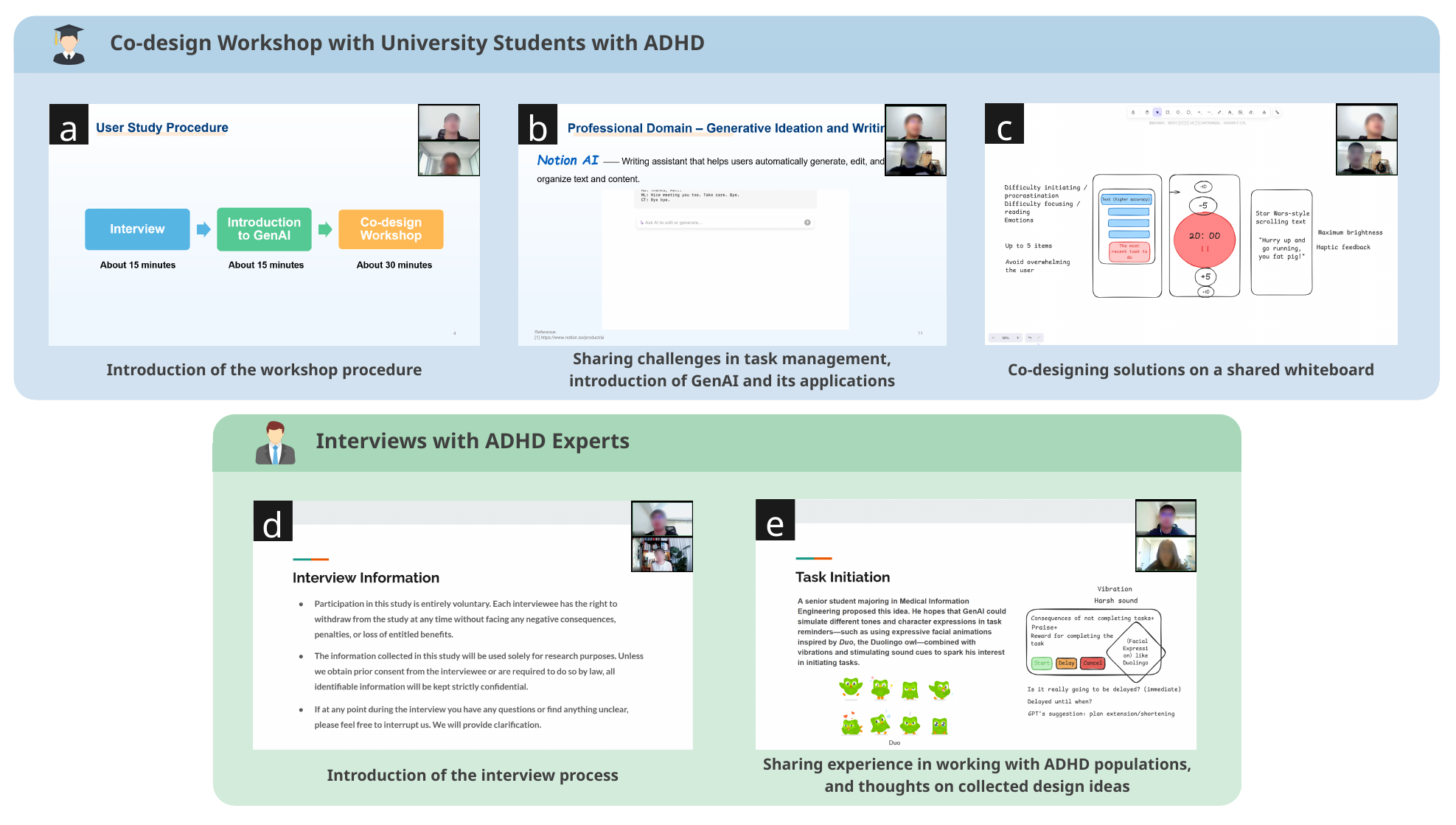}
    \Description{A methodology overview diagram showing the study process, including an individual co-design workshop with university students with ADHD and interviews with ADHD experts. The diagram depicts the main stages of the co-design workshop and the expert interviews, illustrating how participant insights and expert feedback informed the exploration of generative AI design ideas.}
    \caption{Overview of the individual co-design workshop with university students with ADHD (top) and interviews with ADHD experts (bottom). The individual co-design workshop process included (a) the introduction of the workshop process, (b) an interview about the daily academic task management challenges faced by participants, and (c) the co-design of GenAI solutions with the participants on an online collaborative whiteboard. The interview process with ADHD experts included (d) the introduction of the interview process, (e) discussions on the daily task management challenges faced by ADHD university students based on the experts' intervention experience, and the presentation of design ideas from student participants to seek advice and identify potential risks. For ease of reading, the original text in the screenshots has been translated into English and visually enhanced (e.g., larger font size and boldface). The original images are available in the Appendix.}
    \label{fig:method_overview}
\end{figure*}

\subsection{Co-Design Sessions}
Each co-design session started with a warm-up activity for mutual learning where participants shared their task management challenges in academic life, and the researcher facilitator provided an overview of existing GenAI technologies. Next, each participant focused on a specific task management challenge mentioned earlier to create GenAI-powered technology solutions, while thinking aloud their design rationale. Lastly, participants were asked to reflect on their designs and elaborate on additional thoughts regarding how they envision using the support in daily life.
The entire session was conducted remotely via the Tencent Meeting software and lasted between 70 and 100 minutes. All communications were in Mandarin Chinese and later translated into English for the paper writing, and we acknowledge the possibility of minor differences in expression~\cite{sharma2025lost, vigh2025lost}. The upper panel of Figure~\ref{fig:method_overview} depicts representative stages of the co‑design procedure. 

It is noteworthy that we adopted an open-ended co-design stance to minimize framing effects. Rather than instructing participants to design for ``metacognition'' (which could bias ideas toward researcher language and canonical constructs), we invited them to envision any ways GenAI could support their academic task management. This choice preserved ecological validity and allowed needs and strategies to emerge in participants’ own terms, while still enabling us to derive features related to metacognitive strategies afterwards.
% We organized the study into three stages: First, we conducted individual semi-structured interviews to understand the daily task management challenges faced by university students with ADHD. Next, we introduced GenAI to ensure participants had a basic understanding of the technology. Finally, we organized a co-design activity to explore participants' design ideas on how they envision GenAI can assist with their task management challenges. 

\subsubsection{Warm-up Activity}
To contextualize participants within their academic activities, we first invited them to introduce themselves, including their study major and daily routines. We then asked participants to describe how they manage multiple academic tasks, the challenges they faced, and strategies they had employed to navigate these challenges, such as tools or resources they rely on.
%allowing both the participants themselves and the researcher facilitator to gain a better understanding of their current task management situations. 
%Specifically, we asked them about their experiences with ADHD, how they approached managing academic tasks, and the types of support they had explored to improve their academic task management. %These concrete contexts and examples helped us better understand each participant's situation, enabling us to tailor the structure of subsequent design activities accordingly. For example, P4 mentioned that they often arrived late or missed class because they struggled to get out of bed. 
%In subsequent co-design sessions, we guided them intentionally in ideating to address this challenge.
%and what led them to seek a diagnosis. Following this, we asked about the specific task management challenges they face in their daily lives. As participants shared their experiences, we posed additional questions to gain a deeper understanding of the contexts and causes of these challenges. We also asked participants to describe the strategies or tools they currently use to manage these challenges, along with their perspectives on the effectiveness, strengths, and limitations of these approaches.

Next, we provided participants with an overview of GenAI's key capabilities. While most participants were already familiar with GenAI (e.g., ChatGPT~\cite{chatgpt2024}, and ERNIE Bot~\cite{yiyan2024}) and had been actively using it for productivity support, we prepared a shared slide highlighting its ability to understand and generate multimedia content, along with several application areas, ranging from programming assistant (e.g., Copilot~\cite{githubCopilot}), writing and data analysis support (e.g., ChatGPT~\cite{openaiDataAnalysis}), image generation (e.g., DALL·E 3~\cite{openai2023dalle3} and Midjourney~\cite{midjourneyWebsite}), meeting documentation (e.g., Notion AI~\cite{notionAI} and Tencent Meeting AI~\cite{tencentMeetingAI}), and social companion (e.g., Replika~\cite{replikaWebsite} and character.ai~\cite{characterAIWebsite}). 
This brief introduction served two purposes. First, it established a common understanding of GenAI’s functional possibilities, ensuring that all participants had access to the same foundational knowledge regardless of their prior experience. 
Second, by presenting the tools in a neutral, descriptive way rather than prescribing specific uses or solutions, we provided inspiration for participants without biasing their ideas toward particular applications. 
This is a common approach in co-design studies, where researchers present application examples or provide participants with ready-made tools~\cite{qi2025participatory, xue2025toward}, allowing them to quickly become familiar with the target design platform and generate diverse and creative concepts. 
%This approach emphasizes uncovering participants’ underlying needs and motivations, rather than treating their initial ideas as fixed design guidelines.
%for how GenAI could support their academic task management, enabling richer and more grounded co-design outcomes.
%During this process, participants were also free to share their experience with any GenAI-powered applications and how they used them in their daily lives.
% \subsubsection{GenAI Introduction} 
%After the interview, to ensure that users have a foundational understanding of GenAI, we spent about 10 minutes introducing the definition of GenAI and how it can be applied in different areas. We explained how GenAI can be applied in chatbots to customize different roles and communication styles with different prompts. Then, we introduced several applications of GenAI in task management scenarios, including its use in coding, data analysis, writing support, and meeting summaries \cite{githubCopilot, openaiDataAnalysis, notionAI, tencentMeetingAI}. We also discuss how GenAI is used to create emotional companion agents, such as Replika and character.ai \cite{replikaWebsite, characterAIWebsite}, as well as image generation tools like MidJourney and DALL·E 3 \cite{midjourneyWebsite, openai2023dalle3}.

\subsubsection{Design Activity}
Focusing on one or two task management challenges they mentioned in the warm-up activity, we encouraged each participant to envision how GenAI could help address these challenges by illustrating their ideas on Excalidraw~\footnote{https://excalidraw.com}, a collaborative whiteboard that supports commonly used digital tools (e.g., shape frames, text typing, stylus, and image uploading). With the multi-user collaboration feature, Excalidraw allows the researcher to assist participants in creating frames, adding notes, etc. 
% After the introduction of GenAI, we used Excalidraw\footnote{https://excalidraw.com}, an online collaborative whiteboard, to engage participants in co-design activities. We asked the participants to present their design ideas for addressing the daily task management challenges they had previously identified with PCs or tablets with stylus pens. 

During the co-design session, we encouraged participants to think aloud throughout the process, and sometimes prompted them to answer questions such as \textit{``How do you envision this feature to function? What is your motivation to create this design?''} and \textit{``Compared to existing solutions, what do you think makes this design stand out?''} 
%We allowed sufficient time for each participant to complete their designs. 
Although some participants initially hesitated to start due to unfamiliarity with the process and concerns about the quality of their designs, they quickly became engaged once the researcher clarified that our goal was to explore their expectations of how GenAI could assist with task management challenges in their academic life, rather than evaluating the creativity of their designs or the aesthetics of their digital sketches.
%Note that to create a supportive design space, we highlighted that participants could create designs without restriction, as the purpose of the study is to understand their needs rather than evaluate their ideas.

\subsection{Interviews with ADHD Experts}
After organizing the design ideas generated from the above co-design sessions, we interviewed five ADHD experts to understand their perspectives on these design ideas. First, we introduced the research background and goals, and invited the experts to talk about their background and experience in supporting individuals with ADHD, particularly university students. %discuss the challenges ADHD students face in academic settings, particularly in task management, drawing from their intervention experiences. 
Next, we presented the collected design ideas on a shared screen (with notes explaining the corresponding challenges and contexts provided by the participants who created the design) and asked the experts to share their thoughts on these ideas regarding the underlying needs of the creators, feasibility, use cases, and clinical relevance. We also took these opportunities to discuss with the experts on potential pitfalls that might be overlooked by the participants in their design ideas and alternative solutions. On average, each interview lasted 86 minutes. The lower part of Figure~\ref{fig:method_overview} illustrates selected moments from the expert interview sessions.

\subsection{Data and Analysis}
% basic demographic and background information collected during recruitment, 
% The first author conducted the primary analysis, supported by iterative discussions with two other researchers to refine interpretations and enhance reliability.
Our dataset includes participants' design ideas captured on the shared whiteboards, video recordings of co-design sessions (including participants' sharing of everyday task management challenges and ideation processes), and video recordings of interviews with ADHD experts. All video recordings were transcribed into text for analysis. We conducted a bottom-up Thematic Analysis following Braun and Clarke's guidelines to analyze (1) the specific academic task management challenges and (2) design ideas that integrate expert perspectives~\cite{braun2006using}. This process consisted of five steps:

\begin{itemize}
\item[] \textbf{(1) Familiarization with the data}: The first author repeatedly read all transcripts while reviewing the corresponding design artifacts created by participants to gain a comprehensive understanding of participants’ academic task management challenges, design ideas, and how they are related. %When the meaning of a specific excerpt was unclear, we revisited the associated video segment or whiteboard content to ensure grounded interpretation.}

\item[] \textbf{(2) Initial coding of co-design}: The first and the last author independently coded transcripts from three randomly selected co-design sessions following a bottom-up, open-ended approach. We then met to discuss these initial codes, refining the code naming and resolving those with ambiguity or discrepancies. After establishing a consistent interpretation of the data, the first author continued coding more co-design sessions. Through this process, the whole team met regularly to discuss the coded data, merge similar codes, and note how they are related to metacognition (i.e., knowledge, monitoring, and control of one's cognitive activities).  %we developed an initial codebook that guided subsequent analysis. 
In total, we generated 764 initial codes. Examples of these codes covering participants' metacognitive challenges (e.g., \textit{``difficulty distinguishing priorities across different tasks''} and \textit{``struggle with sensing the passage of time during tasks''})), and design ideas (e.g., \textit{``identifying tasks from fragmented chat logs''} and \textit{``assisting students in plan adjustment when task progress does not meet expectations''}). %The first author then coded the remaining co-design sessions following the refined codebook. In total, we generated 764 codes, allowing multiple codes to be assigned to a single excerpt when appropriate.}

\item[] \textbf{(3) Initial coding of expert interviews}: Based on the initial codes generated from the co-design sessions, the first author coded the expert interviews focusing on three aspects: (1) the applicability and practical feasibility of participants' proposed ideas, (2) the key metacognitive knowledge or ability involved, and (3) the edge cases and potential risks that those ADHD participants may overlook. This analysis led to 245 initial codes, which we later integrated with the codes from the co-design sessions. 
%during the expert interview process, bringing the total to 1,009 codes across both the co-design sessions with ADHD participants and the expert interviews.

\item[] \textbf{(4) Integrating codes \& searching themes}: Building upon the initial codes, the whole team collaborated to organize them into a hierarchical structure while searching for predominant themes. Through iterative discussions, it became evident that participants' metacognitive challenges and design ideas are associated with their awareness of daily tasks, strategies to complete these tasks, and assessment of their own capabilities, as well as their ability to monitor the task progress and adaptively adjust strategies. In addition, codes related to emotional barriers (e.g., ``\textit{anxiety caused by continuous procrastination of tasks}'') frequently appeared, which we found also to be a crucial component of metacognition~\cite{spada2008metacognition}. 
As we searched and reviewed the themes, we moved back and forth between raw data and codes to ensure a focused narrative on metacognition, and thus discarded those deemed irrelevant (e.g., using GenAI to generate reading summaries).
%We examined themes related to emotion regulation and organized them as ``emotional barriers to task engagement.'' Some codes considered not related to metacognition were excluded, such as those focused solely on using GenAI to assist with cognitive tasks (e.g., generate reading summaries). Throughout the process, we moved back and forth between raw data, codes, and emerging themes to ensure that only those codes relevant to metacognitive challenges and GenAI support were incorporated into the final thematic groups.} 

\item[] \textbf{(5) Naming themes \& developing narratives}: In the end, our analysis resulted in four themes characterizing participants’ metacognitive challenges, including \textit{(1) Lack of Awareness}, \textit{(2) Barriers to Task initiation}, \textit{(3) Difficulty in Attention Control}, and \textit{(4) Struggles to Regulate Emotions}. Centered on these challenges, we gathered relevant design ideas and developed three themes of GenAI support: \textit{(1) Metacognitive Scaffolding to Enhance Task and Self-Awareness}, \textit{(2) Reflective Task Execution for Developing Metacognitive Abilities}, and \textit{(3) Emotional Regulation for Sustaining Engagement}. 
\end{itemize}

\section{Findings}

\begin{figure*}[t]
    \centering
    \includegraphics[width=\linewidth]{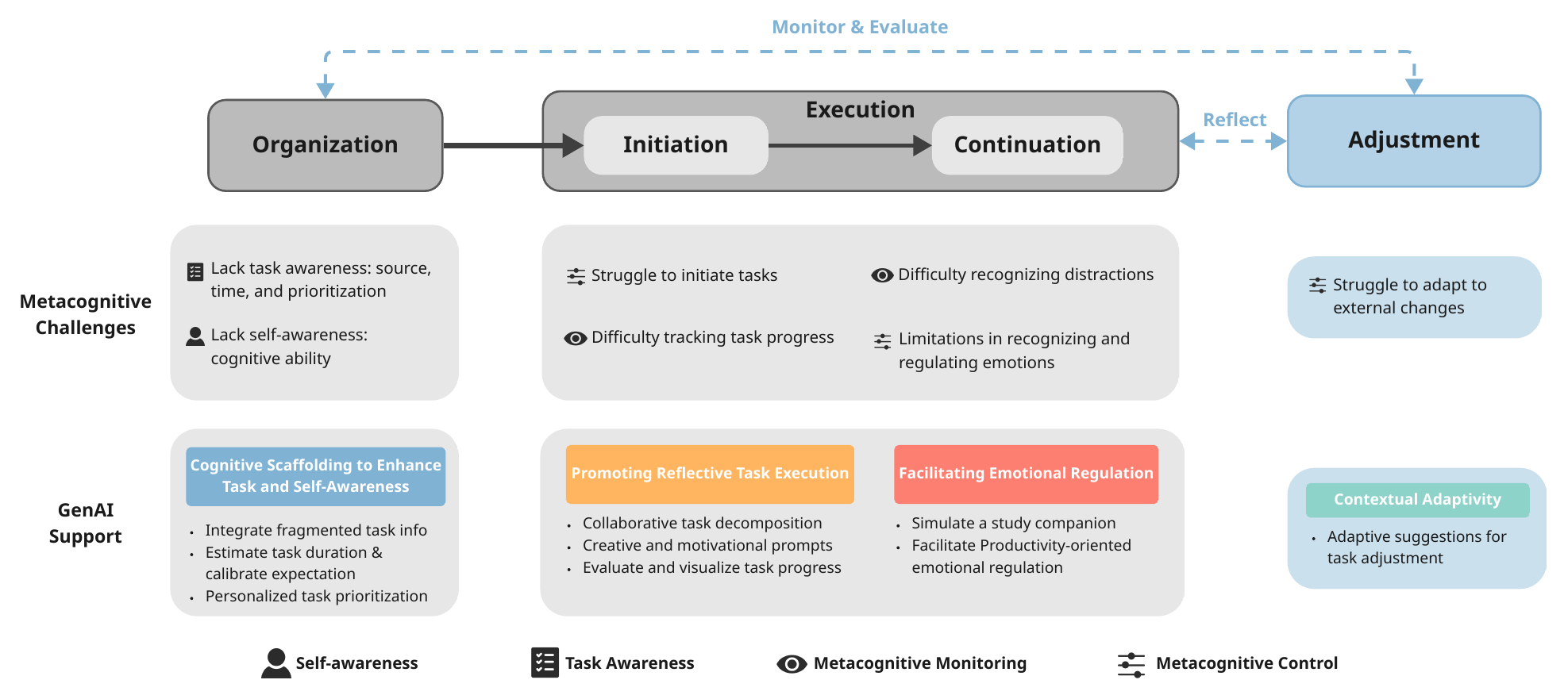}
    \Description{An overview diagram illustrating metacognitive challenges faced by university students with ADHD across three stages of task management—organization, execution, and adjustment—and corresponding opportunities for generative AI support. The diagram highlights challenges in task awareness and self-awareness during organization, limitations in monitoring and control during execution, and difficulties in emotion regulation, along with potential AI-supported scaffolding strategies.}
    \caption{Overview of the metacognitive challenges faced by university students with ADHD across different stages of task management (Organization, Execution, Adjustment), and the opportunities for GenAI to provide support. The metacognitive challenges include a lack of task awareness (source, time, and prioritization) and self-awareness (cognitive ability) during task organization, limited metacognitive monitoring and control (task initiation, attention control, progress tracking, and task adjustment), and difficulties with emotion regulation during task execution. Participant design ideas suggest GenAI could support students in navigating these challenges by scaffolding task awareness and self-awareness, prompting reflection to strengthen monitoring and control, and facilitating productivity-oriented emotional regulation.}
    \label{fig:results_overview}
\end{figure*}

In this section, we first present the themes generated from our analysis of participants' metacognitive challenges---specifically lacking \textit{metacognitive information} (knowledge and experience about the task and oneself) and \textit{metacognitive ability} (monitoring and controlling oneself), and emotional experiences that disrupt metacognitive engagement. Next, we present participants' design ideas of leveraging GenAI to address these challenges, and experts' perspectives.
%and discuss how participants envision GenAI could help them address these challenges from a metacognitive perspective.

\subsection{Metacognitive Challenges in Task Management}
\label{finding-metacognitive-challenge}
%Our analysis reveals that participants’ difficulties with academic task management stemmed from three interconnected metacognitive challenges: what they know (\textit{Metacognitive Knowledge}), how they act (\textit{Metacognitive Ability}), and how they feel (\textit{Emotion}).

Following Flavell's definition of metacognition~\cite{flavell1976metacognitive}, we organized the challenges that participants reported through three aspects of metacognitive functioning: metacognitive knowledge, metacognitive abilities, and emotional barriers. Their limited metacognitive knowledge manifested in three key areas: a lack of awareness of the task's nature (task awareness), strategies to navigate the tasks (strategy awareness), and their own abilities and mental states (self-awareness). Additionally, we observed challenges in metacognitive abilities manifested as difficulties with task initiation, attention control, and adapting to unexpected changes. Participants also described emotional barriers such as disengagement and task avoidance that interacted with and compounded these metacognitive challenges. We elaborate on these findings below.

\subsubsection{\textbf{Lack of Awareness: Source, Time, and Prioritization}}~\label{task-awareness}
First, when confronted with large volumes of fragmented information across multiple sources, such as course syllabi, emails, and messages from instant messaging apps, participants struggled to organize a comprehensive list of tasks they needed to complete or even identify the information related to their tasks (P2, P3, P5, P9). This lack of task clarity often made them unsure about what needed to be done or where to start.
%participants demonstrated a lack of awareness 
In particular, participants faced difficulties in estimating how long a task would take (P3, P7, P8, P9, P12, P18). Similar observations have been reported in prior research, where individuals with ADHD struggle to accurately encode, store, and recall temporal information~\cite{nejati2020time,radonovich2004duration}. In our study, we found this challenge escalated when participants were juggling multiple tasks, affecting their understanding of the importance, urgency, and prerequisites of the tasks. 
%time perception–related tasks (e.g., task time estimation and time reproduction)~\cite{nejati2020time,radonovich2004duration}}. 
Consequently, the most direct influence is their ability to set clear priorities (P7, P15, P16). As P15 described:

\begin{quote}
    ``\textit{I do not have a clear sense of task priority [...] For example, with an assignment I was not interested in, I would end up doing designs that I like on my computer, but never start relevant tasks, like programming}.''
\end{quote}

Moreover, due to a lack of awareness of their own abilities, participants often set overly ambitious goals and misjudge the time needed to complete tasks, resulting in poorly structured plans (P14). This mismatch between expectation and reality created a cycle of inefficiency and frustration.

This finding aligned with experts' perspectives: E1, E2, and E3 all pointed out that many students with ADHD tend to be overly optimistic about the number of tasks they can finish. As noted by E2, the students they had worked with often create lengthy to-do lists, trying to ``\textit{complete everything at once}.'' Such unrealistic expectations reflected participants' limited awareness of their actual capacity to complete a task effectively. 
Furthermore, E2 added, when students attempt to take on too many tasks at once, it not only delays their task completion but also creates unnecessary mental stress.

%This misalignment between self-perception and reality underscores the need for strategies that better support ADHD students in gaining task clarity, realistic planning, and self-regulation.

% \subsubsection{\textbf{Difficulties in Attention Monitoring and Control}}~\label{attention}

% The lack of metacognitive knowledge further affected participants' metacognitive abilities, particularly in monitoring and controlling their ongoing attentions.
%specifically in updating or gaining new understanding of task demands, strategies, and their own abilities (P1, P2, P4, P5, P7, P11, P13, P16). For instance, participants struggled to adapt their approaches as they encountered new challenges ...
%As metacognition is an ongoing process of reflection and adaptation, these challenges then hindered participants' ability to monitor their own and regulate their task progress.

\subsubsection{\textbf{Barriers to Initiation: Task Overload, Lack of Motivation, and Perfectionism}}~\label{initiation}
A recurring challenge across participants was difficulty initiating tasks, which reflects a breakdown in metacognitive control at the very first stage of task management. Many participants described themselves feeling overwhelmed in front of a stack of tasks (P1, P2, P3, P7, P13, P16). Others reported a lack of motivation that prevented them from transitioning from intention to action (P1, P13, P15, P16). 
For some participants, such as P16, this challenge was exacerbated obsessive-compulsive perfectionism, where unrealistic standards of performance created an inflated sense of the gap between their work and the ``ideal'', These accounts highlight how initiation difficulties in ADHD are not only about a lack of motivation but also about maladaptive self-monitoring and evaluation: 

\begin{quote}
    ``\textit{I feel like I need to do something perfectly, and aiming for perfection makes it much harder to get started—not just harder to begin, but harder to carry through. It feels like there are many obstacles I have to overcome before I can even begin}.''
\end{quote}

To overcome these barriers, participants frequently turned to external scaffolds that could help them initiate work. A common approach was the use of alarm apps (e.g., Pomodoro timer) to create a structured sense of urgency and break large tasks into smaller intervals (P5, P8, P9, P11, P12, P13, P16, P20). While these reminders provided a temporary motivation boost, their effectiveness diminished over time as participants either habituated to the alerts or began to ignore them. 
% In addition, participants experimented with environmental regulation strategies. For instance, they sought out physical spaces with fewer distractions and a stronger ``study atmosphere,'' such as a library or classrooms (P2, P13, P20).

Additionally, E3 noted that students with ADHD often set tasks that are broad and abstract, such as ``lose 20 pounds'' while failing to include concrete, actionable tasks like ``do 10 jumping jacks on the balcony''. This lack of specificity raises the threshold for task initiation and makes it more difficult to initiate the task (E3).

% Experts further noted that students with ADHD often set overly high standards when making task plans, which not only makes tasks difficult to complete but also creates significant pressure on them during execution and can ultimately lead them to abandon the tasks (E1, E2). 

\subsubsection{\textbf{Attention Control Difficulty: Distraction, Hyperfocus, and Unexpected Changes}}~\label{attention}
Even when they were able to initiate the task, participants frequently encountered difficulties in monitoring and regulating their attention. For some, this meant drifting off and losing focus during activities such as lectures, as P3 described their experience of losing focus during lectures, ``\textit{I can barely focus during lectures, and sometimes I don't even know which chapter the professor is talking about. I just end up pretending to listen}.'' Others reported experiencing the opposite extreme---hyperfocus, in which they became so deeply absorbed that they struggled to disengage, leading them to neglect other important tasks or deadlines (P1, P2, P3, P5, P7, P11, P15, P16, P18, P19, P20). Compounding this challenge, participants emphasized that university life is filled with unexpected disruptions, such as urgent tasks or the extended duration of planned activities. These disruptions often broke their sense of order and created reluctance to adjust their behaviors to adapt to changes, sometimes leading them to give up their plans altogether (P1, P2, P4, P5, P7, P11, P13, P16). Experts echoed this difficulty, noting that while adjusting plans in response to environmental change may seem straightforward for most people, it can be particularly taxing for individuals with ADHD, as the process requires metacognition of re-prioritizing or shifting strategies (E5).

%note that, when implementing behavioral interventions for children with ADHD, to help their children adapt to changes, parents are encouraged to develop multiple plans in advance. For example, they need to inform their children in advance that if it rains on the way to the amusement park, the plan might need to be adjusted to plan B to return home. However, due to the complexity and unpredictability of university students' schedules, this approach is less feasible, as planning in advance becomes challenging.

To cope with attention lapses and disruptions due to unexpected changes, participants experimented with external supports. Some sought out study companions like friends or classmates to create mutual accountability and maintain focus (P1, P2, P5, P11, P13, P18, P19). While this strategy was seen as helpful, it was often constrained by scheduling conflicts or difficulties in finding suitable partners.
%Additionally, when facing disruptions, such as unexpected events or falling behind schedule, participants struggled to adjust their actions to changes in the environment. As a result, they may give up the entire plan 
Others turned to environmental adjustments, such as relocating to libraries or classrooms to minimize distractions and foster a stronger study atmosphere (P1, P2, P4, P5, P7, P11, P13, P16).
While these strategies provided short-term relief, participants emphasized that they did not address the underlying difficulties in flexibly monitoring attention and adapting to unexpected changes.

\subsubsection{\textbf{Emotional Barriers: Disengagement and Task Avoidance}}~\label{emotion}
%which we see as part of lacking self-awareness on emotional aspects~\cite{greenberg2017social}. 
%Similar to their struggles with self-monitoring task progress, participants also reported difficulties with emotion regulation, which impeded their ability to make consistent progress on tasks
Beyond awareness, participants also described difficulties in regulating emotions, which further impeded them from continuously engaging with the tasks (P3, P6, P7, P10, P13, P16). Negative affect not only delayed initiation but also sustained cycles of avoidance and procrastination. %\modify{Prior research has also shown that individuals with ADHD often experience emotion dysregulation, such as irritability and low frustration tolerance, which is closely associated with poorer academic and occupational performance as well as difficulties in self-esteem and daily functioning~\cite{soler2023evidence, shaw2014emotion}}.
For example, P7 described their experience of repeatedly delaying submission of the thesis draft due to fear of disappointing their advisor:
\begin{quote}
    ``\textit{Back in late November, the professor told me it was time to start revising the paper. I said okay, but I kept procrastinating [...] He really values proactive students, and I could tell he appreciated it. That made me feel anxious. I was afraid of letting him down}.''
\end{quote}

Such accounts demonstrate how emotions interact with metacognitive processes, shaping not only motivation but also the capacity to monitor and adjust behaviors. Unfortunately, participants did not identify any reliable strategies or external supports that effectively helped them recognize and regulate these emotions.
%These anxious feelings not only undermined their engagement in the task but also created a continuous cycle of avoidance, indicating how emotional states can disrupt the execution of plans.

%In summary, participants faced significant barriers in academic task management due to gaps in metacognitive knowledge, difficulties in metacognitive regulation, and emotional struggles. These challenges were deeply interconnected, collectively forming a persistent and multifaceted obstacle to effective task management for university students with ADHD.

\subsection{Co-design Ideas for GenAI to Support Metacognitive Process}

Focusing on the above challenges, participants proposed design ideas for how GenAI could support them in navigating task management in university life. In the following, we organized these ideas together with experts' perspectives around the three themes derived from our thematic analysis. Each of the themes corresponds to metacognitive knowledge (Section~\ref{task-awareness}), metacognitive abilities (Section~\ref{initiation} and Section~\ref{attention}), and emotional barriers (Section~\ref{emotion}), respectively.

\subsubsection{\textbf{Providing Cognitive Scaffolding to Enhance Task and Self-Awareness}}
%Instead of simply listing tasks, effective academic task management requires a deeper level of metacognitive knowledge that includes awareness of one's abilities and emotional states (e.g., knowing when to adjust strategies, strengths, and weaknesses), awareness of the task (e.g., difficulty and the time required to complete it), and an awareness of strategies for completing tasks (e.g., induction and deduction). 
%Participants report that they often have unrealistic expectations regarding task completion, difficulties in accurately estimating task duration, and struggles to prioritize multiple tasks. These challenges stem not from a lack of understanding their own abilities, tasks, and strategies for completing tasks. In this section, we will present participants' design ideas on how GenAI could assist them in addressing these challenges, alongside expert feedback on these design ideas informed by their previous intervention experiences.
To enhance their understanding of different tasks and better align their expectations of personal abilities with reality (metacognitive knowledge---task awareness and self-awareness), participants envisioned multiple ideas for utilizing GenAI as a cognitive scaffold. These ideas focused on helping them navigate fragmented task information and better assess their task expectations, enabling them to gain a clearer sense of the tasks at hand and the steps needed to complete the tasks, as illustrated below.

% belongs to 4.2 %Promoting Metacognitive Control by 
\paragraph{\textbf{Integrating Fragmented Task Information}}
\label{identifytask}

\begin{figure*}[tbp]
    \centering

    \begin{subfigure}[b]{0.37\linewidth}
        \centering
        \includegraphics[width=\linewidth]{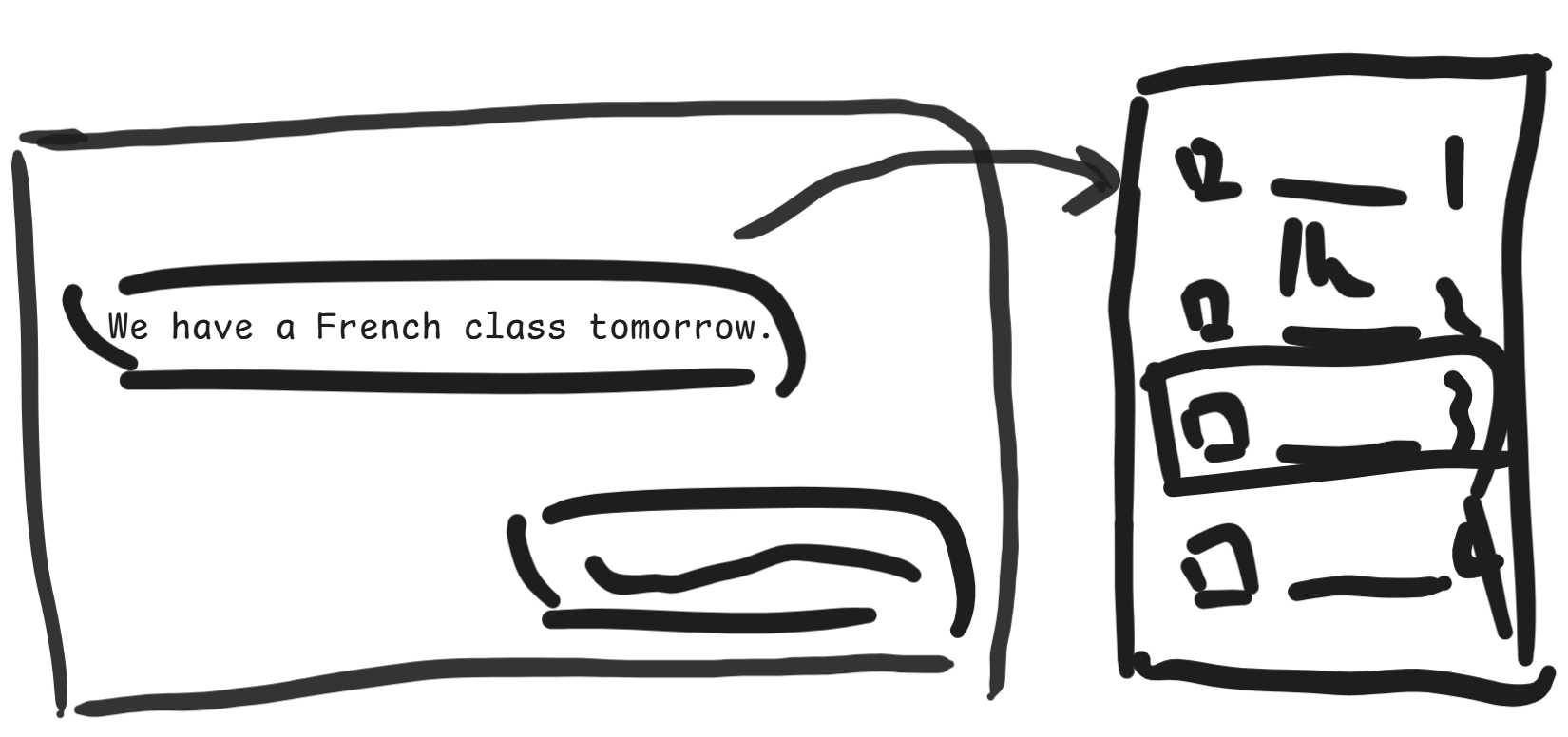}
        \Description{A participant-created sketch illustrating how generative AI could identify tasks by extracting actionable items from chat logs and clipboard content (P2).}
        \label{taskidentification}
        \caption{}
    \end{subfigure}
    \hspace{0.08\linewidth}
    \begin{subfigure}[b]{0.32\linewidth}
        \centering
        \includegraphics[width=\linewidth]{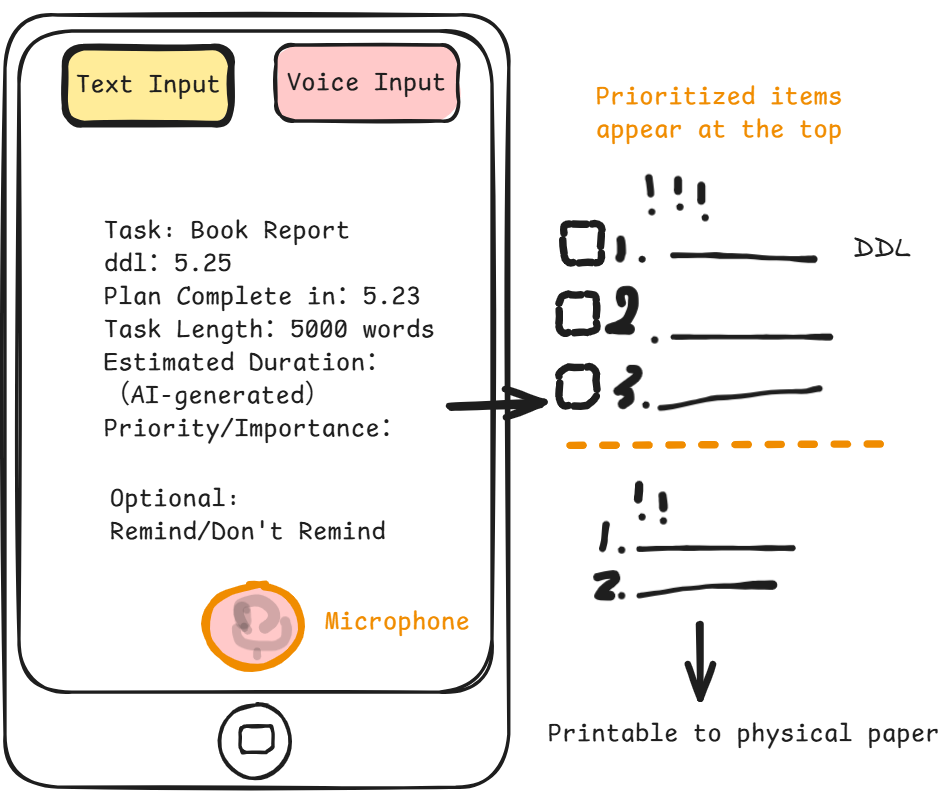}
        \Description{A participant-created sketch showing how generative AI could analyze task priorities and generate an editable task list that users can manually adjust later (P7).}
        \label{taskpriorization}
        \caption{}
    \end{subfigure}

    \caption{Examples of participants' design ideas for leveraging GenAI to enhance task and self-awareness: (a) identifying tasks from chat logs and the clipboard (P2); (b) conducting priority analysis and generating task lists that can be manually adjusted later (P7). For ease of reading, the original design ideas presented in Chinese have been translated into English. The original images are available in the Appendix.}
    \label{fig:codesign_awareness}
\end{figure*}

To address their lack of task awareness mentioned in Section \ref{task-awareness}, participants proposed ideas that leverage GenAI's information understanding and generation ability to identify and extract task information from multiple fragmented data sources (e.g., course management systems, emails, screenshots, messaging apps) and organize them into structured lists (P2, P3, P5, P7). For example, as shown in Figure \ref{fig:codesign_awareness} (a), P2 found that lots of their task-related information was buried in instant messaging apps, which included reminders from classmates and discussions about the group projects. Rather than switching between scheduling and message apps to manually record the tasks, they hoped GenAI could streamline the creation of the task list:

%extract tasks from her phone's chat logs or the clipboard and organize them into a structured task list: 

\begin{quote}
    ``\textit{I just need to long-press the message, and it can read the information and automatically generate a to-do list displayed in the notification ordered by timeline or on my desktop widget (...)}''
\end{quote}

\paragraph{\textbf{Experts’ Perspectives}}
E4 recognized the value of the task-identification design concept and noted that ADHD is often accompanied by reading and writing difficulties, which make it hard for many students to fully record assignments given by their teachers \cite{willcutt2010etiology, molitor2016written}. Additionally, E4 expressed concern that students might forget to use this feature and suggested adding active reminders. For example, when a user opens the learning platform or when GenAI detects relevant content on the clipboard, the tools could proactively prompt the students to add the task, thereby supporting timely engagement with the feature.

% Raising Self-Awareness Through Adaptive Time Estimation
% \paragraph{Supporting Task Allocation through Time Prediction}
\paragraph{\textbf{Rationalizing \& Calibrating Time Allocation for Tasks}}
\label{timeestimation}

% \begin{figure}[tbp]
%     \centering
%     \begin{subfigure}[t]{0.45\textwidth}
%         \centering
%         \includegraphics[width=\textwidth]{Figure/Codesign1_2.png}
%         \caption{}
%         \label{fig:task_organize_a}
%     \end{subfigure}
%     \hfill
%     \begin{subfigure}[t]{0.45\textwidth}
%         \centering
%         \includegraphics[width=\textwidth]{Figure/organize new.png}
%         \caption{}
%         \label{fig:task_organize_b}
%     \end{subfigure}
%     \caption{Participants' design ideas for using GenAI to assist in task organization by promoting metacognitive knowledge: (a) P8 proposed leveraging GenAI to break down tasks into actionable subtasks, estimate workload based on common knowledge and historical user behavior; (b) P7 envisioned a mobile interface enabling GenAI to estimate the task duration and priority and generate an ordered task list, users would have the flexibility to adjust the list based on their own situation.}
%     \label{fig:task_organize}
% \end{figure}

Estimating the time required for each task is a crucial step in creating effective plans. However, several participants reported difficulty in accurately assessing how long it would take for different tasks, due to a lack of task awareness and self-knowledge about their own capabilities (P3, P7, P8, P9, P12, P18). Experts echoed this challenge: E1 and E2 observed that some students they had worked with often created unrealistically long to‑do lists, which not only hindered task completion but also added unnecessary stress. %Such poor time estimates often stem from inaccurate self-assessment of one’s own capabilities and misjudgment of the task’s overall demands. To enhance awareness of one’s own capabilities and task demands, 
To address this challenge, participants brought up that GenAI could help them estimate the time needed for each task by incorporating contextual information, which could be user input (e.g., have prepared an outline in advance for the task of making the slides), conventional benchmarks (e.g., the typical time spent on reading a paper's introduction), and historical data of the users' past performance (e.g., quiz duration) (P7, P8, P9, P18). %P8 suggested that GenAI could assist them in estimating the time required for tasks. Similarly, 
P9 envisioned that GenAI could proactively offer suggestions that help them calibrate the estimated duration in their task plans, particularly when they tend to procrastinate or lack confidence:
%if it seemed misaligned with typical task durations:

\begin{quote}
    ``\textit{For example, if you say you're writing a paper abstract and set aside two days for it, the AI would remind you, like `This is a two-hour task, not two days.' It helps you analyze whether your plan makes sense}.''
\end{quote}

\paragraph{\textbf{Experts’ Perspectives}}
E2 indicated that many individuals with ADHD spend excessive time on the first step of a task, leaving insufficient time for the subsequent steps. Estimating task duration can help students reflect on their progress and task strategies while making timely adjustments to their behavior, which can mitigate perfectionism. Furthermore, E1 suggested that when estimating the task time, GenAI should not only reference the student’s past completion times but also include a buffer. Allowing extra time can help the student perceive the task as less demanding and be more willing to initiate the task.

\begin{quote}
    \textit{``For example, it may take a student 30 minutes to finish the task. In the plan, though, we can allocating a full hour for this task. Beyond accounting for possible distraction or procrastination, the extra buffer was intended to `cheat' the student’s brain into perceiving the task as easier, thereby making them more willing to start.''} (E1)
\end{quote}

In addition, E3 emphasized that, as some students with ADHD tend to experience learned helplessness\footnote{\textit{Learned helplessness} refers to a psychological condition in which individuals stop trying to improve their situation after repeated exposure to uncontrollable events~\cite{seligman1967failure, seligman1975helplessness}.} when encountering difficulties, GenAI should gradually build an understanding of their ability levels through long-term interaction. When organizing tasks, GenAI should evaluate whether a task might exceed the student’s capabilities and proactively prompt them to seek external support (e.g., join peer discussions, seek academic tutoring resources).

Similarly, P14 proposed that GenAI could help them adjust expectations during planning stages by highlighting potential delays:

\begin{quote}
    ``\textit{For the plan set for the next day, I hope AI can help me accept the possibility that tasks may not be completed on time, especially during the planning phase, where AI could remind me that the task might not be finished. This is important because when I truly can’t complete a task, anxiety overwhelms me, and by then, I have no energy left to accept any interventions}.''
\end{quote}

% Although participants noted that certain tasks requiring deep thinking and creativity (e.g., electing a thesis topic and deriving formulas) were challenging to estimate the time duration (P13, P14). After reviewing participants’ design ideas, E2 noted that many individuals with ADHD tend to pursue overly high standards when completing tasks, which can prevent them from keeping up with schedules. GenAI-assisted time estimation can help students make more accurate progress assessments and encourage timely reflection on their task-completion strategies:

% \begin{quote}
%     ``\textit{Many individuals with ADHD tend to be perfectionistic, often getting stuck on the first part of a task because they feel it’s not good enough, leading to spending a lot of time on the first part. If they could estimate the time for each step, they would be able to assess whether they have enough time left to finish the remaining tasks on time}.''
% \end{quote}

\paragraph{\textbf{Personalizing Task Prioritization}}
% Effective task prioritization enhances resource utilization and workflow efficiency during task management. It requires metacognitive knowledge on identifying the nature, urgency, and importance of tasks, as well as understanding one's own progress and status. This is especially crucial for university students with ADHD, who often navigate multiple competing academic demands.

% For example, P15 described that they do not have a clear sense of task priority and usually choose to start with tasks found enjoyable, such as design, rather than those that were more cognitively demanding but essential, like programming:

% \begin{quote}
%     ``\textit{I do not have a clear sense of task priority (...) For example, with an assignment I was not interested in, I would probably end up doing designs that I like on my computer, but never start relevant tasks, like programming. This difficulty was also shared by others, who struggled to navigate competing academic pressures effectively}.''
% \end{quote}

In managing multiple academic tasks and deadlines (e.g., assignments and reviewing for different courses during the final week), participants often felt overwhelmed and were uncertain about which tasks to prioritize (P7, P15, P16). To support decision-making, participants hoped GenAI could assist them in analyzing factors such as importance, urgency, and workload of tasks, as well as their own interests and preferred completion strategies. As such, they could work with GenAI to develop a logical and practical order for executing the tasks. As shown in Figure \ref{fig:codesign_awareness} (b), P7 designed a system that could provide an initial order of the tasks based on prioritization results, which they could then adjust as needed:

\begin{quote}
    ``\textit{I would just pour out everything I needed to do, and the system could intelligently sense the importance of the task. It might say, ‘I recommend giving this five stars,’ and then I’d respond, ‘How about four stars?’—which would make me reflect a bit. If the deadline was really tight, like on the 25th, I’d probably agree with five stars. So I would slightly adjust the priority based on its suggestion}.''
\end{quote}

\paragraph{\textbf{Experts’ Perspectives}}
Notably, experts highlighted that while general populations commonly prioritize tasks based on task importance and urgency (e.g., the Eisenhower Matrix~\cite{eisenhower1954address}), this approach may not be as effective for those with ADHD (E1, E2, E4) as they often struggle to distinguish which tasks are important (E1, E4). In part, their lower baseline dopamine levels may hinder task initiation, even though they could recognize the importance of a task (E2). Thus, experts recommended interest-driven task prioritization to make task initiation easier. As E1 explained, starting with tasks that align with an individual's interests may help engage reward pathways implicated in ADHD, including dopamine-related processes, thereby creating a more favorable mental state for transitioning to important tasks~\cite{williams2005dopamine,locke2023cracks}:

\begin{quote}
    ``\textit{They usually have lower dopamine levels. Dopamine is also part of the energy and nutrients. When they already don't have much of it, being asked to do a task drains them even more. So, I would suggest they start with something enjoyable to build up some dopamine, and then use that bit of joy to tackle the more difficult and important task}.''
\end{quote}

% 像我刚才就是疯狂输出一番，就是我跟他说这个事情我就要描述一下，然后他能智能的感受到我这个事就衡量到这个事儿的重要性。然后他说那我推荐你的这个重要程度是，比如说五颗星，然后就说。四颗星可以吗？会让我考虑一下可以。对，然后再比如说像他25号的话，就假如说25号就很急，然后我就觉得我觉得那五颗星吧，就我会在他这个程度上稍微调整。

% separate emotion: 4.3.3, 4.3.5, 4.3.4?
% \subsubsection{\textbf{Promoting Cognitive Reflection for Metacognitive Abilities Development}}
\subsubsection{\textbf{Promoting Reflective Task Execution for Building Metacognitive Abilities}}
% Metacognitive abilities refer to an individual's capacity to monitor, regulate, and adjust cognitive processes to adapt strategies and responses according to the task demands and environmental changes. 
%From participants' feedback, we found university students with ADHD often face challenges in metacognitive abilities when managing limited cognitive resources, inconsistent motivation, distractible attention, and progress tracking during task execution. In the following sections, we will explore how participants envision GenAI enhancing their metacognitive abilities to address these challenges effectively.
Participants' limited metacognitive ability in monitoring and regulating behaviors often manifested as low motivation, progress management difficulty, and frequent distraction during task execution (see Section~\ref{attention}). Recognizing these challenges, they expressed a desire to become more mindful and reflective about their behaviors and performance while working on tasks. As such, participants proposed ideas for GenAI to support them in actively observing and evaluating their actions and strategies in real time, specifically through collaborative task decomposition, creative and motivational reminders, visualized progress, and adaptive plan adjustment.

\begin{figure*}[tbp]
    \centering

    \begin{subfigure}[b]{0.21\linewidth}
        \centering
        \includegraphics[width=\linewidth]{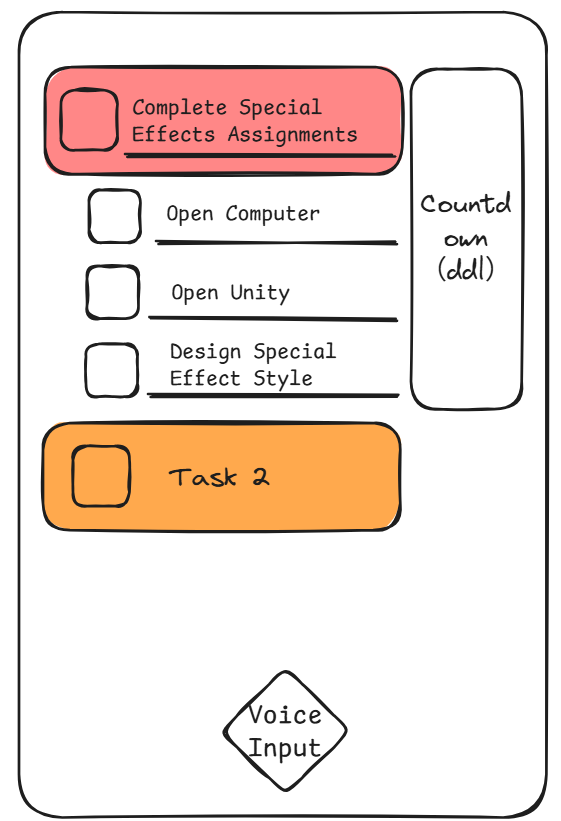}
        \Description{A participant-created sketch illustrating how generative AI could break down a task into actionable subtasks and estimate workload using conventional benchmarks and the individual’s past behavior (P15).}
        \label{taskdecomposition}
        \caption{}
    \end{subfigure}
    \hfill
    \begin{subfigure}[b]{0.31\linewidth}
        \centering
        \includegraphics[width=\linewidth]{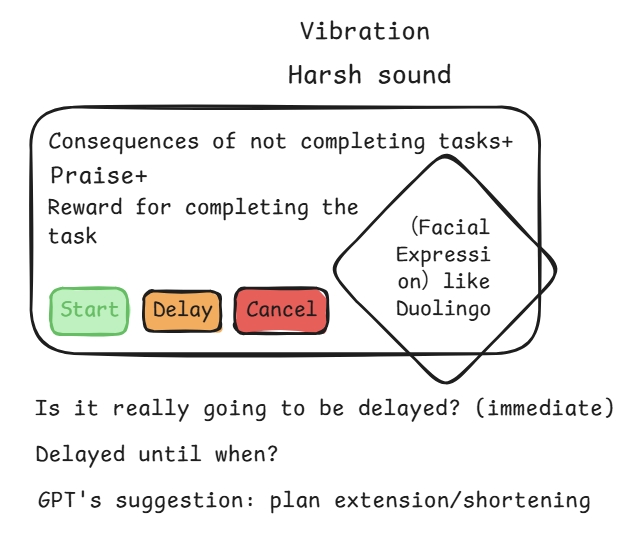}
        \Description{A participant-created sketch showing how generative AI could deliver multimodal motivational prompts (e.g., visual cues, audio reminders, or vibrations) to support task initiation (P16).}
        \label{taskmotivation}
        \caption{}
    \end{subfigure}
    \hfill
    \begin{subfigure}[b]{0.31\linewidth}
        \centering
        \includegraphics[width=\linewidth]{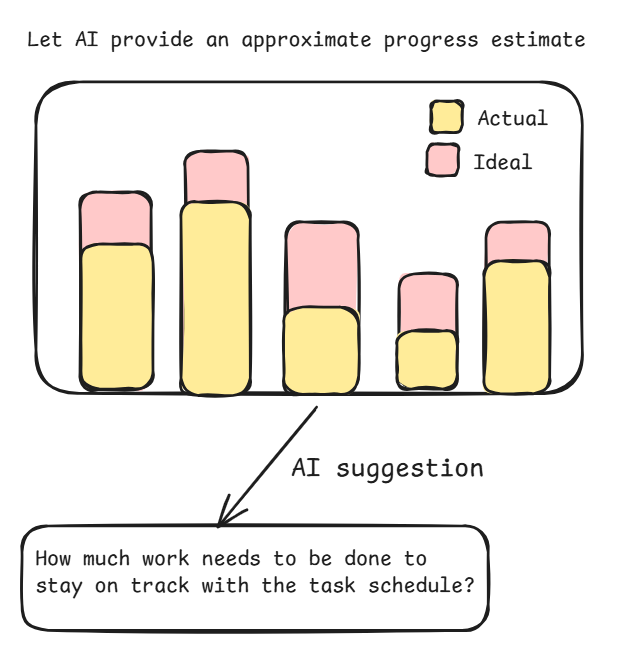}
        \Description{A participant-created sketch depicting how generative AI could compare planned versus actual reading progress and suggest the amount of work needed to reach the goal (P11).}
        \label{taskmonitoring}
        \caption{}
    \end{subfigure}

    \caption{Examples of participants' design ideas for leveraging GenAI to foster metacognitive abilities: (a) breaking down tasks into actionable subtasks and estimating workload based on conventional benchmarks and individuals' historical behavior (P15); (b) delivering multiple forms of motivational prompts (e.g., visual, audio, or vibration) to facilitate task initiation (P16); and (c) tracking the gap between actual vs.\ planned reading progress and suggesting how much work is required to achieve the goal (P11). For ease of reading, the original design ideas presented in Chinese have been translated into English. The original images are available in the Appendix.}
    \label{fig:codesign_abilities}
\end{figure*}

% \paragraph{Facilitating Adjustment of Cognitive Resources through Task Decomposition}
\paragraph{\textbf{Collaborative Task Decomposition}}
\label{taskbreakdown}
Participants reported feeling overwhelmed in front of overarching tasks without concrete execution steps, such as writing a thesis or preparing for exams, which led to repeated procrastination (P1, P2, P3, P7, P13, P16). In particular, students with ADHD typically have lower working memory, making them prone to cognitive overload when handling complicated tasks~\cite{kasper2012moderators, kofler2011working, kofler2018working}. This could lead to misjudgment of the efforts needed to achieve a goal (e.g., perceiving the task as time-consuming and repeatedly procrastinating, although it actually takes little time).
%This sense of overwhelm arises from a misperception of the effort needed and the steps required to complete the task. 
In this case, participants hoped to decompose a task into smaller, actionable subtasks so that they could have a clearer and more realistic understanding of what needs to be done, which can reduce the mental stress of task initiation. Several participants highlighted that the multimodal information processing capabilities of GenAI be utilized to support task decomposition by analyzing materials such as lecture slides, rubrics, and previous submissions (P1, P2, P3, P7, P8, P13, P16, P17). As shown in Figure \ref{fig:codesign_abilities} (a), P15 hoped GenAI could break down the special effect assignment into manageable steps to facilitate task initiation. Additionally, P17 noted that displaying too many subtasks at once may overwhelm them, and they suggest limiting the number of subtasks shown at once and progressively presenting subsequent tasks based on their task progress.

%During academic tasks, students need to continuously monitor, adjust, and optimize their strategies, which requires substantial cognitive resources. 

In the meantime, participants highlighted that they did not want GenAI to fully take over task decomposition, not only because it lacks contextual knowledge but also because they hoped to exert control by themselves. Thus, the role of GenAI should be encouraging participants to reflect on their goals and guiding them to think about additional context for decision-making (P13). For example, P13 hoped that GenAI could prompt them to input information such as the prerequisite steps, people, and resources required to complete the task:

\begin{quote}
    ``\textit{Guiding users through a conversational approach? The required actions should be defined by the users themselves, (GenAI can) encourage them to clearly think through why each step matters. For example, a long-term project could be divided into several shorter tasks, and those shorter tasks could then be further decomposed into smaller steps, guiding users gradually and systematically}.''
\end{quote}

\paragraph{\textbf{Creative \& Motivational Reminders}}
\label{reminders}
To address the barriers to initiating tasks, participants expressed a need for motivational reminders that are interesting, constantly changing, and tailored to their preferences. While participants had attempted to create such reminders in practice, such as by changing the sounds of alarms or the voice of the reminders, they found this process time-consuming, which still failed to engage them over time. In response, they envisioned using GenAI to deliver such reminders with its ability to generate various forms of text, image, and even animations (P5, P6, P8, P10, P12, P13, P15, P16, P18). Ideating and designing personalized reminders enhances self-awareness, as it encourages individuals to reflect on their specific needs and preferences for task initiation. As shown in Figure \ref{fig:codesign_abilities} (b), P16 drew inspiration from Duolingo’s design \footnote{\textit{Duolingo} is a language-learning app that features a green owl mascot and other cartoon characters to engage users in learning~\cite{duolingo}.} of different cartoon characters that engage users in learning. They believed that varying visual forms of the reminder, complemented by vibrations and sound cues, could boost their motivation to initiate and continue a task:

\begin{quote}
    ``\textit{The little (Duolingo) logo changes, and sometimes I find it fun. For someone like me who really dislikes boring things, this kind of visual element makes the experience more engaging}.''
\end{quote}

Some participants proposed using GenAI to generate avatars and voices of their familiar characters to make reminders more engaging. For example, P20 imagined receiving the task reminder in the voice and appearance of their favorite anime character, Hatsune Miku\footnote{\textit{Hatsune Miku} is a Japanese virtual idol~\cite{hatsunemiku}.}:

\begin{quote}
    `` \textit{I’m into anime, so I was thinking, what if users could set a character they like when sending reminders, the system could show an image of that character or something similar. For example, I like Hatsune Miku. It wouldn’t be too hard to mimic her tone, either}.''
\end{quote}

\paragraph{\textbf{Experts’ Perspectives}}
Our expert participant E2 acknowledged the value of this idea, adding that individuals with ADHD have different ways to process information. For example, some are more audio sensitive while others prefer visual feedback. Thus, E2 suggested that reminders should be tailored to individual preferences:
%Given the differences in information processing modalities among ADHD individuals, 

\begin{quote}
    ``\textit{Individuals have different processing modes. For example, some are verbal processors, needing to speak in order to think. Others are kinesthetic, requiring physical movement to facilitate thinking. Some rely more on visual stimuli, while others do not adhere to a single mode. We need to support them in recognizing their different processing styles and help them apply their past successful experiences to their current tasks}.''
\end{quote}

To further motivate individuals following the reminders, E2 suggested that GenAI can prompt students with some reflective questions to kickstart, such as ``\textit{what ideas do you have for starting your task in a fun and engaging way?}'' This design idea shifts the student's focus from the task's content (``what to do'') to their learning process (``how to do it''), fostering self-reflection and encouraging them to plan personalized strategies proactively. This brief moment of thinking is the essence of metacognitive regulation—planning and self-awareness.

\paragraph{\textbf{Evaluation and Visualization of Task Progress}}
\label{taskvis}
One main difficulty that participants faced during task execution was the unclear sense of task progress, as they often struggled to perceive how much progress they had made and how much work still remained (P3, P9, P11, P16).  %Although participants try to understand the passage of time through methods such as the Pomodoro timer (P3) and countdowns (P11), the elapsed time does not always correlate with the actual progress made on the task. As a result, participants still lack a concrete sense of actual task progress. Therefore, they hope to leverage the multimodal information understanding and generative capabilities of GenAI 
To enhance their awareness of task progress, participants highlighted the importance of situated reflection, which would help them stay on track. This requires GenAI to process information from multiple sources, such as an individual's text inputs in a document, interaction records across various apps, and the time elapsed. Additionally, GenAI should have the capability to generate dynamic visualizations that enable individuals to better understand their progress in real-time, making the reflection process more effective and engaging.
%Participants hope that GenAI can help them gain a deeper understanding of their current task progress. 
As shown in Figure \ref{fig:codesign_abilities} (c), P11 envisioned during reading tasks, GenAI could compare the number of pages read with the daily reading goal and visualize the progress:

\begin{quote}
    ``\textit{For example, if I need to finish a book in a week, and it is divided into several chapters, (GenAI) will compare the actual number of pages I’ve read each day with the planned daily reading goal, displaying the progress in a bar chart, and provide suggestions on how much additional effort I need to put to reach daily goals}.''
\end{quote}

\paragraph{\textbf{Experts’ Perspectives}}
Echoing this idea, E3 suggested that adding some peer pressure (e.g., by visualizing the average progress of others) could help motivate individuals to stay committed to their goals. However, this approach may also risk fostering social comparison, which could lead to feelings of frustration or inadequacy. Therefore, it should be implemented with care, ensuring that the focus remains on personal growth and progress rather than creating undue competition among users:
%although the potential impact of frustration should be considered when designing the tool, GenAI can visualize the typical progress of most people in completing tasks, which may motivate students to monitor and adjust their behavior to align with their desired progress:
\begin{quote}
    ``\textit{Individuals with ADHD often consult others about the status of their tasks. If GenAI could provide a reference for the typical progress metrics of most people, driven by a competitive mindset, they might adjust their behavior to align more closely with this benchmark...}'' (E3)
\end{quote}

% 比方说我这一个星期我需要把那个看完。然后它里面？可能细分了几张，几张几张内容，这是我，我给他设定的计划，就是每天他差不多的工作量，然后就是。我填完就是当天实际上看了多少之后，他会跟原本那个计划内容进行比较，然后然后再拉一个条条出来

%P11 从动机来讲的话，做到后面可能感觉跟原本的进度差的有点多。比方说我需要这一段时间把这一本书的内容给看完，但是看几天我不想看了，然后这个东西后面不是越差越多，但是但是这个我看这本书它不是服务一个更大的主题，比方说我要什么考研，我要准备什么什么考试。就是我实际上我这这这些东西差这么一点点，后面还是可以通过其他内容来补救的，或者说时间长了一点，然后他会把后面那部分那个计划给论出来，比方说。我已经差这么多了，然后但但是他后面列一个那个清单出来，你下面需要做到哪些哪些哪些，然后可能最后实现效果跟原来的那个目标差不太多这样。

% \paragraph{Keeping Orderliness and Flexibility through Adaptive Rescheduling and Suggestions}
% University life often presents unexpected disruptions, such as the emergence of urgent tasks or the prolonged execution of planned activities, which can significantly impact students' ability to follow their schedules. For students with ADHD, who depend on structured routines, such interruptions exacerbate their difficulties in tracking and adjusting their actions in response to feedback.

\paragraph{
\textbf{Adaptive Plan Adjustment}}
\label{planadjust}
% Participants reported that university life often involves unexpected events, such as urgent tasks or the extended duration of planned activities. These events frequently led to disruptions and a loss of order, making participants reluctant to adjust their behaviors to adapt to changes, and even leading them to give up continuing their tasks (P1, P2, P4, P5, P7, P11, P13, P16). Experts note that, when implementing behavioral interventions for children with ADHD, to help their children adapt to changes, parents are encouraged to develop multiple plans in advance (E4). For example, they need to inform their children in advance that if it rains on the way to the amusement park, the plan might need to be adjusted to plan B to return home. However, due to the complexity and unpredictability of university students' schedules, this approach is less feasible, as planning in advance becomes challenging.

% As P16 shared, they missed a friend's appointment due to underestimating the time needed for their task and being unable to stop once they began:

% \begin{quote}
%     ``\textit{Before heading to my appointment, I had a one-and-a-half-hour gap and planned to find something to do. But as I worked, I realized it would take me two hours. Even though I should have left, I would lose track of time and couldn't stop. As a result, it's common for me to be scheduled for 6 p.m., but I end up arriving at 6:30 p.m. (...)}''
% \end{quote}

To enhance their ability to adapt to changes, participants envisioned GenAI providing suggestions tailored to their situations, such as skipping or simplifying lower-priority tasks to help them stay on track (P2, P4, P5, P7, P11, P13, P16). Regular scheduling tools rely on predefined rules and static configurations, which are insufficient for handling these rapidly changing and context-specific scenarios. GenAI, with its information understanding and generation capabilities, holds potential for addressing these situations. For example, P2 explained that when time slots are already allocated but something urgent comes up, the disruption may lead to incomplete tasks. In such situations, they envisioned GenAI assisting in rescheduling the original task to reduce the effort required to revise the plan manually. Similarly, P4 imagined GenAI offering alternative suggestions to help users stay on track when facing minor disruptions:

\begin{quote}
    ``\textit{If you plan to go to school but wake up late, it can provide adjustment suggestions for less important tasks, like skipping breakfast or not applying makeup, to help you stay on track and meet your goal on time}.''
\end{quote}

\paragraph{\textbf{Experts’ Perspectives}}
Experts emphasized that AI-assisted task suggestions are crucial for maintaining a sense of order in individuals with ADHD, as they often experience confusion and struggle to focus on finding solutions when confronted with sudden and urgent situations (E2, E3, E5). E2 mentioned that when plans change, students with ADHD often fall into rigid thinking patterns, which limit their flexibility. Alternative suggestions from GenAI could provide them with more options and flexibility, helping them better cope with changes:

\begin{quote}
    ``\textit{When faced with sudden changes, they often fall into black-and-white thinking, limiting their flexibility. However, when they allow for some flexibility, they can find better solutions. For example, they might feel they must wear a certain pair of earrings or dress a certain way in the morning, but these are not requirements and they don’t necessarily have to. Often, the issue isn’t the task itself, but how they perceive it. AI suggestions can provide flexibility and help them realize their available options}.'' (E2)
\end{quote}

\subsubsection{\textbf{Facilitating Emotional Regulation to Sustain Task Engagement}}
\label{emotionsupport}
% Participants often experience emotional fluctuations during task execution, leading to inconsistent attention and slow task progress. 

During the co-design session, participants highlighted the importance of tackling their emotional barriers. They envisioned that GenAI could provide various support to enhance their emotional awareness and regulation, which could in turn help them sustain engagement during task execution (P1--P3, P5, P7, P10, P11, P13, P16, P18, P19). These supports mainly centered on \textbf{\textit{virtual agents that served as a study companion}}, either silent or interactive via text or speech (P1, P2, P5, P11, P13, P18, P19). 

\begin{figure*}[tbp]
    \centering

    \begin{subfigure}[b]{0.35\linewidth}
        \centering
        \includegraphics[width=\linewidth]{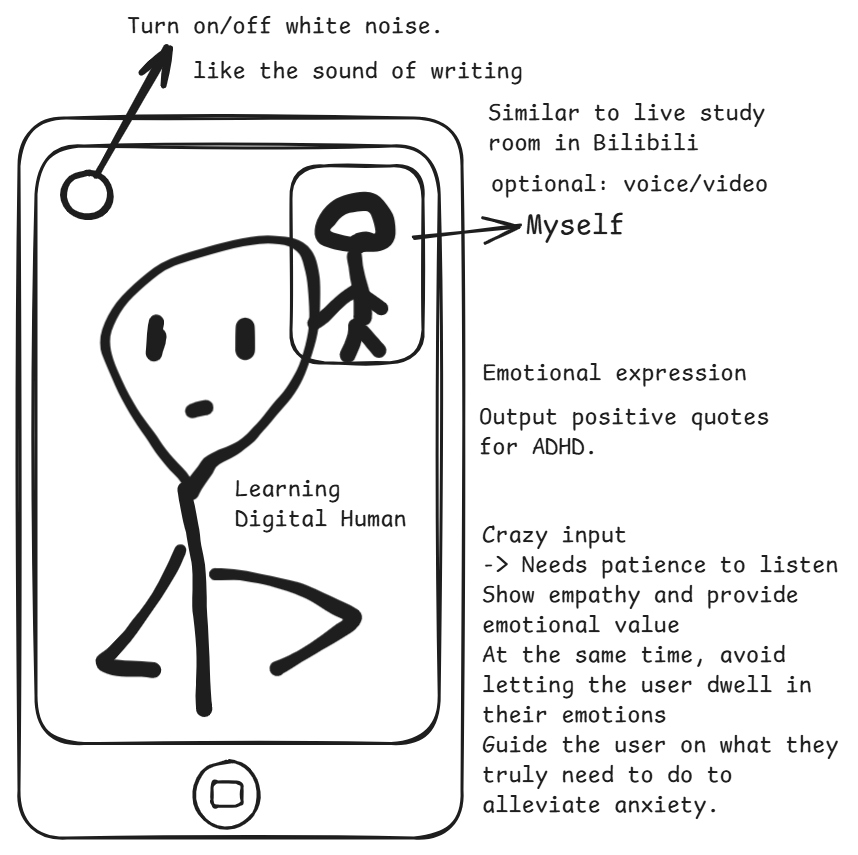}
        \Description{A participant-created sketch illustrating how generative AI could act as a study companion to help an individual stay focused during task engagement (P7).}
        \caption{}
        \label{studycompanion}
    \end{subfigure}
    \hspace{0.08\linewidth}
    \begin{subfigure}[b]{0.44\linewidth}
        \centering
        \includegraphics[width=\linewidth]{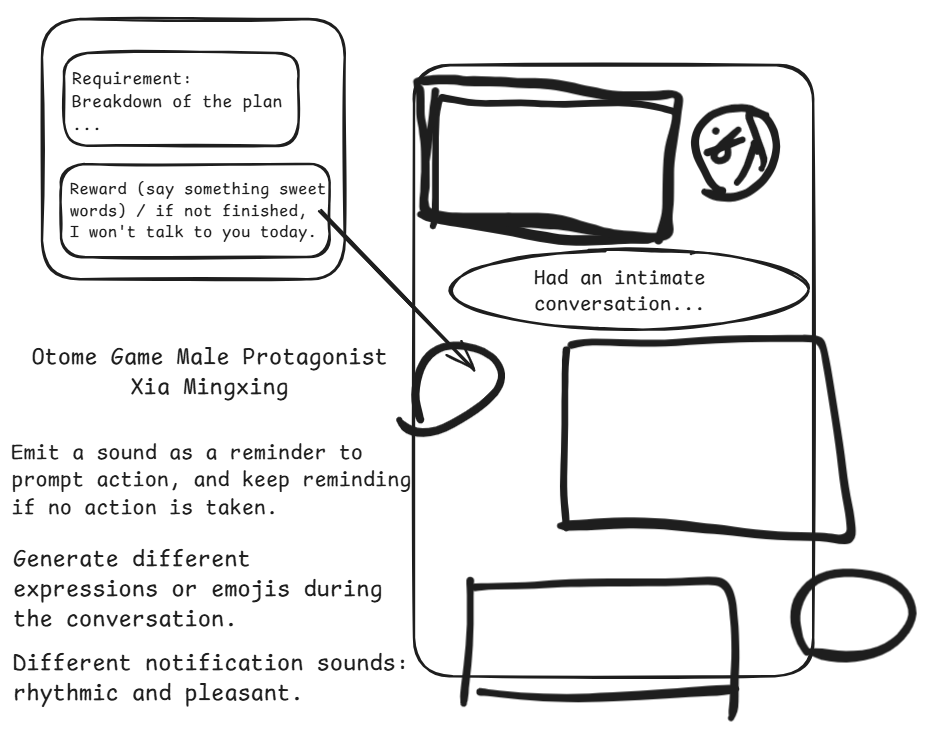}
        \Description{A participant-created sketch showing how generative AI could provide emotional support through interactive, multimodal conversations with a favorite fictional character as part of a reward mechanism (P10).}
        \caption{}
        \label{emotionalsupport}
    \end{subfigure}

    \caption{Examples of participants' design ideas for leveraging GenAI to facilitate emotional regulation for sustained task engagement: (a) serving as a study companion to help individuals stay focused (P7); and (b) providing emotional support through interactive, multimodal conversations with one's favorite fictional character as part of a reward mechanism (P10). For ease of reading, the original design ideas presented in Chinese have been translated into English. The original images are available in the Appendix.}
    \label{fig:codesign_emotion}
\end{figure*}

%verbal feedback (e.g., role-playing their favorite characters and offering verbal reward) and non-verbal feedback (e.g., the presence of a virtual avatar) to facilitate self-reflection and regulation of emotion, thereby enhancing the sustainability and efficiency of overall academic productivity.
%Frequent distractions, from both external environment (e.g., phone notifications, social media) or intrusive thoughts (e.g., ruminations on past mistakes, worries about personal failures) are barriers to staying engaged in a task (P1, P3, P5, P7, P11). 
This finding aligned with what participants mentioned in the warm-up session in describing their challenges (see Section~\ref{attention}), where they found that one of the most effective strategies to prevent external distractions was to have a study partner who could prompt them to stay focused. For example, as shown in Figure \ref{fig:codesign_emotion} (a), P7 envisions a GenAI-driven virtual avatar that can serve as a study companion, promoting self-reflection on her current state through its presence while also assisting with reminders of her schedule:

\begin{quote}
    ``\textit{There is a virtual avatar that pretends to study and supervise me. I can occasionally vent to it, and it responds with understanding and empathy (...) It can also remind me of urgent tasks, as sometimes I tend to forget the tasks when I get caught up in a conversation}.''
\end{quote}

Additionally, P18 suggested that the appearance of the study partner could also be the animal they like (e.g., Garfield\footnote{\textit{Garfield} is a fictional orange cat character from a popular comic and media franchise~\cite{garfield1978}.}, British Shorthair), and they also suggested that the partner's background could be adjusted by capturing or scanning their own surroundings to align with the study environment (e.g., library, home) and enhance the sense of companionship.

Furthermore, to address their declining efficiency in task execution caused by intrusive thoughts and negative emotions, participants envisioned GenAI to be an empathetic, emotionally responsive companion that could offer personalized emotional support (P7, P10, P13, P16). Unlike general emotional support, which primarily focuses on comfort and connection, participants sought productivity-oriented emotional support that not only regulates negative emotions but also facilitates task initiation and sustains productive momentum. As shown in Figure \ref{fig:codesign_emotion} (b), P10 imagined GenAI taking the role of their favorite male protagonist from the otome game \textit{Light and Night} to help them organize tasks, with different emojis matching the conversation dynamics and offering small, personalized rewards (e.g., romantic dialogue) for their continued task engagement, while delivering punishment they were unable to adhere to the plan:

\begin{quote}
    ``\textit{I hope it can be the male character I like, Xia Mingxing\footnote{Xia Mingxing is one of the male protagonists in the female-oriented romance mobile game \textit{Light and Night}~\cite{lightandnight2021}.}. I could ask him to help me break down and prioritize my plan. If I can follow through it, it can reward me with some intimate words. If I don't stick to the plan, then he shouldn't talk to me at all night}.''
\end{quote}

\paragraph{\textbf{Experts’ Perspectives}}
Experts acknowledged the importance of emotional support in the lives of those with ADHD because it helps to improve their task engagement and overall well-being. Specifically, they believe that individuals can benefit from P10's design by providing continuous positive feedback and rewards throughout the task process, and noted that these should be specific to help students identify which actions are effective. However, they pointed out the risks of incorporating penalty-based feedback, as they may trigger negative emotions, such as frustration, and lead to resistance in students, making students with ADHD even less likely to engage with the task:

% While some participants suggested penalty-based strategies for procrastination or delays in task progress, for example, P10 mentioned withholding communication as a punishment for not following through on a plan, and P17 envisioned using insulting reminders such as ``Hurry up and go running, you fat pig.'' However, experts advised against such approaches (E2, E4, E5), as 

\begin{quote}
    ``\textit{Punishment leads to negative emotions, such as frustration, because it feels like saying, 'See, you didn’t do it again.' However, we could view each failure as an experiment. I may not have succeeded this time, but that doesn’t mean I can’t succeed in the future}.'' (E5)
\end{quote}

In addition to managing negative emotions, experts suggested that GenAI should help to identify and strengthen students' positive emotions. As E2 noted, GenAI should recognize users' positive emotions during interactions and engage them in exploring the underlying reasons to further strengthen their positive emotions:

\begin{quote}
    ``\textit{For example, when noticing that their tone rises or they are laughing, this often reflects a positive emotion. At such moments, GenAI should explore with them what thoughts lie behind this positive emotion. Often, these are linked to their past successes and to positive experiences associated with self-recognition. By discussing these with the students, GenAI can help reinforce these positive aspects in their mind}.'' (E2)
\end{quote}

In parallel, 
experts also suggested that GenAI could help students regulate their emotions by guiding them to view their difficulties from different perspectives (E1, E2). For instance, E2 observed that many students with ADHD feel discouraged because they perceive themselves as fundamentally flawed. GenAI should assist them in separating personal traits from the impacts of ADHD, underscoring that these challenges arise from the cognitive characteristics of ADHD rather than from personal shortcomings, thereby encouraging students to confront difficulties and explore coping and action strategies tailored to their own needs. Similarly, E1 emphasized that GenAI should guide students to see the situations from different perspectives, which may help them realize that what is before them is not necessarily a genuine ``difficulty'':

\begin{quote}
    ``\textit{For example, some students may feel frustrated when they cannot recall what they have read. GenAI can remind them that what matters is not how much they remember, but the extent of change they bring about, and making such changes is an achievement in itself}.''
\end{quote}

Furthermore, experts highlighted that GenAI's tendency to agree with users' statements can unintentionally reinforce recurring inaccurate or unhelpful ways of interpreting their tasks or performance (e.g., inaccurate self-evaluations and overgeneralization), and emphasized that GenAI should demonstrate critical thinking to better support students’ self-reflection:

\begin{quote}
    ``\textit{I feel like current (GenAI) models always tend to agree with what the user says and constantly offer praise. It makes me feel like it's just trying to make me feel better, rather than providing real feedback. This makes me feel a bit skeptical, and I’d prefer if AI's responses showed more critical thinking.}'' (E3)
\end{quote}

\section{Discussion}
% Drawing from the metacognitive challenges and design ideas shared by the participants and expert suggestions. In this section, we examine the productivity challenges faced by university students with ADHD through a metacognitive lens, situating them within the context of existing literature. We also explore design opportunities for leveraging GenAI to support metacognitive processes, considering both its strengths and potential pitfalls in assisting individuals with ADHD.

Adopting a metacognitive lens, our findings revealed several task management challenges faced by university students with ADHD and highlighted how they envisioned GenAI could enhance both their metacognitive knowledge and abilities. In this section, we reflect on how the metacognitive perspective helped deepen and expand current understandings of ADHD in task management contexts. We also discuss how GenAI emerges as a double-edged sword: while it can scaffold awareness and reflection to strengthen metacognitive knowledge, it also risks fostering dependency and undermining metacognitive monitoring if not carefully designed.

% the underlying needs in the participants' design ideas and potential design opportunities for leveraging GenAI to support academic task management for university students with ADHD.

% \subsection{Reflecting Upon the Findings in Time Management for University Students with ADHD}

% \subsection{Reconsidering Time Management Tool Design for Individuals with ADHD in the Era of GenAI}

% summary of interesting findings

\subsection{Deepening the Understanding of Task Management Practice of ADHD Individuals Through Metacognition}
Research on ADHD has a long history, documenting a wide range of behavioral challenges related to task management~\cite{prevatt2011time, owens2007critical, deprenger2010feasibility, alderson2013attention, brown2005attention}, but understandings of this population through the lens of metacognition have only gained attention in the past decade~\cite{butzbach2021metacognition, solanto2010efficacy}.
%such as biased time estimation\cite{prevatt2011time, owens2007critical}, unrealistic self-expectations\cite{owens2007critical, hoza2004self, hoza2002boys}, and difficulties maintaining focus due to frequent distractions~\cite{deprenger2010feasibility, alderson2013attention, brown2005attention}. These studies have advanced the understanding of how individuals with ADHD differ from their peers in day-to-day academic performance.
From this perspective, difficulties like inaccurate estimation of task duration or procrastination are not isolated performance, but rather manifestations of challenges in metacognitive knowledge (e.g., miscalibrating task demands and one's own ability) and metacognitive abilities (e.g., difficulties in controlling and monitoring one's behaviors)~\cite{butzbach2021metacognition, solanto2010efficacy}.

Our findings extend this perspective by showing how common struggles reported by university students with ADHD can be explained by their metacognitive challenges.
As reported in Section~\ref{finding-metacognitive-challenge}, mismatched self-exception and task performance was not merely a reflection of poor time management skills, but also limited awareness of both tasks and themselves. Similarly, perfectionism represented not simple avoidance but maladaptive self-monitoring and evaluation.
%These interpretations aligned with recent neuropsychological assessments conducted by Butzbach et al., which showed that adults with ADHD exhibit wider discrepancy between their self-evaluation and actual cognitive performance compared to those without ADHD---an indication of challenges in self-awareness of attentional functions ~\cite{butzbach2021metacognition}. 
By delving into participants' metacognitive knowledge and abilities, our findings also revealed several underexplored task management barriers in university settings. While previous literature primarily focused on difficulties in task prioritization and biased time estimation~\cite{kofler2018working, nejati2020time, american2013diagnostic, frutos2014adaptive}, we found that the struggles started even earlier when students were identifying and extracting task-related information in front of the fragmented information sources, such as syllabi, emails, and chat messages. This task information is not only distributed across multiple platforms but also presented in different formats (e.g., text, image, and sometimes videos). Drawing on the Cognitive Load Theory~\cite{sweller1988cognitive}, we interpret this as a case where frequent context switching across platforms and modalities adds extraneous cognitive load. For students with ADHD, this burden is further magnified by limited working memory capacity and difficulties in inhibitory control, which make it harder to filter out irrelevant details and retain task-relevant information~\cite{gropper2009pilot, rubia2011disorder}. 
In other words, what might appear to neurotypical students as a relatively simple organizational step (e.g., checking the syllabus or compiling assignments from multiple channels) may become cognitively taxing for those with ADHD.
This finding highlights the need for supports that can reduce extraneous cognitive load at the very beginning of the task cycle, as reflected in the co-design ideas of our participants, which involved integrating information across channels and organizing relevant task attributes. 
%frequent context switching can increase extraneous cognitive load, and for university students with ADHD, limited working memory capacity and difficulties in inhibitory control over irrelevant information further exacerbate this burden \cite{sweller1988cognitive, gropper2009pilot, rubia2011disorder}. As a result, the act of locating and integrating the necessary information constitutes a complex subtask in its own right. In addition, we found that students not only set unrealistically high expectations for the number of tasks to be completed within a given timeframe, but also often set high standards for their work, which led them to anticipate excessive effort to complete it and created additional barriers to initiation (P16). Consistent with our findings, Smith et al. conducted a daily diary study and found that perfectionistic concerns generate a persistent gap between the actual and the ideal self that contributes to procrastinatory behavior \cite{smith2017clarifying}. We also found that negative emotions such as anxiety and fear of failure (P3, P6, P7, P9) interacted with reduced task efficiency during task execution, undermining both metacognitive monitoring (e.g., awareness of one's emotional state) and control processes required for sustained task progress. Moran found that anxiety reduces working memory capacity by competing with task-relevant processes \cite{moran2016anxiety}. Extending this view, our findings show that the impact is not confined to the cognitive level but also manifests in the disruption of metacognitive monitoring and control processes.

Furthermore, emotional disruptions formed a critical barrier to self-regulation among our participants. While research on metacognition has traditionally focused on cognitive processes such as planning, monitoring, and evaluation, it has rarely considered emotional awareness and regulation as integral parts of metacognition~\cite{jacobs1987children, schraw1998promoting, schraw1995metacognitive}. Yet, emotions are inseparable from cognition: awareness of one's affective states is itself a form of metacognitive awareness, and regulating those emotions directly influences the ability to sustain attention and monitor progress~\cite{merkebu2023metacognitive, moran2016anxiety, wiswede2009negative}.
Our findings showed that the emotional dimension of metacognition is crucial for sustaining task engagement, as participants frequently described how anxiety, frustration, or fear of failure not only slowed their task progress but also led to task avoidance. Unlike difficulties with time estimation or prioritization, which can be addressed through external structure, emotional vulnerability often leaves students without effective coping strategies. In response, many participants envisioned supports that went beyond purely cognitive scaffolding: they imagined companionship-based GenAI designs, including virtual study partners or even fictional characters, that could provide encouragement, empathy, and positive reinforcement.
These co-design ideas highlighted a gap in current productivity tools, which typically prioritize scheduling, reminders, or information organization~\cite{lund2021less,desrochers2019evaluation, wiese2023adding}, but overlook the affective regulation that underpins task engagement. 

Taken together, our findings reframe ``poor task management'' in ADHD as a metacognitive problem space that spans \textit{knowledge} (judging task demands and one’s capacity), \textit{abilities} (monitoring and control), and \textit{emotion} (awareness and regulation). University contexts that fragment task information across platforms and formats amplify extraneous cognitive load, while emotional vulnerability further erodes self-regulation, jointly producing avoidance and inconsistent follow-through. Recognizing these intertwined mechanisms points to support that front-loads cognitive offloading, integrates dispersed task cues, and scaffolds emotion regulation alongside planning and monitoring. By centering metacognition in both its cognitive and affective facets, we offer a coherent account of why everyday academic tasks become disproportionately effortful for students with ADHD and a roadmap for interventions that target the right levers of change. In the next section, we reflect on how these findings and participants' design ideas can inform the use of a powerful everyday technology, GenAI, in university students’ lives to support their productivity. 

\subsection{Optimizing the Benefits of GenAI to Support Metacognition in Task Management}
%- How to use GenAI to promote metacognition (knowledge, ability); The fluidity between knowledge and ability; How GenAI can better support these metacognitive aspects
% Design Implications
% What broader design principles emerge across the ideas (e.g., personalization, transparency, scaffolding, co-adaptation)?
% How might these principles guide future systems beyond your specific context?
% Are there tensions (e.g., between simplicity and flexibility, autonomy and guidance) that future designers should consider?
% How feasible are these ideas for real-world implementation?
% What kinds of technologies, infrastructures, or resources would be needed?
% What user adoption or ethical issues might arise?
% Critical Reflection
% What limitations or blind spots may remain in the design ideas?
% Whose perspectives might not have been captured?
% How might future research address these omissions?

% From participants' design ideas, we found that while GenAI introduces many opportunities for providing metacognitive support by clarifying tasks, tracking progress, and supporting attention and emotion, both participants and experts shared concerns about the potential risk, such as over-reliance, distraction, and mismatched support. In this section, drawing on our findings, we will discuss the strengths and pitfalls of GenAI in providing metacognitive support for students with ADHD.

In our co-design study, participants shared ideas on how GenAI could enhance their metacognitive knowledge and abilities. However, experts pointed out several pitfalls of relying on GenAI, including the risk of reinforcing biased beliefs, fostering over-reliance, and introducing new distractions rather than alleviating them.
In the following, we discuss how GenAI can act as a double-edged sword in assisting academic task management for individuals with ADHD, and how to best leverage it by mitigating potential risks.

%by addressing not just surface-level performance, but also their underlying cognitive processes of monitoring, control, and regulating behaviors. At the same time, we recognize that GenAI can function as a double-edged sword and discuss how to mitigate potential pitfalls brought up by experts.

\subsubsection{Promote Reflection, Not Automation}
To enhance metacognitive knowledge, an important step is to promote self-reflection---an ability to critically examine one's own thoughts, intentions, and strategies to complete the goals~\cite{desautel2009becoming}. Participants' design ideas suggested that GenAI’s advanced context understanding capabilities have the potential to facilitate reflection in two aspects.

First, as our participants envisioned, GenAI can enhance their task awareness and help them reflect on possible sources containing task information and collaborate with them to decompose tasks into manageable steps. By extracting and organizing such information into a structured task list, GenAI could not only reduce students' cognitive load for processing task information, but also prompt students to reflect on what constitutes a complete task representation~\cite{zimmerman1986becoming}. 
Previous systems have demonstrated that automatic synthesis of information across sources could reduce time spent on planning~\cite{rahman2020mixtape, Grover2020VirtualAgent}. Yet, these systems primarily functioned as an automated planner for a narrow context (e.g., web design projects) and imposed additional burden on project managers to resolve conflicts among the sources.
With GenAI, such applications could be extended through its flexible information-understanding and reasoning abilities across heterogeneous sources. Importantly, consistent with the mixed-initiative approach emphasized in prior work~\cite{rahman2020mixtape}, we advocate GenAI should go beyond simply producing a finished plan. This concern is also underscored by prior studies showing that when users fully rely on GenAI outputs to complete their work and develop a dependency on the system, it could undermine their ability of critical thinking~\cite{lee2025impact}. Instead, GenAI should act as a reflective partner—engaging students in dialogue, encouraging them to validate and refine the structured information, and helping them cultivate a stronger awareness of what their tasks involve.

%This aligns with the participants' design ideas to leverage GenAI as an external scaffold to collect, organize, and prompt reflection of task information, thereby promoting the formation of students' task awareness. Future tool design should not only automate the integration and structuring of multimodal task-related information for students but also guide them in reflecting on the completeness of tasks, such as prompting students to define task criteria and check for any missing key points.

Second, GenAI can act as an interactive calibration tool, prompting students to reflect on the relationship between task demands and their own abilities. Incorporating contextual cues (e.g., historical performance data, conventional benchmarks, or user-provided inputs) can help students reconsider unrealistic expectations about task completion time, difficulty, or priority.
As shown in our study, participants hoped that AI could proactively remind them when they overestimated or underestimated the time required for specific tasks and help them reflect on task priority. 
Beyond reducing misjudgments, calibration can also strengthen self-awareness and protect self-efficacy: by setting more realistic expectations, students are more likely to experience success and sustain motivation. Prior work on metacognitive monitoring highlights the importance of accurate self-calibration for effective learning and self-regulation~\cite{dunlosky2008metacognition, buehler1994exploring}. Similarly, research on the planning fallacy shows that optimism bias often drives underestimation of time requirements~\cite{buehler1994exploring}. By surfacing historical patterns and prompting reflection, GenAI could help students with ADHD overcome these biases, adapt strategies when plans fail, and ultimately develop more flexible and resilient task management habits.

\subsubsection{Balance Interests with Reality, Not Reinforce Bias}
Although participants suggested that GenAI can provide personalized feedback and progress visualization, experts pointed out that the effectiveness of the intervention depended on whether the support balanced students' intrinsic interests with realistic task demands. 
They noted that for individuals with ADHD, interest-driven strategies can be powerful motivators to overcome initiation barriers, which aligned with findings in previous studies~\cite{morsink2017motivates, scheres2008temporal, bertilsdotter2023intensity}. However, if task prioritization is guided only by personal preference, important deadlines or high-stakes assignments may be neglected.
Another critical concern is that current GenAI systems are often designed to be agreeable, mirroring majority perspectives or simply validating users' inputs~\cite{dohnany2025technological}, for example, OpenAI publicly rolled back a GPT-4o update after acknowledging sycophantic tendencies in the model~\cite{openaisycophancy2025}. This communication style may inadvertently reinforce the biases and avoidance patterns that students with ADHD struggle with. Our expert participants echoed this worry: by defaulting to validation rather than challenge, GenAI risks amplifying procrastination. For instance, if a student wishes to spend hours perfecting a creative project instead of preparing for an imminent exam, an overly agreeable GenAI might endorse this choice rather than prompting reflection on its long-term consequences.

Such reinforcement is not a neutral misstep; it undermines metacognitive growth. Students may recognize a discrepancy between effort and outcomes but remain unable to identify its root causes, further entrenching distorted self-knowledge (e.g., ``\textit{I'm just bad at time management}'') rather than helping them refine strategies~\cite{parsakia2023effect, harpin2016long, carik2025exploring}. Over time, this risks fostering dependence on AI feedback without building transferable self-regulatory skills~\cite{qadir2023engineering}.
To avoid this pitfall, GenAI systems must adopt a balanced stance: supporting interest-driven engagement while also surfacing realistic considerations such as deadlines, workload, and task importance. Such a balance could involve presenting the benefits of following one's interest and the risks of neglecting important tasks, thereby scaffolding students' metacognitive reflection without imposing rigid prescriptions. By guiding students to weigh their interests alongside external demands, GenAI can help foster more adaptive decision-making rather than simply echoing users' biases.

% as GenAI is trained on massive datasets in which neurodivergent perspectives are underrepresented, it introduces risks that go beyond strategy misalignment, potentially undermining the development of metacognitive knowledge. 

% Previous research highlights that GenAI may propagate historical patterns of prejudice encoded within limited training datasets and improperly restrict opportunities for minority or marginalized groups \cite{iloanusi2024ai}. Similarly, Carik et al. analyzed discussions in neurodivergent communities on Reddit and found that AI judgments and responses are predominantly shaped by the majority's perspective, which fails to capture the divergent thinking processes of individuals with ADHD, often making these responses ineffective for them \cite{carik2025exploring}. When students adopt such mismatched strategies and encounter failure, it undermines their metacognitive monitoring. Although students may notice a discrepancy between effort and progress, they still find it difficult to determine the underlying cause. Over time, this can reinforce inaccurate self-knowledge (e.g., ``I am poor at time management'') instead of supporting more accurate reflections (e.g., ``That strategy does not suit my needs''), ultimately diminishing their ability to identify and internalize effective strategies. 

% \subsubsection{Foster Emotional Regulation, Not Emotional Dependence}
\subsubsection{Encourage Emotional Growth, Not Dependency}
Our study corroborated that emotional struggles were common among individuals with ADHD, which further complicated their task management routines~\cite{bodalski2023adhd, unver2022metacognitive}. Prior research has also shown that individuals with ADHD often experience emotion dysregulation, such as irritability and low frustration tolerance, which is closely associated with poorer academic and occupational performance as well as difficulties in self-esteem and daily functioning~\cite{soler2023evidence, shaw2014emotion}. In response, participants hoped GenAI could simulate a virtual study companion that created a psychologically safe environment, providing an outlet to cope with negative emotions without introducing the social pressure of a human partner. 
Beyond stress-free companionship, some participants hoped the AI could take on the persona of their favorite game or anime characters, offering personalized rewards or even intimate interactions. However, when GenAI primarily offers external rewards or validation, it risks encouraging reliance on affective support rather than cultivating sustainable emotional regulation skills~\cite{ryan2000self}. Our expert participants also shared this perspective, noting that overly intimate or flattering responses could be counterproductive.

To ensure that GenAI support remains aligned with the goal of enhancing emotional awareness and regulation, it is important to position these systems as productivity-oriented tools rather than affective companions. This emphasizes that emotional regulation is not an end in itself but a means to sustain engagement with academic tasks and foster metacognitive growth. 
Building on recent work on AI-powered productivity support~\cite{hu2025ai, li2024stayfocused}, it is feasible for GenAI to facilitate emotional regulation while avoiding the risks of over-comforting or reinforcing avoidance behaviors. For instance, Hu et al. framed AI-powered venting tools as a journaling assistant, which more effectively reduced negative emotions with high and medium arousal (e.g., anger, frustration, and fear), compared to traditional journaling methods~\cite{hu2025ai}. Likewise, Li et al. demonstrated that by highlighting the AI's role as a focus companion, it could help college students stay focused on their goals and reduce compulsive smartphone use~\cite{li2024stayfocused}. Overall, the evidence points to framing GenAI as a focus- and reflection-oriented partner that helps students notice, label, and regulate emotions in service of task progress, rather than as an endlessly soothing companion. Concretely, this entails brief, structured check-ins, goal-linked reframing, and graduated prompts that return attention to the task after emotional validation. By aligning emotional support with action, GenAI can cultivate metacognitive resilience while minimizing dependency.

\section{Limitation and Future Work}
% the method itself, the design ideas are not validated
% potential risks
% In this study, participants' specific design ideas are not validated. However, the focus of this study was not on evaluating the specific effectiveness of GenAI tools in enhancing the productivity of ADHD university students. As an exploratory step to understand, we aimed to explore the potential of GenAI in addressing the unique daily productivity challenges faced by ADHD university students through co-design. Our research gathered rich empirical insights into the productivity challenges encountered by participants and their expectations of how GenAI could address these challenges, which helped answer the RQ.

In this research, we focused on exploring the potential of GenAI in scaffolding metacognition, which may overlook potential design solutions that could be more easily implemented using traditional techniques. However, this exploration was motivated by the growing prevalence of GenAI in daily productivity contexts and the likelihood that such systems will become increasingly difficult to separate from students' academic lives in the near future.
Notably, we focused on extracting design ideas and rationales from a metacognitive perspective, which has been underexplored in both HCI and ADHD task management research.
This perspective not only deepened our understanding of students’ task management challenges but also opened up new design opportunities for GenAI to scaffold awareness, monitoring, and regulation in ways traditional tools have overlooked.

Second, participants in our study were all recruited through Chinese social media platforms and shared similar Mandarin-speaking cultural backgrounds,  which may limit the generalizability of their design ideas to other cultural settings. However, as a starting point, this focused sampling aligns with common practices in prior ADHD research (e.g., ~\cite{li2016association, wang2025effects}) and offers a solid foundation for examining our research questions. Future studies should broaden the scope to include culturally diverse populations, as cultural norms may influence both metacognitive processes and the perceived role of GenAI in task management.

Going forward, we plan to develop a GenAI-based task management tool to support metacognition among individuals with ADHD. This requires careful consideration of potential risks, including hallucinations~\cite{huang2025survey} and over-reliance~\cite{zhong2024personal}. One way to mitigate these risks is to prioritize user agency with more open-ended, reflective prompts rather than providing direct suggestions or solutions. Examples of such prompts can be evaluating the feasibility of their proposed task timelines, identifying potential distractions in their planned workflow, and reflecting on how past task outcomes can inform current goal-setting. %One risk stems from its probabilistic nature, which can lead to hallucinations, such as suggesting metacognitive strategies that do not align with the user's cognitive goals~\cite{huang2025survey}. Another concern is the persistent user modeling in GenAI, which may reinforce miscalibrated self-representations~\cite{he2025human, wang2025mental}. For example, initial assumptions, such as labeling a student as easily distracted, potentially limit their engagement with alternative strategies or more challenging tasks~\cite{kearney2025echoes}. Future GenAI tool designs could consider incorporating mechanisms that, when the AI's confidence in an answer falls below a threshold, prevent a direct response and instead prompt users with clarifying questions or suggest alternative sources to reduce hallucinations~\cite{xureducing}. Additionally, the GenAI tools could enable users to examine and revise their representations, thereby reducing the risk of self-reinforcing personalization~\cite{broklyn2024ai}.

Additionally, future work can explore the integration of personal and contextual data to improve the tools' adaptive support. Building upon prior research, physiological data such as eye movements and heart rate, can signal users' attention levels and emotional states, thereby informing the timing and form of support (e.g., whether to interrupt ongoing work, when to initiate reflection, or how to adjust the tone of feedback to align with one's affective needs)~\cite{griffiths2017sustained, mauri2010psychophysiological, porges1992autonomic}. Individuals' previous task records or system interaction logs can also enhance this adaptive framework by capturing behavioral patterns (e.g., task abandonment, repetitive queries) that indicate unmet metacognitive needs~\cite{segura2018ethical, abel2013cross}. However, such personal and physiological signals are highly sensitive and may raise privacy concerns. Future tool designs should follow privacy-by-design principles by minimizing data collection, prioritizing on-device processing when possible, and clearly communicating how data are collected and used~\cite{cavoukian2009privacy, schaub2015design}. Furthermore, physiological sensing and behavioral traces can be noisy, missing, and context-dependent. To mitigate the risk of inappropriate system responses, tools should provide users with granular controls, such as independent on/off options and clear manual override mechanisms~\cite{amershi2019guidelines,lee2024priviaware}.
%By closely examining how AI-driven systems can be designed to account for neurodivergent needs, we aim to make solid knowledge contributions to accessibility and the HCI community.  %Given that individuals with ADHD may handle sensitive personal information when interacting with GenAI, future designs should prioritize transparency and user control, incorporating measures such as data anonymization and encryption to safeguard users' personal and academic information while mitigating privacy risks.

%intends to delve into the core needs of this group rather than making assumptions about universal design principles.
% Even with various cultural backgrounds, university students with ADHD often exhibit similar symptoms~\cite{}, the findings of this study are still applicable to most populations.
%Second, we did not involve experts in ADHD, such as professional psychiatrists and counsellors, who are also important stakeholders~\cite{kildea2011making}. 

\section{Conclusion}
In this study, we adopted a metacognitive lens to explore how GenAI can help university students with ADHD in addressing daily productivity challenges. We conducted individual co-design sessions with 20 university students with ADHD, followed by expert interviews to assess and validate students' design ideas. 
Drawing on participants' design ideas and expert insights, we identified three directions for GenAI to scaffold metacognition: (1) providing cognitive scaffolding to enhance task and self-awareness, (2) promoting reflective task execution to strengthen monitoring and control, and (3) facilitating emotional regulation to sustain task engagement. At the same time, experts cautioned against potential pitfalls such as reinforcing users' biases and fostering dependency.
Taken together, our study underscores both the opportunities and risks of applying GenAI to enhance metacognition in task management. By positioning GenAI as a reflective partner, future designs can better cultivate sustainable metacognitive growth and inclusive productivity support.
%Based on these challenges, participants proposed solutions leveraging GenAI to enhance task organisation and adjustment, focus maintenance, information processing, and emotional regulation in relation to productivity. Building on our findings, we discuss the design implications of using GenAI to help students with ADHD navigate complexity in less structured work environments, address inattention in independent work settings, and handle negative emotions in productivity. We hope this work will inspire future research on AI-driven productivity tools tailored to the neurodivergent community.

\begin{acks}
We thank our participants for their time and valuable contributions to this study. We also thank the reviewers for their thoughtful and constructive feedback. This research was supported by City University of Hong Kong (\# 7020106).
We used GPT-5 and Gemini 2.5 Pro to correct grammatical issues and enhance the clarity of expressions.
\end{acks}

\balance{}

\bibliographystyle{ACM-Reference-Format}
\bibliography{reference}

@book{american2013diagnostic,
  title={Diagnostic and statistical manual of mental disorders: DSM-5},
  author={{American Psychiatric Association}},
  year={2013},
  edition={5},
  publisher={American Psychiatric Association},
  address={Washington, DC},
  url = {https://www.psychiatry.org/psychiatrists/practice/dsm}
}

@article{dupaul2009college,
  title={College students with ADHD: Current status and future directions},
  author={DuPaul, George J. and Weyandt, Lisa L. and O'Dell, Sean M. and Varejao, Michael},
  journal={Journal of Attention Disorders},
  volume={13},
  number={3},
  pages={234--250},
  year={2009},
  publisher={SAGE Publications},
  doi = {10.1177/1087054709340650},
  url = {https://doi.org/10.1177/1087054709340650}
}

@inproceedings{imai2022github,
  title={Is GitHub Copilot a substitute for human pair-programming? An empirical study},
  author={Imai, Saki},
  booktitle={Proceedings of the ACM/IEEE 44th International Conference on Software Engineering: Companion Proceedings},
  year={2022},
  publisher={Association for Computing Machinery},
  address={New York, NY, USA},
  pages={319--321},
  doi = {10.1145/3510454.3522684},
  url = {https://doi.org/10.1145/3510454.3522684}
}

@article{weyandt2006adhd,
  title     = {ADHD in College Students},
  author    = {Weyandt, Lisa L. and DuPaul, George},
  journal   = {Journal of Attention Disorders},
  volume    = {10},
  number    = {1},
  pages     = {9--19},
  year      = {2006},
  publisher = {SAGE Publications},
  doi       = {10.1177/1087054705286061},
  url       = {https://doi.org/10.1177/1087054705286061}
}

@article{meaux2009adhd,
  title   = {ADHD in the College Student: A Block in the Road},
  author  = {Meaux, Julie B. and Green, Angela and Broussard, Lisa},
  journal = {Journal of Psychiatric and Mental Health Nursing},
  volume  = {16},
  number  = {3},
  pages   = {248--256},
  year    = {2009},
  doi     = {10.1111/j.1365-2850.2008.01349.x},
  url     = {https://doi.org/10.1111/j.1365-2850.2008.01349.x}
}

@article{weyandt2008adhd,
  title     = {ADHD in College Students: Developmental Findings},
  author    = {Weyandt, Lisa L. and DuPaul, George J.},
  journal   = {Developmental Disabilities Research Reviews},
  volume    = {14},
  number    = {4},
  pages     = {311--319},
  year      = {2008},
  publisher = {Wiley},
  doi       = {10.1002/ddrr.38},
  url       = {https://doi.org/10.1002/ddrr.38}
}

@inproceedings{spiel2022adhd,
  title     = {ADHD and Technology Research--Investigated by Neurodivergent Readers},
  author    = {Spiel, Katta and Hornecker, Eva and Williams, Rua Mae and Good, Judith},
  booktitle = {Proceedings of the 2022 CHI Conference on Human Factors in Computing Systems},
  articleno = {547},
  numpages  = {21},
  year      = {2022},
  publisher = {Association for Computing Machinery},
  address   = {New York, NY, USA},
  doi       = {10.1145/3491102.3517592},
  url       = {https://doi.org/10.1145/3491102.3517592}
}

@article{kirino2015sociodemographics,
  title={Sociodemographics, comorbidities, healthcare utilization and work productivity in Japanese patients with adult ADHD},
  author={Kirino, Eiji and Imagawa, Hideyuki and Goto, Taro and Montgomery, William},
  journal={PLOS ONE},
  volume={10},
  number={7},
  pages={e0132233},
  year={2015},
  publisher={Public Library of Science},
  doi = {10.1371/journal.pone.0132233},
  url = {https://doi.org/10.1371/journal.pone.0132233}
}

@article{boland2020literature,
  title={A literature review and meta-analysis on the effects of ADHD medications on functional outcomes},
  author={Boland, Heidi and DiSalvo, Maura and Fried, Ronna and Woodworth, K. Yvonne and Wilens, Timothy and Faraone, Stephen V. and Biederman, Joseph},
  journal={Journal of Psychiatric Research},
  volume={123},
  pages={21--30},
  year={2020},
  publisher={Elsevier},
  doi = {10.1016/j.jpsychires.2020.01.006},
  url = {https://doi.org/10.1016/j.jpsychires.2020.01.006}
}

@inproceedings{sonne2016changing,
  title     = {Changing Family Practices with Assistive Technology: MOBERO Improves Morning and Bedtime Routines for Children with ADHD},
  author    = {Sonne, Tobias and M{\"u}ller, J{\"o}rg and Marshall, Paul and Obel, Carsten and Gr{\o}nb{\ae}k, Kaj},
  booktitle = {Proceedings of the 2016 CHI Conference on Human Factors in Computing Systems},
  pages     = {152--164},
  year      = {2016},
  publisher = {Association for Computing Machinery},
  address   = {New York, NY, USA},
  doi       = {10.1145/2858036.2858157},
  url       = {https://doi.org/10.1145/2858036.2858157}
}

@inproceedings{barriga2023design,
  title={Design of a Mobile Application Prototype Focused on Physical Activity Management in University Students to Compensate for the Effects of ADHD},
  author={Barriga, Nadia Jimenez},
  booktitle={Proceedings of the XI Latin American Conference on Human Computer Interaction},
  year={2024},
  publisher={Association for Computing Machinery},
  address={New York, NY, USA},
  articleno={31},
  numpages={3},
  doi = {10.1145/3630970.3631071},
  url = {https://doi.org/10.1145/3630970.3631071}
}

@inproceedings{moroyoqui2022smartasko,
  title     = {SmarTasko: Supporting Short and Spontaneous Activities of Daily Living of ADHD Individuals},
  author    = {Moroyoqui, Marcos and Espina, David and Osuna, Hector and Escobedo, Lizbeth},
  booktitle = {Proceedings of the 9th Mexican International Conference on Human-Computer Interaction},
  series    = {MexIHC '22},
  pages     = {1--6},
  year      = {2022},
  publisher = {Association for Computing Machinery},
  address   = {New York, NY, USA},
  doi       = {10.1145/3565494.3565500},
  url       = {https://doi.org/10.1145/3565494.3565500}
}

@inproceedings{deprenger2010feasibility,
  title={Feasibility study of a smart pen for autonomous detection of concentration lapses during reading},
  author={DePrenger, M. and Shao, Y. and Lu, F. and Fleming, N. and Sikdar, S.},
  booktitle={2010 Annual International Conference of the IEEE Engineering in Medicine and Biology Society},
  year={2010},
  publisher={IEEE},
  address={Piscataway, NJ, USA},
  pages={1864--1867},
  doi = {10.1109/IEMBS.2010.5626256},
  url = {https://doi.org/10.1109/IEMBS.2010.5626256}
}

@inproceedings{desrochers2019evaluation,
  title={Evaluation of why individuals with ADHD struggle to find effective digital time management tools},
  author={Desrochers, Breanna and Tuson, Ella and Magee, John},
  booktitle={Proceedings of the 21st International ACM SIGACCESS Conference on Computers and Accessibility},
  publisher = {Association for Computing Machinery},
  address = {New York, NY, USA},
  url = {https://doi.org/10.1145/3308561.3354622},
  doi = {10.1145/3308561.3354622},
  pages={603--605},
  year={2019}
}

@incollection{ulfsnes2024transforming,
  title     = {Transforming Software Development with Generative AI: Empirical Insights on Collaboration and Workflow},
  author    = {Ulfsnes, Rasmus and Moe, Nils Brede and Stray, Viktoria and Skarpen, Marianne},
  booktitle = {Generative AI for Effective Software Development},
  pages     = {219--234},
  year      = {2024},
  publisher = {Springer},
  address   = {Cham},
  doi       = {10.1007/978-3-031-55642-5_10},
  url       = {https://doi.org/10.1007/978-3-031-55642-5_10}
}

@techreport{harjamaki2024report,
  title={The Report of 85 AI Tools: GenAI Content Production: Enhancing Repeatability and Automation with ChatGPT},
  author={Harjam{\"a}ki, Janne and Rantanen, Petri and Lahtinen, Daniel and Sillberg, Pekka and Saari, Mika and Gr{\"o}nman, Jere and Rasheed, Zeeshan and Sami, Abdul Malik and Abrahamsson, Pekka},
  institution={Tampere University},
  year={2024},
  url = {https://urn.fi/URN:ISBN:978-952-03-3502-1}
}

@article{jiang2024artificial,
  title={Artificial Intelligence Exploring the Patent Field},
  author={Jiang, Lekang and Goetz, Stephan},
  journal={arXiv preprint},
  volume={arXiv:2403.04105},
  pages={arXiv:2403.04105},
  year={2024},
  doi={10.48550/arXiv.2403.04105}
}

@misc{openai2023dalle3,
  author       = {OpenAI},
  title        = {DALLE-3},
  howpublished = {\url{https://openai.com/index/dall-e-3/}},
  year         = {2024},
  note         = {Accessed: 2024-09-25}
}

@book{rosqvist2020neurodiversity,
  title     = {Neurodiversity Studies: A New Critical Paradigm},
  editor    = {Bertilsdotter Rosqvist, Hanna and Chown, Nick and Stenning, Anna},
  year      = {2020},
  publisher = {Routledge},
  address   = {Abingdon, UK},
  doi       = {10.4324/9780429322297},
  url       = {https://www.taylorfrancis.com/books/edit/10.4324/9780429322297/neurodiversity-studies-hanna-rosqvist-nick-chown-anna-stenning},
  isbn      = {9780429322297}
}

@misc{CDC_ADHD_Symptoms,
  author       = {Centers for Disease Control and Prevention},
  title        = {Signs and Symptoms of ADHD},
  howpublished = {\url{https://www.cdc.gov/adhd/signs-symptoms/?CDC_AAref_Val=https://www.cdc.gov/ncbddd/adhd/diagnosis.html}},
  year         = {2024},
  note         = {Accessed: 2024-09-25}
}

@article{gropper2009pilot,
  title={A pilot study of working memory and academic achievement in college students with ADHD},
  author={Gropper, Rachel J. and Tannock, Rosemary},
  journal={Journal of Attention Disorders},
  volume={12},
  number={6},
  pages={574--581},
  year={2009},
  publisher={SAGE Publications},
  doi = {10.1177/1087054708320390},
  url = {https://doi.org/10.1177/1087054708320390}
}

@article{niermann2014relation,
  title   = {The Relation between Procrastination and Symptoms of Attention-Deficit Hyperactivity Disorder (ADHD) in Undergraduate Students},
  author  = {Niermann, Hannah C. M. and Scheres, Anouk},
  journal = {International Journal of Methods in Psychiatric Research},
  volume  = {23},
  number  = {4},
  pages   = {411--421},
  year    = {2014},
  doi     = {10.1002/mpr.1440},
  url     = {https://doi.org/10.1002/mpr.1440}
}

@article{palmini2008professionally,
  title   = {Professionally Successful Adults with Attention-Deficit/Hyperactivity Disorder (ADHD): Compensation Strategies and Subjective Effects of Pharmacological Treatment},
  author  = {Palmini, Andre},
  journal = {Dementia \& Neuropsychologia},
  volume  = {2},
  number  = {1},
  pages   = {63--70},
  year    = {2008},
  doi     = {10.1590/S1980-57642009DN20100013},
  url     = {https://doi.org/10.1590/S1980-57642009DN20100013}
}

@book{brown2005attention,
  title={Attention deficit disorder: The unfocused mind in children and adults},
  author={Brown, Thomas E.},
  year={2005},
  publisher={Yale University Press},
  address={New Haven, CT},
  url = {https://www.amazon.com/Attention-Deficit-Disorder-Unfocused-University/dp/0300119895}
}

@article{kessler2006prevalence,
  title={The prevalence and correlates of adult ADHD in the United States: results from the National Comorbidity Survey Replication},
  author={Kessler, Ronald C and Adler, Lenard and Barkley, Russell and Biederman, Joseph and Conners, C Keith and Demler, Olga and Faraone, Stephen V and Greenhill, Laurence L and Howes, Mary J and Secnik, Kristina and others},
  journal={American Journal of Psychiatry},
  volume={163},
  number={4},
  pages={716--723},
  year={2006},
  publisher={American Psychiatric Association},
  doi={10.1176/appi.ajp.163.4.716},
  url={https://doi.org/10.1176/appi.ajp.163.4.716}
}

@misc{githubCopilot,
  author = {GitHub},
  title = {GitHub Copilot},
  year = {2024},
  howpublished = {\url{https://github.com/features/copilot}},
  note = {Accessed: 2024-08-30}
}

@misc{notionAI,
  author = {Notion},
  title = {Notion AI},
  year = {2024},
  howpublished = {\url{https://www.notion.so/product/ai}},
  note = {Accessed: 2024-08-30}
}

@misc{tencentMeetingAI,
  author = {Tencent},
  title = {Tencent Meeting AI},
  year = {2024},
  howpublished = {\url{https://meeting.tencent.com/ai/index.html}},
  note = {Accessed: 2024-08-30}
}

@misc{openaiDataAnalysis,
  author = {OpenAI},
  title = {Improvements to Data Analysis in ChatGPT},
  year = {2024},
  howpublished = {\url{https://openai.com/index/improvements-to-data-analysis-in-chatgpt/}},
  note = {Accessed: 2024-08-30}
}

@misc{replikaWebsite,
  author = {Replika},
  title = {Replika: The AI Companion Who Cares},
  year = {2024},
  howpublished = {\url{https://replika.com/}},
  note = {Accessed: 2024-08-30}
}

@misc{characterAIWebsite,
  author = {Character AI},
  title = {Character AI},
  year = {2024},
  howpublished = {\url{https://character.ai/}},
  note = {Accessed: 2024-08-30}
}

@misc{midjourneyWebsite,
  author = {MidJourney},
  title = {MidJourney: AI Art Generator},
  year = {2024},
  howpublished = {\url{https://www.midjourney.com/home}},
  note = {Accessed: 2024-08-30}
}

@article{kofler2011working,
  title={Working memory deficits and social problems in children with ADHD},
  author={Kofler, Michael J and Rapport, Mark D and Bolden, Jennifer and Sarver, Dustin E and Raiker, Joseph S and Alderson, R Matt},
  journal={Journal of Abnormal Child Psychology},
  volume={39},
  number={6},
  pages={805--817},
  year={2011},
  publisher={Springer},
  doi = {10.1007/s10802-011-9492-8},
  url = {https://doi.org/10.1007/s10802-011-9492-8}
}

@article{roberts2012constraints,
  title   = {Constraints on Information Processing Capacity in Adults with ADHD},
  author  = {Roberts, Walter and Milich, Richard and Fillmore, Mark T.},
  journal = {Neuropsychology},
  volume  = {26},
  number  = {6},
  pages   = {695--703},
  year    = {2012},
  doi     = {10.1037/a0030296},
  url     = {https://doi.org/10.1037/a0030296},
  pmid    = {23106116},
  pmcid   = {PMC6996017}
}

@article{morsink2017motivates,
  title   = {What Motivates Individuals with ADHD? A Qualitative Analysis from the Adolescent’s Point of View},
  author  = {Morsink, Sarah and Sonuga-Barke, Edmund and Mies, Gabry and Glorie, Nathalie and Lemiere, Jurgen and Van der Oord, Saskia and Danckaerts, Marina},
  journal = {European Child \& Adolescent Psychiatry},
  volume  = {26},
  number  = {8},
  pages   = {923--932},
  year    = {2017},
  doi     = {10.1007/s00787-017-0961-7},
  url     = {https://doi.org/10.1007/s00787-017-0961-7}
}

@inproceedings{cibrian2020supporting,
  title     = {Supporting Self-Regulation of Children with ADHD Using Wearables: Tensions and Design Challenges},
  author    = {Cibrian, Franceli L. and Lakes, Kimberley D. and Tavakoulnia, Arya and Guzman, Kayla and Schuck, Sabrina and Hayes, Gillian R.},
  booktitle = {Proceedings of the 2020 CHI Conference on Human Factors in Computing Systems},
  pages     = {1--13},
  year      = {2020},
  publisher = {Association for Computing Machinery},
  address   = {New York, NY, USA},
  doi       = {10.1145/3313831.3376837},
  url       = {https://doi.org/10.1145/3313831.3376837}
}

@article{fusch2015we,
  title={Are we there yet? Data saturation in qualitative research},
  author={Fusch, Patricia I and Ness, Lawrence R},
  journal={The Qualitative Report},
  volume={20},
  number={9},
  pages={1408--1416},
  year={2015},
  doi={10.46743/2160-3715/2015.2281}
}

@article{kasper2012moderators,
  title={Moderators of working memory deficits in children with attention-deficit/hyperactivity disorder (ADHD): A meta-analytic review},
  author={Kasper, Lisa J. and Alderson, R. Matt and Hudec, Kristen L.},
  journal={Clinical Psychology Review},
  volume={32},
  number={7},
  pages={605--617},
  year={2012},
  publisher={Elsevier},
  doi = {10.1016/j.cpr.2012.07.001},
  url = {https://doi.org/10.1016/j.cpr.2012.07.001}
}

@article{kofler2018working,
  title={Working memory and organizational skills problems in ADHD},
  author={Kofler, Michael J and Sarver, Dustin E and Harmon, Sherelle L and Moltisanti, Allison and Aduen, Paula A and Soto, Elia F and Ferretti, Nicole},
  journal={Journal of Child Psychology and Psychiatry},
  volume={59},
  number={1},
  pages={57--67},
  year={2018},
  publisher={Wiley},
  doi={10.1111/jcpp.12773},
  url={https://doi.org/10.1111/jcpp.12773}
}

@article{song2021prevalence,
  title   = {The Prevalence of Adult Attention-Deficit Hyperactivity Disorder: A Global Systematic Review and Meta-Analysis},
  author  = {Song, Peige and Zha, Mingming and Yang, Qingwen and Zhang, Yan and Li, Xue and Rudan, Igor},
  journal = {Journal of Global Health},
  year    = {2021},
  volume  = {11},
  pages   = {04009},
  doi     = {10.7189/jogh.11.04009}
}

@article{ginsberg2014underdiagnosis,
  title     = {Underdiagnosis of Attention-Deficit/Hyperactivity Disorder in Adult Patients: A Review of the Literature},
  author    = {Ginsberg, Ylva and Quintero, Javier and Anand, Ernie and Casillas, Marta and Upadhyaya, Himanshu P.},
  journal   = {Primary Care Companion to CNS Disorders},
  volume    = {16},
  number    = {3},
  pages     = {PCC.13r01600},
  year      = {2014},
  publisher = {Physicians Postgraduate Press},
  doi       = {10.4088/PCC.13r01600},
  url       = {https://doi.org/10.4088/PCC.13r01600}
}

@article{adamis2022adhd,
  title={ADHD in adults: A systematic review and meta-analysis of prevalence studies in outpatient psychiatric clinics},
  author={Adamis, Dimitrios and Flynn, Caroline and Wrigley, Margo and Gavin, Blanaid and McNicholas, Fiona},
  journal={Journal of Attention Disorders},
  volume={26},
  number={12},
  pages={1523--1534},
  year={2022},
  publisher={SAGE Publications Sage CA: Los Angeles, CA},
  doi = {10.1177/10870547221085503},
  url = {https://doi.org/10.1177/10870547221085503}
}

@article{quinn2004perceptions,
  title     = {Perceptions of Girls and ADHD: Results from a National Survey},
  author    = {Quinn, Patricia and Wigal, Sharon},
  journal   = {Medscape General Medicine},
  volume    = {6},
  number    = {2},
  pages     = {2},
  year      = {2004},
  url       = {https://www.ncbi.nlm.nih.gov/pmc/articles/PMC1395774/}
}

@article{blase2009self,
  title={Self-reported ADHD and adjustment in college: Cross-sectional and longitudinal findings},
  author={Blase, Stacey L. and Gilbert, Adrianne N. and Anastopoulos, Arthur D. and Costello, E. Jane and Hoyle, Rick H. and Swartzwelder, H. Scott and Rabiner, David L.},
  journal={Journal of Attention Disorders},
  volume={13},
  number={3},
  pages={297--309},
  year={2009},
  publisher={SAGE Publications Sage CA: Los Angeles, CA},
  doi = {10.1177/1087054709334446},
  url = {https://doi.org/10.1177/1087054709334446}
}

@article{wolf2001college,
  title     = {College Students with ADHD and Other Hidden Disabilities: Outcomes and Interventions},
  author    = {Wolf, Lorraine E.},
  journal   = {Annals of the New York Academy of Sciences},
  volume    = {931},
  number    = {1},
  pages     = {385--395},
  year      = {2001},
  publisher = {Wiley},
  doi       = {10.1111/j.1749-6632.2001.tb05792.x},
  url       = {https://doi.org/10.1111/j.1749-6632.2001.tb05792.x}
}

@article{mirsky1999model,
  title   = {A Model of Attention and Its Relation to ADHD},
  author  = {Mirsky, Allan F. and Pascualvaca, Daisy M. and Duncan, Connie C. and French, Louis M.},
  journal = {Mental Retardation and Developmental Disabilities Research Reviews},
  volume  = {5},
  number  = {3},
  pages   = {169--176},
  year    = {1999},
  doi     = {10.1002/(SICI)1098-2779(1999)5:3<169::AID-MRDD2>3.0.CO;2-K},
  url     = {https://doi.org/10.1002/(SICI)1098-2779(1999)5:3<169::AID-MRDD2>3.0.CO;2-K}
}

@online{nhsadhd,
  author       = {{NHS}},
  title        = {Symptoms of Attention Deficit Hyperactivity Disorder (ADHD)},
  year         = {2021},
  organization = {National Health Service},
  url          = {https://www.nhs.uk/conditions/attention-deficit-hyperactivity-disorder-adhd/symptoms/},
  urldate      = {2024-12-01}
}

@article{sedgwick2018university,
  title   = {University Students with Attention Deficit Hyperactivity Disorder (ADHD): A Literature Review},
  author  = {Sedgwick, J. A.},
  journal = {Irish Journal of Psychological Medicine},
  volume  = {35},
  number  = {3},
  pages   = {221--235},
  year    = {2018},
  doi     = {10.1017/ipm.2017.20},
  url     = {https://doi.org/10.1017/ipm.2017.20}
}

@article{spencer2006adhd,
  title     = {ADHD and Comorbidity in Childhood},
  author    = {Spencer, Thomas J.},
  journal   = {Journal of Clinical Psychiatry},
  volume    = {67},
  number    = {Suppl. 8},
  pages     = {27--31},
  year      = {2006},
  publisher = {Physicians Postgraduate Press},
  url       = {https://www.psychiatrist.com/jcp/adhd-comorbidity-childhood-2/}
}

@article{kwon2018difficulties,
  title={Difficulties faced by university students with self-reported symptoms of attention-deficit hyperactivity disorder: a qualitative study},
  author={Kwon, Soo Jin and Kim, Yoonjung and Kwak, Yeunhee},
  journal={Child and Adolescent Psychiatry and Mental Health},
  volume={12},
  number={1},
  pages={12},
  year={2018},
  publisher={BioMed Central},
  doi={10.1186/s13034-018-0218-3},
  url={https://doi.org/10.1186/s13034-018-0218-3}
}

@article{braun2006using,
  title={Using thematic analysis in psychology},
  author={Braun, Virginia and Clarke, Victoria},
  journal={Qualitative Research in Psychology},
  volume={3},
  number={2},
  pages={77--101},
  year={2006},
  publisher={Taylor \& Francis},
  doi = {10.1191/1478088706qp063oa},
  url = {https://doi.org/10.1191/1478088706qp063oa}
}

@inproceedings{li2024stayfocused,
  author = {Li, Zhuoyang and Liang, Minhui and Lc, Ray and Luo, Yuhan},
  title = {StayFocused: Examining the Effects of Reflective Prompts and Chatbot Support on Compulsive Smartphone Use},
  year = {2024},
  isbn = {9798400703300},
  publisher = {Association for Computing Machinery},
  address = {New York, NY, USA},
  url = {https://doi.org/10.1145/3613904.3642479},
  doi = {10.1145/3613904.3642479},
  booktitle = {Proceedings of the 2024 CHI Conference on Human Factors in Computing Systems},
  articleno = {247},
  numpages = {19},
  location = {Honolulu, HI, USA},
  series = {CHI '24}
}

@article{wills2014implementation,
  title     = {Implementation of a Self-Monitoring Application to Improve On-Task Behavior: A High-School Pilot Study},
  author    = {Wills, Howard P. and Mason, Benjamin A.},
  journal   = {Journal of Behavioral Education},
  volume    = {23},
  number    = {4},
  pages     = {421--434},
  year      = {2014},
  publisher = {Springer},
  doi       = {10.1007/s10864-014-9204-x},
  url       = {https://doi.org/10.1007/s10864-014-9204-x}
}

@inproceedings{van2016deep,
  title     = {DEEP: A Biofeedback Virtual Reality Game for Children At-risk for Anxiety},
  author    = {Van Rooij, Marieke and Lobel, Adam and Harris, Owen and Smit, Niki and Granic, Isabela},
  booktitle = {Proceedings of the 2016 CHI Conference Extended Abstracts on Human Factors in Computing Systems},
  pages     = {1989--1997},
  year      = {2016},
  publisher = {ACM},
  address   = {New York, NY, USA},
  doi       = {10.1145/2851581.2892452},
  url       = {https://doi.org/10.1145/2851581.2892452}
}

@article{jiang2015preliminary,
  title={A preliminary multiple case report of neurocognitive training for children with AD/HD in China},
  author={Jiang, Han and Johnstone, Stuart J.},
  journal={SAGE Open},
  volume={5},
  number={2},
  pages={2158244015586811},
  year={2015},
  publisher={SAGE Publications},
  doi = {10.1177/2158244015586811},
  url = {https://doi.org/10.1177/2158244015586811}
}

@misc{thiergart2021understanding,
  title        = {Understanding Emails and Drafting Responses: An Approach Using GPT-3},
  author       = {Thiergart, Jonas and Huber, Stefan and {\"U}bellacker, Thomas},
  year         = {2021},
  eprint       = {2102.03062},
  archivePrefix= {arXiv},
  primaryClass = {cs.CL},
  url          = {https://arxiv.org/abs/2102.03062}
}

@inproceedings{larsenllm,
  title={LLM-powered conversational AI in customer service: Users’ expectations and anticipated use},
  author={Larsen, Anna Gr{\o}ndahl and Skjuve, Marita and Kvale, Knut and F{\o}lstad, Asbj{\o}rn},
  booktitle={International Symposium on Chatbots and Human-Centered AI},
  pages={217--233},
  year={2025},
  publisher={Springer},
  address={Cham, Switzerland},
  doi={10.1007/978-3-031-88045-2_14},
  url={https://doi.org/10.1007/978-3-031-88045-2_14}
}

@article{zhou2024exploring,
author = {Zhibin Zhou and Yaoqi Li and Junnan Yu},
title = {Exploring the application of LLM-based AI in UX design: an empirical case study of ChatGPT},
journal = {Human–Computer Interaction},
volume = {0},
number = {0},
pages = {1--33},
year = {2024},
publisher = {Taylor \& Francis},
doi = {10.1080/07370024.2024.2420991},
}

@article{prevatt2011time,
  title   = {Time Estimation Abilities of College Students with ADHD},
  author  = {Prevatt, Frances and Proctor, Briley and Baker, Leigh and Garrett, Lori and Yelland, Sherry},
  journal = {Journal of Attention Disorders},
  volume  = {15},
  number  = {7},
  pages   = {531--538},
  year    = {2011},
  doi     = {10.1177/1087054710370673},
  url     = {https://doi.org/10.1177/1087054710370673}
}

@misc{nih_adhd,
  author = {{National Institute of Mental Health}},
  title = {Attention-Deficit/Hyperactivity Disorder (ADHD)},
  year = 2023,
  url = {https://www.nimh.nih.gov/health/topics/attention-deficit-hyperactivity-disorder-adhd},
  note = {Accessed: 2025-01-15}
}

@inproceedings{Grover2020VirtualAgent,
author = {Grover, Ted and Rowan, Kael and Suh, Jina and McDuff, Daniel and Czerwinski, Mary},
title = {Design and Evaluation of Intelligent Agent Prototypes for Assistance with Focus and Productivity at Work},
year = {2020},
isbn = {9781450371186},
publisher = {Association for Computing Machinery},
address = {New York, NY, USA},
url = {https://doi.org/10.1145/3377325.3377507},
doi = {10.1145/3377325.3377507},
booktitle = {Proceedings of the 25th International Conference on Intelligent User Interfaces},
pages = {390–400},
numpages = {11},
location = {Cagliari, Italy},
series = {IUI '20}
}

@article{carik2025exploring,
  title={Exploring Large Language Models Through a Neurodivergent Lens: Use, Challenges, Community-Driven Workarounds, and Concerns},
  author={Carik, Buse and Ping, Kaike and Ding, Xiaohan and Rho, Eugenia H.},
  journal={Proceedings of the ACM on Human-Computer Interaction},
  volume={9},
  number={GROUP},
  pages={1--28},
  year={2025},
  publisher={Association for Computing Machinery},
  doi = {10.1145/3701194},
  url = {https://doi.org/10.1145/3701194}
}

@article{antshel2014cognitive,
  title={Cognitive behavioral therapy for adolescents with ADHD},
  author={Antshel, Kevin M. and Olszewski, Amy K.},
  journal={Child and Adolescent Psychiatric Clinics of North America},
  volume={23},
  number={4},
  pages={825--842},
  year={2014},
  publisher={Elsevier},
  doi = {10.1016/j.chc.2014.05.001},
  url = {https://doi.org/10.1016/j.chc.2014.05.001}
}

@misc{chatgpt2024,
  author = {OpenAI},
  title = {ChatGPT},
  year = {2024},
  url = {https://chatgpt.com/},
  note = {Accessed: 2024-10-19}
}

@misc{yiyan2024,
  author = {baidu},
  title = {yiyan},
  year = {2024},
  url = {https://yiyan.baidu.com/},
  note = {Accessed: 2024-10-19}
}

@inproceedings{qin2024charactermeet,
  title     = {CharacterMeet: Supporting Creative Writers' Entire Story Character Construction Processes Through Conversation with LLM-Powered Chatbot Avatars},
  author    = {Qin, Hua Xuan and Jin, Shan and Gao, Ze and Fan, Mingming and Hui, Pan},
  booktitle = {Proceedings of the 2024 CHI Conference on Human Factors in Computing Systems},
  series    = {CHI '24},
  articleno = {1051},
  numpages  = {19},
  year      = {2024},
  publisher = {Association for Computing Machinery},
  address   = {New York, NY, USA},
  doi       = {10.1145/3613904.3642105},
  url       = {https://doi.org/10.1145/3613904.3642105}
}

@inproceedings{de2024dialogues,
  title={Dialogues with Digital Wisdom: Can LLMs Help Us Put Down the Phone?},
  author={De Russis, Luigi and Monge Roffarello, Alberto and Scibetta, Luca},
  booktitle={Proceedings of the 2024 International Conference on Information Technology for Social Good},
  year={2024},
  publisher={Association for Computing Machinery},
  address={New York, NY, USA},
  pages={56--61},
  doi = {10.1145/3677525.3678640},
  url = {https://doi.org/10.1145/3677525.3678640}
}

@article{rahman2020mixtape,
  title     = {MixTAPE: Mixed-Initiative Team Action Plan Creation Through Semi-Structured Notes, Automatic Task Generation, and Task Classification},
  author    = {Rahman, Sajjadur and Siangliulue, Pao and Marcus, Adam},
  journal   = {Proceedings of the ACM on Human-Computer Interaction},
  volume    = {4},
  number    = {CSCW2},
  articleno = {169},
  numpages  = {26},
  year      = {2020},
  publisher = {Association for Computing Machinery},
  doi       = {10.1145/3415240},
  url       = {https://doi.org/10.1145/3415240}
}

@inproceedings{qi2025participatory,
  title     = {Participatory Design in Human-Computer Interaction: Cases, Characteristics, and Lessons},
  author    = {Qi, Xiang and Yu, Junnan},
  booktitle = {Proceedings of the 2025 CHI Conference on Human Factors in Computing Systems},
  series    = {CHI '25},
  articleno = {804},
  numpages  = {26},
  year      = {2025},
  publisher = {Association for Computing Machinery},
  address   = {New York, NY, USA},
  doi       = {10.1145/3706598.3713436},
  url       = {https://doi.org/10.1145/3706598.3713436}
}

@inproceedings{xue2025toward,
  title     = {Toward Interactive Reading: Co-designing With Adolescents to Explore Opportunities for Overcoming Reading Challenges},
  author    = {Xue, Katie and Mao, Yaxuan and Yu, Junnan and Luo, Yuhan},
  booktitle = {Proceedings of the 24th Annual ACM Conference on Interaction Design and Children (IDC '25)},
  pages     = {904--909},
  year      = {2025},
  publisher = {ACM},
  address   = {New York, NY, USA},
  doi       = {10.1145/3713043.3731503},
  url       = {https://doi.org/10.1145/3713043.3731503}
}

@inproceedings{lund2021less,
  title     = {Less is More: Exploring Support for Time Management Planning},
  author    = {Lund, John R. and Wiese, Jason},
  booktitle = {Proceedings of the 2021 ACM Designing Interactive Systems Conference},
  series    = {DIS '21},
  pages     = {392--405},
  year      = {2021},
  publisher = {Association for Computing Machinery},
  address   = {New York, NY, USA},
  doi       = {10.1145/3461778.3462133},
  url       = {https://doi.org/10.1145/3461778.3462133}
}

@article{fernie2016contribution,
  title={The contribution of metacognitions and attentional control to decisional procrastination},
  author={Fernie, Bruce A. and McKenzie, Ann-Marie and Nik{\v{c}}evi{\'c}, Ana V. and Caselli, Gabriele and Spada, Marcantonio M.},
  journal={Journal of Rational-Emotive \& Cognitive-Behavior Therapy},
  volume={34},
  number={1},
  pages={1--13},
  year={2016},
  publisher={Springer},
  doi = {10.1007/s10942-015-0222-y},
  url = {https://doi.org/10.1007/s10942-015-0222-y}
}

@article{de2017decisional,
  title={Decisional procrastination in academic settings: The role of metacognitions and learning strategies},
  author={de Palo, Valeria and Monacis, Lucia and Miceli, Silvana and Sinatra, Maria and Di Nuovo, Santo},
  journal={Frontiers in Psychology},
  volume={8},
  pages={973},
  year={2017},
  publisher={Frontiers Media SA},
  doi = {10.3389/fpsyg.2017.00973},
  url = {https://doi.org/10.3389/fpsyg.2017.00973}
}

@incollection{flavell1976metacognitive,
  title={Metacognitive aspects of problem solving},
  author={Flavell, John H.},
  booktitle={The Nature of Intelligence},
  editor={Resnick, Lauren B.},
  pages={231--236},
  year={1976},
  publisher={Lawrence Erlbaum Associates},
  address={Hillsdale, NJ, USA},
  doi = {10.4324/9781032646527-16},
  url = {https://doi.org/10.4324/9781032646527-16}
}

@article{eisenhower1954address,
  title={Address at the second assembly of the world council of churches},
  author={Eisenhower, Dwight D and Peters, Gerhard and Woolley, John T},
  journal={Evanston, Illinois},
  volume={19},
  pages={1954},
  year={1954}
}

@article{jacobs1987children,
  title={Children's metacognition about reading: Issues in definition, measurement, and instruction},
  author={Jacobs, Janis E. and Paris, Scott G.},
  journal={Educational Psychologist},
  volume={22},
  number={3--4},
  pages={255--278},
  year={1987},
  publisher={Taylor \& Francis},
  doi = {10.1080/00461520.1987.9653052},
  url = {https://doi.org/10.1080/00461520.1987.9653052}
}

@article{schraw1998promoting,
  title   = {Promoting General Metacognitive Awareness},
  author  = {Schraw, Gregory},
  journal = {Instructional Science},
  volume  = {26},
  number  = {1},
  pages   = {113--125},
  year    = {1998},
  doi     = {10.1023/A:1003044231033},
  url     = {https://doi.org/10.1023/A:1003044231033}
}

@book{wellman1990child,
  title     = {The Child's Theory of Mind},
  author    = {Wellman, Henry M.},
  year      = {1990},
  publisher = {MIT Press},
  address   = {Cambridge, MA},
  url       = {https://mitpress.mit.edu/9780262730990/the-childs-theory-of-mind/}
}

@article{merkebu2023metacognitive,
  title   = {What Is Metacognitive Reflection? The Moderating Role of Metacognition on Emotional Regulation and Reflection},
  author  = {Merkebu, Jerusalem and Kitsantas, Anastasia and Durning, Steven J. and Ma, Tinglan},
  journal = {Frontiers in Education},
  volume  = {8},
  pages   = {1166195},
  year    = {2023},
  doi     = {10.3389/feduc.2023.1166195},
  url     = {https://doi.org/10.3389/feduc.2023.1166195}
}

@inproceedings{tankelevitch2024metacognitive,
  title     = {The Metacognitive Demands and Opportunities of Generative AI},
  author    = {Tankelevitch, Lev and Kewenig, Viktor and Simkute, Auste and Scott, Ava Elizabeth and Sarkar, Advait and Sellen, Abigail and Rintel, Sean},
  booktitle = {Proceedings of the 2024 CHI Conference on Human Factors in Computing Systems},
  year      = {2024},
  publisher = {Association for Computing Machinery},
  address   = {New York, NY, USA},
  pages     = {1--24},
  doi       = {10.1145/3613904.3642902},
  url       = {https://doi.org/10.1145/3613904.3642902}
}

@article{butzbach2021metacognition,
  title={Metacognition in adult ADHD: subjective and objective perspectives on self-awareness of cognitive functioning},
  author={Butzbach, Marah and Fuermaier, Anselm B. M. and Aschenbrenner, Steffen and Weisbrod, Matthias and Tucha, Lara and Tucha, Oliver},
  journal={Journal of Neural Transmission},
  volume={128},
  number={7},
  pages={939--955},
  year={2021},
  publisher={Springer},
  doi = {10.1007/s00702-020-02293-w},
  url = {https://doi.org/10.1007/s00702-020-02293-w}
}

@article{lenartowicz2024training,
  title     = {Training of Awareness in ADHD: Leveraging Metacognition},
  author    = {Lenartowicz, Agatha and DeSchepper, Brett and Simpson, Gregory V.},
  journal   = {Journal of Psychiatry and Brain Science},
  volume    = {9},
  number    = {5},
  pages     = {e240006},
  year      = {2024},
  doi       = {10.20900/jpbs.20240006},
  url       = {https://doi.org/10.20900/jpbs.20240006}
}

@article{solanto2010efficacy,
  title   = {Efficacy of Meta-Cognitive Therapy for Adult ADHD},
  author  = {Solanto, Mary V. and Marks, David J. and Wasserstein, Jeanette and Mitchell, Katherine and Abikoff, Howard and Alvir, Jose Ma. J. and Kofman, Michele D.},
  journal = {American Journal of Psychiatry},
  volume  = {167},
  number  = {8},
  pages   = {958--968},
  year    = {2010},
  doi     = {10.1176/appi.ajp.2009.09081123},
  url     = {https://doi.org/10.1176/appi.ajp.2009.09081123}
}

@article{kajka2023assessment,
  title={The assessment of the impact of training with various metacognitive interventions on the enhancement of verbal fluency in school-age children with ADHD},
  author={Kajka, Natalia and Kulik, Agnieszka},
  journal={Journal of Attention Disorders},
  volume={27},
  number={1},
  pages={89--97},
  year={2023},
  publisher={SAGE Publications},
  doi = {10.1177/10870547221121289},
  url = {https://doi.org/10.1177/10870547221121289}
}

@article{diamond2013executive,
  title={Executive functions},
  author={Diamond, Adele},
  journal={Annual review of psychology},
  volume={64},
  number={1},
  pages={135--168},
  year={2013},
  publisher={Annual Reviews},
  url = {https://doi.org/10.1146/annurev-psych-113011-143750}
}

@article{roebers2017executive,
  title={Executive function and metacognition: Towards a unifying framework of cognitive self-regulation},
  author={Roebers, Claudia M},
  journal={Developmental review},
  volume={45},
  pages={31--51},
  year={2017},
  publisher={Elsevier},
  url = {https://doi.org/10.1016/j.dr.2017.04.001}
}

@article{rosen2017role,
  title={The role of executive functioning and technological anxiety (FOMO) in college course performance as mediated by technology usage and multitasking habits},
  author={Rosen, Larry D and Carrier, L Mark and Pedroza, Jonathan A and Elias, Stephanie and O’Brien, Kaitlin M and Lozano, Joshua and Kim, Karina and Cheever, Nancy A and Bentley, Jonathan and Ruiz, Abraham},
  journal={Psicologia Educativa},
  volume={24},
  number={1},
  pages={14},
  year={2017},
  url = {https://doi.org/10.5093/psed2018a3}
}

@article{casale2021systematic,
  title={A systematic review of metacognitions in Internet Gaming Disorder and problematic Internet, smartphone and social networking sites use},
  author={Casale, Silvia and Music{\`o}, Alessia and Spada, Marcantonio M},
  journal={Clinical Psychology \& Psychotherapy},
  volume={28},
  number={6},
  pages={1494--1508},
  year={2021},
  publisher={Wiley Online Library},
  url = { https://doi.org/10.1002/cpp.2588}
}

@article{spada2008metacognition,
  title={Metacognition, perceived stress, and negative emotion},
  author={Spada, Marcantonio M and Nik{\v{c}}evi{\'c}, Ana V and Moneta, Giovanni B and Wells, Adrian},
  journal={Personality and individual differences},
  volume={44},
  number={5},
  pages={1172--1181},
  year={2008},
  publisher={Elsevier},
  url = {https://doi.org/10.1016/j.paid.2007.11.010}
}

@article{flavell1979metacognition,
  author = {Flavell, J. H.},
  title = {Metacognition and cognitive monitoring: A new area of cognitive--developmental inquiry},
  journal = {American Psychologist},
  volume = {34},
  number = {10},
  pages = {906--911},
  year = {1979},
  doi = {10.1037/0003-066X.34.10.906}
}

@article{re2015effect,
  author = {Re, A. M. and Capodieci, A. and Cornoldi, C.},
  title = {Effect of training focused on executive functions (attention, inhibition, and working memory) in preschoolers exhibiting ADHD symptoms},
  journal = {Frontiers in Psychology},
  year = {2015},
  volume = {6},
  pages = {1161},
  doi = {10.3389/fpsyg.2015.01161}
}

@article{swanson1990influence,
  title   = {Influence of Metacognitive Knowledge and Aptitude on Problem Solving},
  author  = {Swanson, H. Lee},
  journal = {Journal of Educational Psychology},
  volume  = {82},
  number  = {2},
  pages   = {306--314},
  year    = {1990},
  doi     = {10.1037/0022-0663.82.2.306},
  url     = {https://doi.org/10.1037/0022-0663.82.2.306}
}

@article{ertmer1996expert,
  title={The expert learner: Strategic, self-regulated, and reflective},
  author={Ertmer, Peggy A. and Newby, Timothy J.},
  journal={Instructional Science},
  volume={24},
  number={1},
  pages={1--24},
  year={1996},
  publisher={Springer},
  doi = {10.1007/BF00156001},
  url = {https://doi.org/10.1007/BF00156001}
}

@article{schraw1995metacognitive,
  title     = {Metacognitive Theories},
  author    = {Schraw, Gregory and Moshman, David},
  journal   = {Educational Psychology Review},
  volume    = {7},
  number    = {4},
  pages     = {351--371},
  year      = {1995},
  publisher = {Springer},
  doi       = {10.1007/BF02212307},
  url       = {https://doi.org/10.1007/BF02212307}
}

@article{sluiter2020exploring,
  title   = {Exploring Neuropsychological Effects of a Self-Monitoring Intervention for ADHD-Symptoms in School},
  author  = {Sluiter, Maruschka N. and Groen, Yvonne and de Jonge, Peter and Tucha, Oliver},
  journal = {Applied Neuropsychology: Child},
  volume  = {9},
  number  = {3},
  pages   = {246--258},
  year    = {2020},
  doi     = {10.1080/21622965.2019.1575218},
  url     = {https://doi.org/10.1080/21622965.2019.1575218}
}

@article{alderson2013attention,
  title={Attention-deficit/hyperactivity disorder (ADHD) and working memory in adults: a meta-analytic review},
  author={Alderson, R. Matt and Kasper, Lisa J. and Hudec, Kristen L. and Patros, Connor H. G.},
  journal={Neuropsychology},
  volume={27},
  number={3},
  pages={287--302},
  year={2013},
  publisher={American Psychological Association},
  doi = {10.1037/a0032371},
  url = {https://doi.org/10.1037/a0032371}
}

@article{xu2025enhancing,
  title   = {Enhancing Self-Regulated Learning and Learning Experience in Generative AI Environments: The Critical Role of Metacognitive Support},
  author  = {Xu, Xiaoqing and Qiao, Lifang and Cheng, Nuo and Liu, Hongxia and Zhao, Wei},
  journal = {British Journal of Educational Technology},
  year    = {2025},
  volume  = {56},
  number  = {5},
  pages   = {1842--1863},
  month   = sep,
  doi     = {10.1111/bjet.13599}
}

@incollection{lin2024genai,
  title     = {GenAI Tools in Academic Reading: A Study on AI-Assisted Metacognitive Strategies and Emotional Reactions},
  author    = {Lin, Haoming and Chen, Ziqi and Wei, Wei and Lu, Handan},
  booktitle = {Proceedings of the Asia Education Technology Symposium (AETS 2024)},
  series    = {Lecture Notes in Educational Technology},
  pages     = {98--112},
  year      = {2024},
  publisher = {Springer},
  address   = {Singapore},
  doi       = {10.1007/978-981-96-4952-5_7},
  url       = {https://doi.org/10.1007/978-981-96-4952-5_7}
}

@article{li2025generative,
  title={Generative artificial intelligence-supported programming education: Effects on learning performance, self-efficacy and processes},
  author={Li, Siran and Liu, Jiangyue and Dong, Qianyan},
  journal={Australasian Journal of Educational Technology},
  volume={41},
  number={3},
  pages={1--25},
  year={2025},
  doi={10.14742/ajet.9932}
}

@incollection{ligenai2025,
  title={How GenAI Affects Metacognition in Self-Regulated Learning: Between Enhancement and Inhibition},
  author={Li, Zijian and Fan, Yizhou},
  booktitle={Learning with Generative Artificial Intelligence},
  pages={148--177},
  year={2025},
  publisher={Routledge},
  address={London, UK},
  doi={10.4324/9781003632146-6},
  url={https://doi.org/10.4324/9781003632146-6}
}

@misc{deepmind2025gemini2.5pro,
  author       = {Google DeepMind},
  title        = {Gemini 2.5 Pro},
  year         = 2025,
  url          = {https://deepmind.google/models/gemini/pro/},
  note         = {Accessed: 2025-08-24},
}

@article{unver2022metacognitive,
  title   = {Metacognitive awareness and emotional resilience in children with attention deficit hyperactivity disorder},
  author  = {{\"U}nver, Hatice and Arman, Ay{\c{s}}e Rodopman and Akpunar, {\c{S}}erife Nur},
  journal = {Scandinavian Journal of Child and Adolescent Psychiatry and Psychology},
  volume  = {10},
  number  = {1},
  pages   = {33--39},
  year    = {2022},
  doi     = {10.2478/sjcapp-2022-0003},
  url     = {https://doi.org/10.2478/sjcapp-2022-0003}
}

@article{mohammadi2022effectiveness,
  title={Effectiveness of Metacognitive Therapy in Behavioral-Emotional Problem, Cognitive-Emotional Regulation Strategies, and Mind Wandering of 9 to 13-Year-Old Children with ADHD: A Quasi-Experimental Study},
  author={Mohammadi, Zeinab Darehshoori and Bavi, Sassan and Human, Farzaneh},
  journal={Jundishapur Journal of Chronic Disease Care},
  volume={11},
  number={4},
  pages={e123921},
  year={2022},
  publisher={Brieflands},
  doi={10.5812/jjcdc.123921}
}

@inproceedings{zhang2025impact,
  title     = {The Impact of GenAI Assistance on Knowledge Building in Tasks of Different Difficulty Levels},
  author    = {Zhang, Hui and Wang, Qi},
  booktitle = {Proceedings of the 2025 14th International Conference on Educational and Information Technology (ICEIT)},
  pages     = {556--561},
  year      = {2025},
  publisher = {IEEE},
  address   = {Tokyo, Japan},
  doi       = {10.1109/ICEIT64364.2025.10976085},
  url       = {https://doi.org/10.1109/ICEIT64364.2025.10976085}
}

@article{shrestha2020non,
  title   = {Non-Pharmacologic Management of Attention-Deficit/Hyperactivity Disorder in Children and Adolescents: A Review},
  author  = {Shrestha, Mahesh and Lautenschleger, Julianna and Soares, Neelkamal},
  journal = {Translational Pediatrics},
  volume  = {9},
  number  = {Suppl 1},
  pages   = {S114--S124},
  year    = {2020},
  doi     = {10.21037/tp.2019.10.01},
  url     = {https://doi.org/10.21037/tp.2019.10.01}
}

@article{zhong2024personal,
  title     = {How Do Personal Attributes Shape AI Dependency in Chinese Higher Education Context? Insights from Needs Frustration Perspective},
  author    = {Zhong, Wenjun and Luo, Jianghua and Lyu, Ya},
  journal   = {PLOS ONE},
  volume    = {19},
  number    = {11},
  pages     = {e0313314},
  year      = {2024},
  publisher = {Public Library of Science},
  doi       = {10.1371/journal.pone.0313314},
  url       = {https://doi.org/10.1371/journal.pone.0313314}
}

@article{owens2007critical,
  title   = {A Critical Review of Self-Perceptions and the Positive Illusory Bias in Children with ADHD},
  author  = {Owens, Julie Sarno and Goldfine, Matthew E. and Evangelista, Nicole M. and Hoza, Betsy and Kaiser, Nina M.},
  journal = {Clinical Child and Family Psychology Review},
  volume  = {10},
  number  = {4},
  pages   = {335--351},
  year    = {2007},
  doi     = {10.1007/s10567-007-0027-3},
  url     = {https://doi.org/10.1007/s10567-007-0027-3}
}

@article{nejati2020time,
  title   = {Time Perception in Children with Attention Deficit--Hyperactivity Disorder (ADHD): Does Task Matter? A Meta-Analysis Study},
  author  = {Nejati, Vahid and Yazdani, Samira},
  journal = {Child Neuropsychology},
  volume  = {26},
  number  = {7},
  pages   = {900--916},
  year    = {2020},
  doi     = {10.1080/09297049.2020.1712347},
  url     = {https://doi.org/10.1080/09297049.2020.1712347}
}

@article{frutos2014adaptive,
  title={Adaptive tele-therapies based on serious games for health for people with time-management and organisational problems: preliminary results},
  author={Frutos-Pascual, Maite and Garc{\'\i}a Zapirain, Bego{\~n}a and M{\'e}ndez Zorrilla, Amaia},
  journal={International Journal of Environmental Research and Public Health},
  volume={11},
  number={1},
  pages={749--772},
  year={2014},
  publisher={MDPI},
  doi = {10.3390/ijerph110100749},
  url = {https://doi.org/10.3390/ijerph110100749}
}

@article{sweller1988cognitive,
  title   = {Cognitive Load during Problem Solving: Effects on Learning},
  author  = {Sweller, John},
  journal = {Cognitive Science},
  volume  = {12},
  number  = {2},
  pages   = {257--285},
  year    = {1988},
  doi     = {10.1207/s15516709cog1202_4},
  url     = {https://doi.org/10.1207/s15516709cog1202_4}
}

@article{rubia2011disorder,
  title   = {Disorder-specific dysfunctions in patients with attention-deficit/hyperactivity disorder compared to patients with obsessive-compulsive disorder during interference inhibition and attention allocation},
  author  = {Rubia, Katya and Cubillo, Ana and Woolley, James and Brammer, Michael J. and Smith, Anna},
  journal = {Human Brain Mapping},
  volume  = {32},
  number  = {4},
  pages   = {601--611},
  year    = {2011},
  doi     = {10.1002/hbm.21048},
  url     = {https://doi.org/10.1002/hbm.21048}
}

@article{moran2016anxiety,
  title   = {Anxiety and Working Memory Capacity: A Meta-Analysis and Narrative Review},
  author  = {Moran, Tim P.},
  journal = {Psychological Bulletin},
  volume  = {142},
  number  = {8},
  pages   = {831--864},
  year    = {2016},
  doi     = {10.1037/bul0000051},
  url     = {https://doi.org/10.1037/bul0000051}
}

@article{ryan2000self,
  title     = {Self-determination theory and the facilitation of intrinsic motivation, social development, and well-being},
  author    = {Ryan, Richard M. and Deci, Edward L.},
  journal   = {American Psychologist},
  volume    = {55},
  number    = {1},
  pages     = {68--78},
  year      = {2000},
  publisher = {American Psychological Association},
  doi       = {10.1037/0003-066X.55.1.68},
  url       = {https://doi.org/10.1037/0003-066X.55.1.68}
}

@article{zimmerman1986becoming,
  title     = {Becoming a Self-Regulated Learner: Which Are the Key Subprocesses?},
  author    = {Zimmerman, Barry J.},
  journal   = {Contemporary Educational Psychology},
  volume    = {11},
  number    = {4},
  pages     = {307--313},
  year      = {1986},
  publisher = {Elsevier},
  doi       = {10.1016/0361-476X(86)90027-5},
  url       = {https://doi.org/10.1016/0361-476X(86)90027-5}
}

@article{hu2025ai,
  title={AI as your ally: The effects of AI-assisted venting on negative affect and perceived social support},
  author={Hu, Meilan and Chua, Xavier Cheng Wee and Diong, Shu Fen and Kasturiratna, K. T. A. Sandeeshwara and Majeed, Nadyanna M. and Hartanto, Andree},
  journal={Applied Psychology: Health and Well-Being},
  volume={17},
  number={1},
  pages={e12621},
  year={2025},
  publisher={Wiley},
  doi = {10.1111/aphw.12621},
  url = {https://doi.org/10.1111/aphw.12621}
}

@article{wiswede2009negative,
  title     = {Negative Affect Induced by Derogatory Verbal Feedback Modulates the Neural Signature of Error Detection},
  author    = {Wiswede, Daniel and M{\"u}nte, Thomas F. and R{\"u}sseler, Jascha},
  journal   = {Social Cognitive and Affective Neuroscience},
  volume    = {4},
  number    = {3},
  pages     = {227--237},
  year      = {2009},
  publisher = {Oxford University Press},
  doi       = {10.1093/scan/nsp015},
  url       = {https://doi.org/10.1093/scan/nsp015}
}

@inproceedings{wiese2023adding,
  title     = {Adding Domain-Specific Features to a Text-Editor to Support Diverse, Real-World Approaches to Time Management Planning},
  author    = {Wiese, Jason and Lund, John R. and Kabir, Kazi Sinthia},
  booktitle = {Proceedings of the 2023 CHI Conference on Human Factors in Computing Systems},
  pages     = {1--13},
  year      = {2023},
  publisher = {ACM},
  address   = {New York, NY, USA},
  doi       = {10.1145/3544548.3581536},
  url       = {https://doi.org/10.1145/3544548.3581536}
}

@article{desautel2009becoming,
  title={Becoming a thinking thinker: Metacognition, self-reflection, and classroom practice},
  author={Desautel, Daric},
  journal={Teachers College Record},
  volume={111},
  number={8},
  pages={1997--2020},
  year={2009},
  publisher={SAGE Publications},
  doi = {10.1177/016146810911100803},
  url = {https://doi.org/10.1177/016146810911100803}
}

@book{dunlosky2008metacognition,
  title={Metacognition},
  author={Dunlosky, John and Metcalfe, Janet},
  year={2008},
  publisher={SAGE Publications},
  address={Thousand Oaks, CA, USA},
  url = {https://uk.sagepub.com/en-gb/eur/metacognition/book229322}
}

@article{buehler1994exploring,
  title={Exploring the ``planning fallacy'': Why people underestimate their task completion times},
  author={Buehler, Roger and Griffin, Dale and Ross, Michael},
  journal={Journal of Personality and Social Psychology},
  volume={67},
  number={3},
  pages={366--381},
  year={1994},
  publisher={American Psychological Association},
  doi = {10.1037/0022-3514.67.3.366},
  url = {https://doi.org/10.1037/0022-3514.67.3.366}
}

@article{dohnany2025technological,
  title={Technological folie{\`a} deux: Feedback Loops Between AI Chatbots and Mental Illness},
  author={Dohn{\'a}ny, Sebastian and Kurth-Nelson, Zeb and Spens, Eleanor and Luettgau, Lennart and Reid, Alastair and Summerfield, Christopher and Shanahan, Murray and Nour, Matthew M},
  journal={arXiv preprint},
  volume={arXiv:2507.19218},
  pages={arXiv:2507.19218},
  year={2025},
  doi={10.48550/arXiv.2507.19218}
}

@article{scheres2008temporal,
  title   = {Temporal reward discounting and ADHD: Task and symptom specific effects},
  author  = {Scheres, Anouk and Lee, Amy and Sumiya, Motofumi},
  journal = {Journal of Neural Transmission},
  volume  = {115},
  number  = {2},
  pages   = {221--226},
  year    = {2008},
  doi     = {10.1007/s00702-007-0813-6},
  url     = {https://doi.org/10.1007/s00702-007-0813-6}
}

@article{bertilsdotter2023intensity,
  title     = {Intensity and Variable Attention: Counter Narrating ADHD, from ADHD Deficits to ADHD Difference},
  author    = {Bertilsdotter Rosqvist, Hanna and Hultman, Lill and {\"O}sterborg Wiklund, Sofia and Nygren, Anna and Storm, Palle and Sandberg, Greta},
  journal   = {The British Journal of Social Work},
  volume    = {53},
  number    = {8},
  pages     = {3647--3664},
  year      = {2023},
  publisher = {Oxford University Press},
  doi       = {10.1093/bjsw/bcad138},
  url       = {https://doi.org/10.1093/bjsw/bcad138}
}

@article{bodalski2023adhd,
  title={ADHD symptoms and procrastination in college students: The roles of emotion dysregulation and self-esteem},
  author={Bodalski, Elizabeth A. and Flory, Kate and Canu, Will H. and Willcutt, Erik G. and Hartung, Cynthia M.},
  journal={Journal of Psychopathology and Behavioral Assessment},
  volume={45},
  number={1},
  pages={48--57},
  year={2023},
  publisher={Springer},
  doi = {10.1007/s10862-022-09996-2},
  url = {https://doi.org/10.1007/s10862-022-09996-2}
}

@article{harpin2016long,
  title={Long-term outcomes of ADHD: a systematic review of self-esteem and social function},
  author={Harpin, Val and Mazzone, Luigi and Raynaud, Jean-Philippe and Kahle, Jennifer and Hodgkins, Paul},
  journal={Journal of Attention Disorders},
  volume={20},
  number={4},
  pages={295--305},
  year={2016},
  publisher={SAGE Publications},
  doi = {10.1177/1087054713486516},
  url = {https://doi.org/10.1177/1087054713486516}
}

@article{parsakia2023effect,
  title   = {The Effect of Chatbots and AI on the Self-Efficacy, Self-Esteem, Problem-Solving and Critical Thinking of Students},
  author  = {Parsakia, Kamdin},
  journal = {Health Nexus},
  volume  = {1},
  number  = {1},
  pages   = {71--76},
  year    = {2023},
  doi     = {10.61838/kman.hn.1.1.11},
  url     = {https://doi.org/10.61838/kman.hn.1.1.11}
}

@inproceedings{qadir2023engineering,
  title     = {Engineering Education in the Era of ChatGPT: Promise and Pitfalls of Generative AI for Education},
  author    = {Qadir, Junaid},
  booktitle = {2023 IEEE Global Engineering Education Conference (EDUCON)},
  pages     = {1--9},
  year      = {2023},
  publisher = {IEEE},
  address   = {Kuwait City, Kuwait},
  doi       = {10.1109/EDUCON54358.2023.10125121},
  url       = {https://doi.org/10.1109/EDUCON54358.2023.10125121}
}

@incollection{nelson1990metamemory,
  title     = {Metamemory: A Theoretical Framework and New Findings},
  author    = {Nelson, Thomas O.},
  booktitle = {Psychology of Learning and Motivation},
  editor    = {Bower, Gordon H.},
  volume    = {26},
  pages     = {125--173},
  year      = {1990},
  publisher = {Academic Press},
  address   = {San Diego, CA},
  doi       = {10.1016/S0079-7421(08)60053-5},
  url       = {https://doi.org/10.1016/S0079-7421(08)60053-5}
}

@book{tarricone2011taxonomy,
  title     = {The Taxonomy of Metacognition},
  author    = {Tarricone, Pina},
  year      = {2011},
  publisher = {Psychology Press},
  address   = {London}
}

@techreport{tobias1996assessing,
  title       = {Assessing Metacognitive Knowledge Monitoring},
  author      = {Tobias, Sigmund and Everson, Howard T.},
  institution = {College Entrance Examination Board},
  year        = {1996},
  number      = {Report No. 96-01},
  note        = {ERIC Document ED562584},
  url         = {https://files.eric.ed.gov/fulltext/ED562584.pdf}
}

@misc{lightandnight2021,
  title        = {Light \& Night},
  author       = {{Tencent Games}},
  year         = {2021},
  howpublished = {\url{https://love.qq.com/}},
  note         = {Official website, accessed 2025-11-08.}
}

@misc{garfield1978,
  title        = {Garfield},
  author       = {Jim Davis},
  year         = {1978},
  howpublished = {Syndicated comic strip created by Jim Davis, first published by United Feature Syndicate.},
  note         = {See also: Garfield Wikipedia entry, \url{https://en.wikipedia.org/wiki/Garfield}, accessed 2025-11-08.}
}

@misc{duolingo,
  title        = {Duolingo},
  author       = {{Duolingo, Inc.}},
  year         = {2011},
  howpublished = {\url{https://www.duolingo.com/}},
  note         = {Accessed 2025-11-08}
}

@misc{hatsunemiku,
  title        = {Hatsune Miku},
  author       = {{Crypton Future Media, Inc.}},
  year         = {2007},
  howpublished = {\url{https://piapro.net/intl/en.html}},
  note         = {Accessed 2025-11-08}
}

@article{seligman1967failure,
  title   = {Failure to Escape Traumatic Shock},
  author  = {Seligman, Martin E. P. and Maier, Steven F.},
  journal = {Journal of Experimental Psychology},
  volume  = {74},
  number  = {1},
  pages   = {1--9},
  year    = {1967},
  doi     = {10.1037/h0024514},
  url     = {https://doi.org/10.1037/h0024514}
}

@book{seligman1975helplessness,
  title     = {Helplessness: On Depression, Development, and Health},
  author    = {Seligman, Martin E. P.},
  year      = {1975},
  publisher = {W. H. Freeman},
  address   = {San Francisco, CA},
  url       = {https://books.google.com/books/about/Helplessness.html?id=t_c1wAEACAAJ}
}

@article{willcutt2010etiology,
  title     = {Etiology and Neuropsychology of Comorbidity between RD and ADHD: The Case for Multiple-Deficit Models},
  author    = {Willcutt, Erik G. and Betjemann, Rebecca S. and McGrath, Lauren M. and Chhabildas, Nomita A. and Olson, Richard K. and DeFries, John C. and Pennington, Bruce F.},
  journal   = {Cortex},
  volume    = {46},
  number    = {10},
  pages     = {1345--1361},
  year      = {2010},
  publisher = {Elsevier},
  doi       = {10.1016/j.cortex.2010.06.009},
  url       = {https://doi.org/10.1016/j.cortex.2010.06.009}
}

@article{molitor2016written,
  title   = {The Written Expression Abilities of Adolescents with Attention-Deficit/Hyperactivity Disorder},
  author  = {Molitor, Stephen J. and Langberg, Joshua M. and Evans, Steven W.},
  journal = {Research in Developmental Disabilities},
  volume  = {51--52},
  pages   = {49--59},
  year    = {2016},
  doi     = {10.1016/j.ridd.2016.01.005},
  url     = {https://doi.org/10.1016/j.ridd.2016.01.005}
}

@article{kutscheidt2019interoceptive,
  title={Interoceptive awareness in patients with attention-deficit/hyperactivity disorder (ADHD)},
  author={Kutscheidt, Katrin and Dresler, Thomas and Hudak, Justin and Barth, Beatrix and Blume, Friederike and Ethofer, Thomas and Fallgatter, Andreas J and Ehlis, Ann-Christine},
  journal={ADHD Attention Deficit and Hyperactivity Disorders},
  volume={11},
  number={4},
  pages={395--401},
  year={2019},
  publisher={Springer},
  doi={10.1007/s12402-019-00299-3},
  url={https://doi.org/10.1007/s12402-019-00299-3}
}

@article{roshani2017comparison,
  title   = {Comparison of Alexithymia in Individuals with and without Attention Deficit/Hyperactivity Disorder},
  author  = {Roshani, F. and Najafi, M. and Naqshbandi, S. and Malekzadeh, P.},
  journal = {Journal of Clinical Psychology},
  volume  = {9},
  number  = {2},
  pages   = {73--82},
  year    = {2017},
  doi     = {10.22075/jcp.2017.11204.1109},
  url     = {https://jcp.semnan.ac.ir/article_2720.html?lang=en}
}

@inproceedings{tcherdakoff2025burnout,
  title     = {Burnout by Design: How Digital Systems Overburden Neurodivergent Students in Higher Education},
  author    = {Tcherdakoff, Nathalie Alexandra and Marshall, Paul and Dowthwaite, Anna and Bird, Jon and Cox, Anna L.},
  booktitle = {Proceedings of the 4th Annual Symposium on Human-Computer Interaction for Work (CHIWORK 2025)},
  pages     = {17:1--17:18},
  year      = {2025},
  publisher = {Association for Computing Machinery},
  address   = {New York, NY, USA},
  url       = {https://dblp.org/rec/conf/chiwork/TcherdakoffMDBC25}
}

@article{soler2023evidence,
  title   = {Evidence of Emotion Dysregulation as a Core Symptom of Adult ADHD: A Systematic Review},
  author  = {Soler-Guti{\'e}rrez, Ana-Mar{\'\i}a and P{\'e}rez-Gonz{\'a}lez, Juan-Carlos and Mayas, Julia},
  journal = {PLOS ONE},
  volume  = {18},
  number  = {1},
  pages   = {e0280131},
  year    = {2023},
  doi     = {10.1371/journal.pone.0280131},
  url     = {https://doi.org/10.1371/journal.pone.0280131}
}

@article{shaw2014emotion,
  title   = {Emotion Dysregulation in Attention-Deficit/Hyperactivity Disorder},
  author  = {Shaw, Philip and Stringaris, Argyris and Nigg, Joel and Leibenluft, Ellen},
  journal = {American Journal of Psychiatry},
  volume  = {171},
  number  = {3},
  pages   = {276--293},
  year    = {2014},
  doi     = {10.1176/appi.ajp.2013.13070966},
  url     = {https://doi.org/10.1176/appi.ajp.2013.13070966}
}

@article{radonovich2004duration,
  title   = {Duration Judgments in Children with ADHD Suggest Deficient Utilization of Temporal Information Rather Than General Impairment in Timing},
  author  = {Radonovich, Krestin J. and Mostofsky, Stewart H.},
  journal = {Child Neuropsychology},
  volume  = {10},
  number  = {3},
  pages   = {162--172},
  year    = {2004},
  doi     = {10.1080/09297040409609807},
  url     = {https://doi.org/10.1080/09297040409609807}
}

@inproceedings{lee2025impact,
  author = {Lee, Hao-Ping and Sarkar, Advait and Tankelevitch, Lev and Drosos, Ian and Rintel, Sean and Banks, Richard and Wilson, Nicholas},
  title = {The Impact of Generative AI on Critical Thinking: Self-Reported Reductions in Cognitive Effort and Confidence Effects from a Survey of Knowledge Workers},
  year = {2025},
  publisher = {Association for Computing Machinery},
  address = {New York, NY, USA},
  booktitle = {Proceedings of the 2025 CHI Conference on Human Factors in Computing Systems},
  articleno = {1121},
  numpages = {22},
  doi = {10.1145/3706598.3713778},
  url = {https://doi.org/10.1145/3706598.3713778}
}

@article{locke2023cracks,
  title   = {In the Cracks of Attention: ADHD, Vernacular Anthropologies and Communities of Care on TikTok},
  author  = {Locke, Toby Austin},
  journal = {Teaching Anthropology},
  volume  = {12},
  number  = {1},
  pages   = {23--35},
  year    = {2023},
  doi     = {10.22582/ta.v12i1.683},
  url     = {https://doi.org/10.22582/ta.v12i1.683}
}

@article{williams2005dopamine,
  title     = {Dopamine, Learning, and Impulsivity: A Biological Account of Attention-Deficit/Hyperactivity Disorder},
  author    = {Williams, Jonathan and Dayan, Peter},
  journal   = {Journal of Child \& Adolescent Psychopharmacology},
  volume    = {15},
  number    = {2},
  pages     = {160--179},
  year      = {2005},
  publisher = {Mary Ann Liebert, Inc.},
  doi       = {10.1089/cap.2005.15.160},
  url       = {https://doi.org/10.1089/cap.2005.15.160}
}

@inproceedings{sharma2025lost,
  title     = {Lost in Translation: Researchers’ Reflections on Writing in English for CHI},
  author    = {Sharma, Sumita and Norouzi, Behnaz and White, Edward Peter Greenwood and Durall Gazulla, Eva and Yousufi, Mohsin Y. K. and Iivari, Netta and Howell, Noura and Klemettil{\"a}, Pauli and Shahid, Suleman and Sarcar, Sayan and Li, Xingyu and Mishra, Wricha},
  booktitle = {Proceedings of the Extended Abstracts of the 2025 CHI Conference on Human Factors in Computing Systems},
  pages     = {1--9},
  year      = {2025},
  publisher = {Association for Computing Machinery},
  address   = {New York, NY, USA},
  doi       = {10.1145/3706599.3716231},
  url       = {https://doi.org/10.1145/3706599.3716231}
}

@inproceedings{vigh2025lost,
  author    = {Vigh, Eszter and Weir, Ellen and Tcherdakoff, Nathalie Alexandra Penglin and Stangroome, Grace Jane and Gu, Yelu and Metatla, Oussama and Watanabe, Mamoru and Sch\"{a}fer, Ren\'{e} and Hahn, Sophie and Krawczyk, Konrad Mikolaj and Godoy, Marcela and Cossovich, Rodolfo and Morin, Randy and Dreaver-Charles, Kristine and Koole, Marguerite and Lewis, Frank B. W.},
  title     = {Lost in Translation: A Cross-Cultural Examination of Linguistic Inaccessibility in HCI},
  year      = {2025},
  isbn      = {9798400713958},
  publisher = {Association for Computing Machinery},
  address   = {New York, NY, USA},
  booktitle = {Proceedings of the Extended Abstracts of the CHI Conference on Human Factors in Computing Systems},
  articleno = {635},
  numpages  = {16},
  series    = {CHI EA '25},
  doi       = {10.1145/3706599.3716229},
  url       = {https://doi.org/10.1145/3706599.3716229}
}

@misc{openaisycophancy2025,
  author       = {OpenAI},
  title        = {Sycophancy in GPT-4o: What happened and what we’re doing about it},
  howpublished = {\url{https://openai.com/index/sycophancy-in-gpt-4o}},
  note         = {Accessed: 2025-11-29},
  year         = {2025}
}

@inproceedings{mauri2010psychophysiological,
  title     = {Psychophysiological Signals Associated with Affective States},
  author    = {Mauri, Maurizio and Magagnin, Valentina and Cipresso, Pietro and Mainardi, Luca and Brown, Emery N. and Cerutti, Sergio and Villamira, Marco and Barbieri, Riccardo},
  booktitle = {Proceedings of the 32nd Annual International Conference of the IEEE Engineering in Medicine and Biology Society},
  series    = {EMBS '10},
  pages     = {3563--3566},
  year      = {2010},
  publisher = {IEEE},
  address   = {Buenos Aires, Argentina},
  doi       = {10.1109/IEMBS.2010.5627465},
  url       = {https://doi.org/10.1109/IEMBS.2010.5627465}
}

@incollection{porges1992autonomic,
  title     = {Autonomic Regulation and Attention},
  author    = {Porges, Stephen W.},
  booktitle = {Attention and Information Processing in Infants and Adults},
  editor    = {Campbell, B. A. and Hayne, H. and Richardson, R.},
  pages     = {201--223},
  year      = {1992},
  publisher = {Lawrence Erlbaum Associates},
  address   = {Hillsdale, NJ},
  doi       = {10.4324/9781315807355-11},
  url       = {https://doi.org/10.4324/9781315807355-11}
}

@article{griffiths2017sustained,
  title={Sustained attention and heart rate variability in children and adolescents with ADHD},
  author={Griffiths, Kristi R. and Quintana, Daniel S. and Hermens, Daniel F. and Spooner, Chris and Tsang, Tracey W. and Clarke, Simon and Kohn, Michael R.},
  journal={Biological Psychology},
  volume={124},
  pages={11--20},
  year={2017},
  publisher={Elsevier},
  doi = {10.1016/j.biopsycho.2017.01.004},
  url = {https://doi.org/10.1016/j.biopsycho.2017.01.004}
}

@article{segura2018ethical,
  title   = {Ethical Implications of User Perceptions of Wearable Devices},
  author  = {Segura Anaya, L. H. and Alsadoon, Abeer and Costadopoulos, N. and Prasad, P. W. C.},
  journal = {Science and Engineering Ethics},
  volume  = {24},
  number  = {1},
  pages   = {1--28},
  year    = {2018},
  doi     = {10.1007/s11948-017-9872-8},
  url     = {https://doi.org/10.1007/s11948-017-9872-8}
}

@article{abel2013cross,
  title={Cross-system user modeling and personalization on the social web},
  author={Abel, Fabian and Herder, Eelco and Houben, Geert-Jan and Henze, Nicola and Krause, Daniel},
  journal={User Modeling and User-Adapted Interaction},
  volume={23},
  number={2},
  pages={169--209},
  year={2013},
  publisher={Springer},
  doi = {10.1007/s11257-012-9131-2},
  url = {https://doi.org/10.1007/s11257-012-9131-2}
}

@article{huang2025survey,
  title={A survey on hallucination in large language models: Principles, taxonomy, challenges, and open questions},
  author={Huang, Lei and Yu, Weijiang and Ma, Weitao and Zhong, Weihong and Feng, Zhangyin and Wang, Haotian and Chen, Qianglong and Peng, Weihua and Feng, Xiaocheng and Qin, Bing and Liu, Ting},
  journal={ACM Transactions on Information Systems},
  volume={43},
  number={2},
  pages={1--55},
  year={2025},
  publisher={Association for Computing Machinery},
  doi = {10.1145/3703155},
  url = {https://doi.org/10.1145/3703155}
}

@article{li2016association,
  title={The association of Internet addiction symptoms with impulsiveness, loneliness, novelty seeking and behavioral inhibition system among adults with attention-deficit/hyperactivity disorder (ADHD)},
  author={Li, Wendi and Zhang, Wei and Xiao, Lin and Nie, Jia},
  journal={Psychiatry Research},
  volume={243},
  pages={357--364},
  year={2016},
  publisher={Elsevier},
  doi={10.1016/j.psychres.2016.06.020},
  url={https://doi.org/10.1016/j.psychres.2016.06.020}
}

@article{wang2025effects,
  title   = {The Effects of Mindfulness and Cognitive Strategy Interventions on Core Symptoms in Children with ADHD: A Randomized Controlled Trial},
  author  = {Wang, Xueke and Chen, Li and Feng, Tingyong},
  journal = {Journal of Affective Disorders},
  year    = {2026},
  volume  = {394},
  pages   = {120513},
  doi     = {10.1016/j.jad.2025.120513}
}

@techreport{cavoukian2009privacy,
  title={Privacy by Design: The 7 Foundational Principles},
  author={Cavoukian, Ann},
  institution={Information and Privacy Commissioner of Ontario},
  year={2009},
  address={Ontario, Canada},
  url = {https://student.cs.uwaterloo.ca/~cs492/papers/7foundationalprinciples_longer.pdf}
}

@inproceedings{schaub2015design,
  title     = {A Design Space for Effective Privacy Notices},
  author    = {Schaub, Florian and Balebako, Rebecca and Durity, Adam L. and Cranor, Lorrie Faith},
  booktitle = {Proceedings of the 11th Symposium on Usable Privacy and Security (SOUPS 2015)},
  pages     = {1--17},
  year      = {2015},
  publisher = {USENIX Association},
  address   = {Ottawa, ON, Canada},
  url       = {https://www.usenix.org/system/files/conference/soups2015/soups15-paper-schaub.pdf}
}

@inproceedings{amershi2019guidelines,
  title={Guidelines for Human-AI Interaction},
  author={Amershi, Saleema and Weld, Dan and Vorvoreanu, Mihaela and Fourney, Adam and Nushi, Besmira and Collisson, Penny and Suh, Jina and Iqbal, Shamsi and Bennett, Paul N. and Inkpen, Kori},
  booktitle={Proceedings of the 2019 CHI Conference on Human Factors in Computing Systems},
  pages={1--13},
  year={2019},
  publisher={Association for Computing Machinery},
  address={Glasgow, Scotland, UK},
  doi={10.1145/3290605.3300233}
}

@inproceedings{lee2024priviaware,
  title={PriviAware: Exploring Data Visualization and Dynamic Privacy Control Support for Data Collection in Mobile Sensing Research},
  author={Lee, Hyunsoo and Jung, Yugyeong and Law, Hei Yiu and Bae, Seolyeong and Lee, Uichin},
  booktitle={Proceedings of the 2024 CHI Conference on Human Factors in Computing Systems},
  year={2024},
  publisher={Association for Computing Machinery},
  address={Honolulu, HI, USA},
  pages={1--17},
  doi = {10.1145/3613904.3642815},
  url = {https://doi.org/10.1145/3613904.3642815}
}

\clearpage
\appendix
\onecolumn

\section*{Appendix}
\textbf{A: Participants’ GenAI Tool Use and Purposes}

\setlength{\tabcolsep}{1.5pt}
{\sffamily
\begin{table*}[htbp]
\small
\centering
\caption{Information About Participants’ GenAI Tool Use and Their Primary Uses.}
\label{participants_genai_usage}

\scalebox{0.92}{%
\begin{minipage}{\linewidth}
\centering
\begin{tabular}{l@{\hspace{4pt}} >{\raggedright\arraybackslash}p{2.8cm} >{\raggedright\arraybackslash}p{7.6cm}}
\toprule
\textbf{ID} & 
\textbf{GenAI Tools} &
\textbf{Primary Uses} \\
\midrule
P1  & ChatGPT & Writing and polishing, role-play for entertainment \\
P2  & ChatGPT & Reading literature, writing and polishing \\
P3  & ChatGPT & Writing and polishing, interpersonal reflection \\
P4  & None & — \\
P5  & \makecell[l]{ChatGPT, ERNIE Bot,\\ Replica, Midjourney, Pi} 
    & Simulated companionship, writing and polishing \\
P6  & ChatGPT, ERNIE Bot & Interpersonal reflection, companionship, quick Q\&A \\
P7  & ChatGPT & English speaking practice, emotional venting (e.g., swearing), information search \\
P8  & ChatGPT & Data processing, organizing lecture notes, information search \\
P9  & ChatGPT & Coursework ideation, prototyping, data processing, health-related inquiries \\
P10 & ChatGPT & Japanese speaking practice, writing and polishing \\
P11 & ChatGPT & Writing and polishing \\
P12 & ChatGPT & Programming learning, writing and polishing \\
P13 & ChatGPT & Translation, writing and polishing \\
P14 & ERNIE Bot & Writing and polishing, information search \\
P15 & ChatGPT & Writing and polishing, role-play for psychological counseling \\
P16 & ChatGPT, Claude & Writing and polishing \\
P17 & ChatGPT & Information search, image generation \\
P18 & ChatGPT & Writing and polishing, information search, programming \\
P19 & ChatGPT & English speaking practice, information search, writing and polishing \\
P20 & ChatGPT & Writing and polishing, quick Q\&A \\
\bottomrule
\end{tabular}
\end{minipage}
}

\end{table*}
}

\textbf{B: Original User Study Process}

\begin{figure}[htbp]
    \centering
    \includegraphics[width=0.75\linewidth]{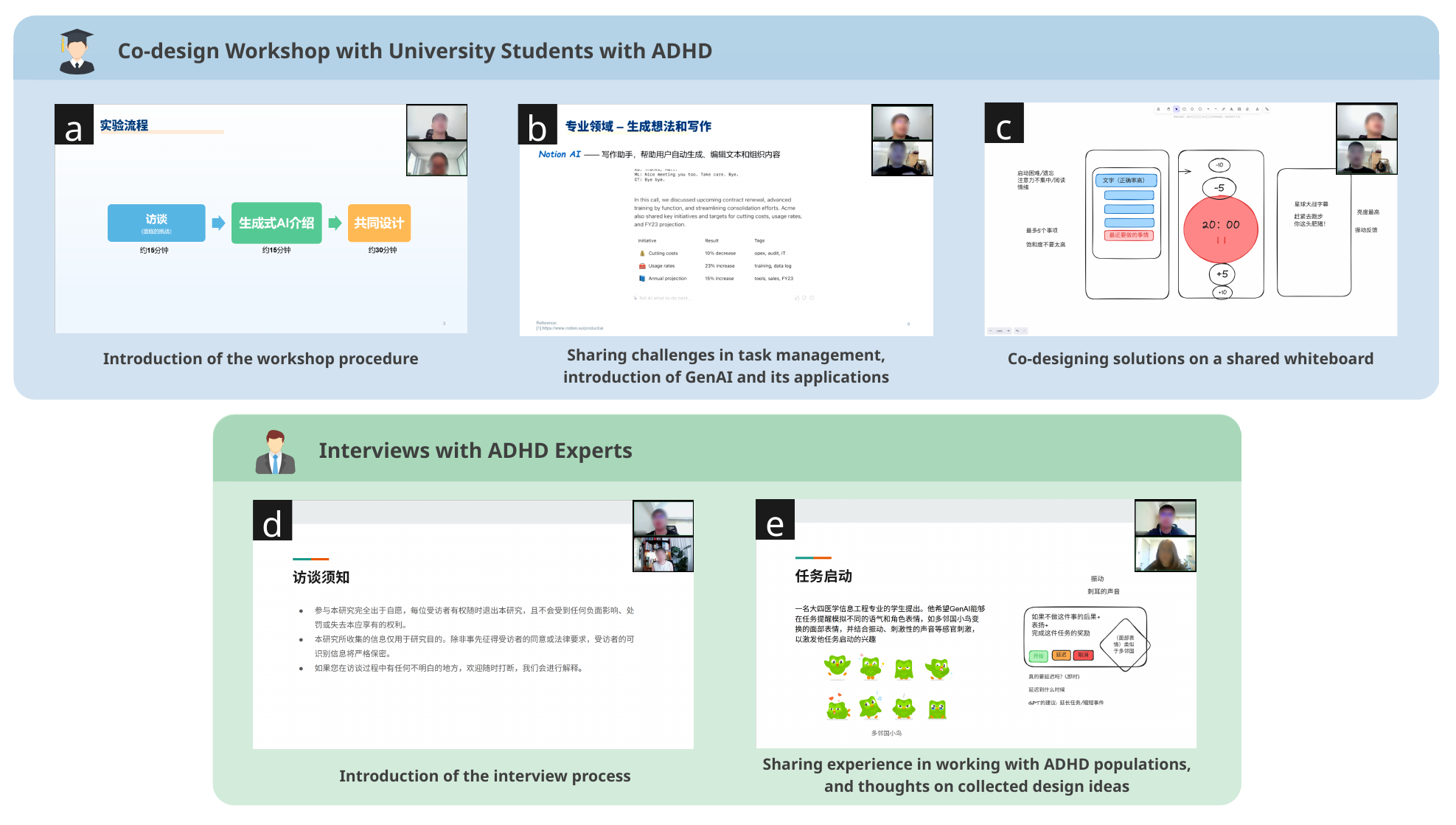}
    \Description{The overview of the study procedure, with instructions in the original language (Chinese). }
    \caption{The overview of the study procedure with instructions in the original language (Chinese).}
    \label{fig:method_overview_original}
\end{figure}

% \vspace{-10pt}

\clearpage

\textbf{C: Co-Design Artifacts in Their Original Language}

\begin{figure*}[ht]
    \centering
    \includegraphics[width=0.3\textwidth]{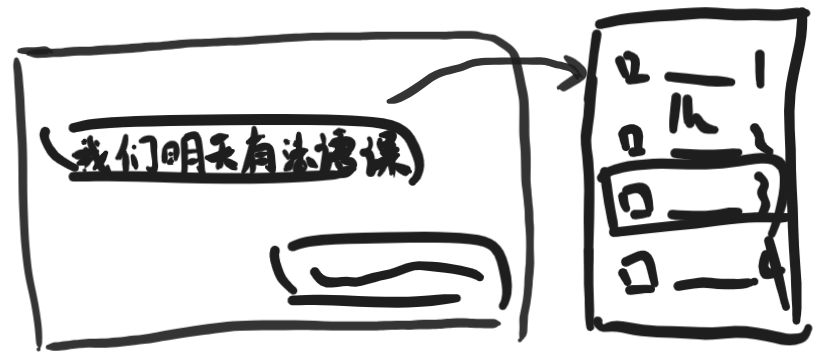} 
    \caption{Leveraging GenAI to identify tasks from chat logs and the clipboard (P2).}
    \Description{A participant-created whiteboard sketch illustrating how participants envision using generative AI to extract and identify task items from chat history and clipboard content.}
    \label{fig:appendix_informationtrain}
\end{figure*}

\begin{figure*}[htbp]
    \centering
    \includegraphics[width=0.3\textwidth]{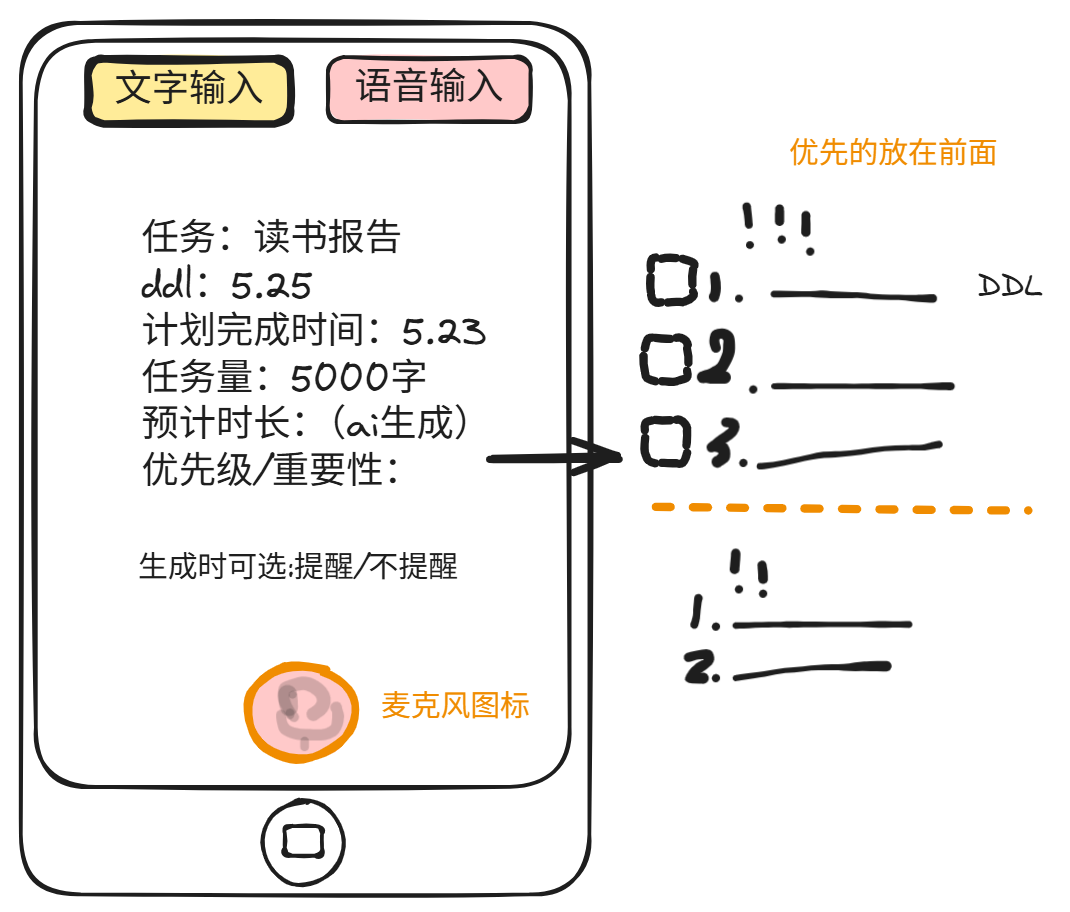} 
    \caption{Priority analysis and task list generation, which can be manually adjusted later (P7).}
    \Description{A participant-created whiteboard sketch illustrating an envisioned workflow for GenAI analyzing task priorities and generating a task list that can be manually adjusted later.}
    \label{fig:appendix_taskprior}
\end{figure*}

\begin{figure*}[htbp]
    \centering
    \includegraphics[width=0.2\textwidth]{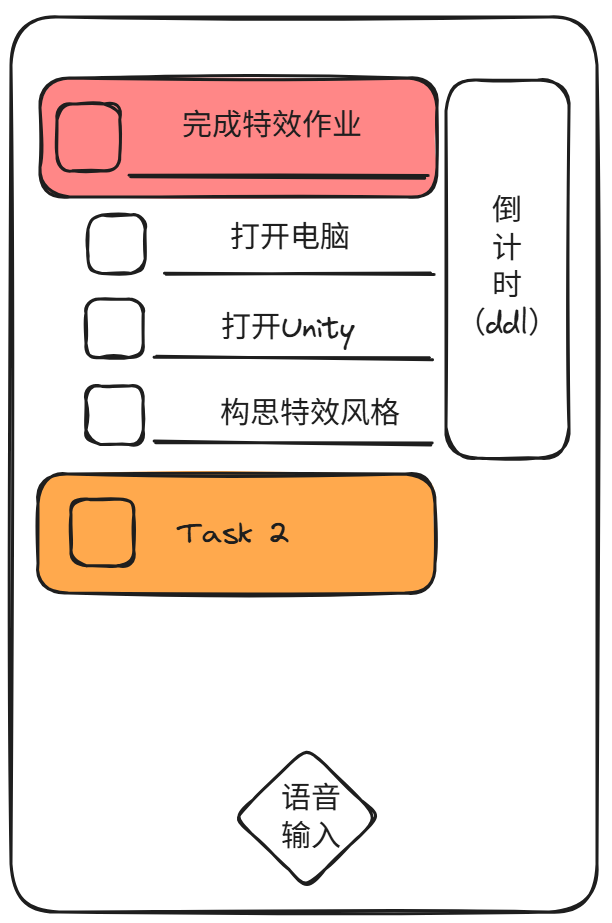} 
    \caption{Breaking tasks into actionable subtasks and estimating workload based on individuals’ historical behavior (P15).}
    \Description{A participant-created whiteboard sketch showing how generative AI could be used to break a task into actionable subtasks and estimate workload using common knowledge and personal historical information.}
    \label{fig:appendix_decompose}
\end{figure*}

\begin{figure*}[htbp]
    \centering
    \includegraphics[width=0.2\textwidth]{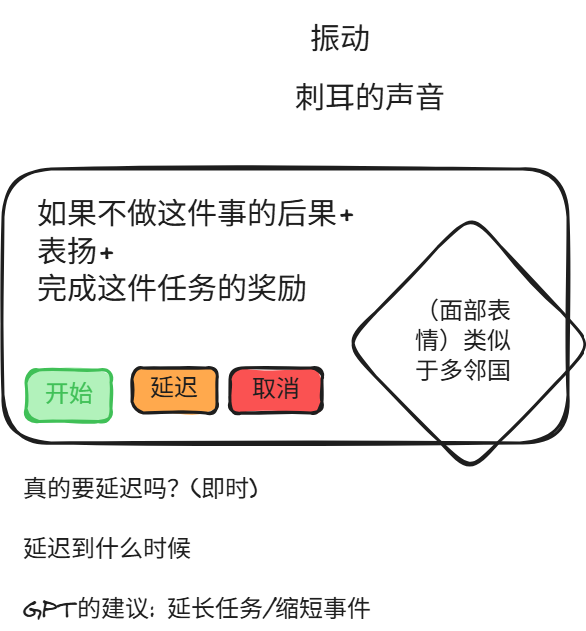} 
    \caption{Delivering multiple forms of motivational prompts (e.g., visual, audio, or vibration) to facilitate task initiation (P16).}
    \Description{A participant-created whiteboard sketch showing how generative AI could be used to deliver multimodal motivational prompts, such as visual cues, audio signals, and vibrations, to support task initiation.}
    \label{fig:appendix_reminder}
\end{figure*}

\begin{figure*}[htbp]
    \centering
    \includegraphics[width=0.25\textwidth]{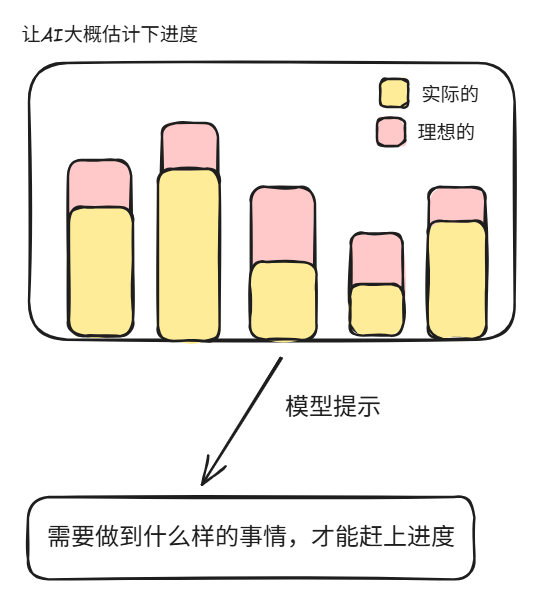} 
    \caption{Tracking the gap between actual vs. planned task progress, with suggestions on the preparations required (P11).}
    \Description{A participant-created whiteboard sketch showing how generative AI could be used to compare planned and actual task progress and provide suggestions on the remaining work needed to reach a goal.}
    \label{fig:appendix_progress}
\end{figure*}

\begin{figure*}[htbp]
    \centering
    \includegraphics[width=0.25\textwidth]{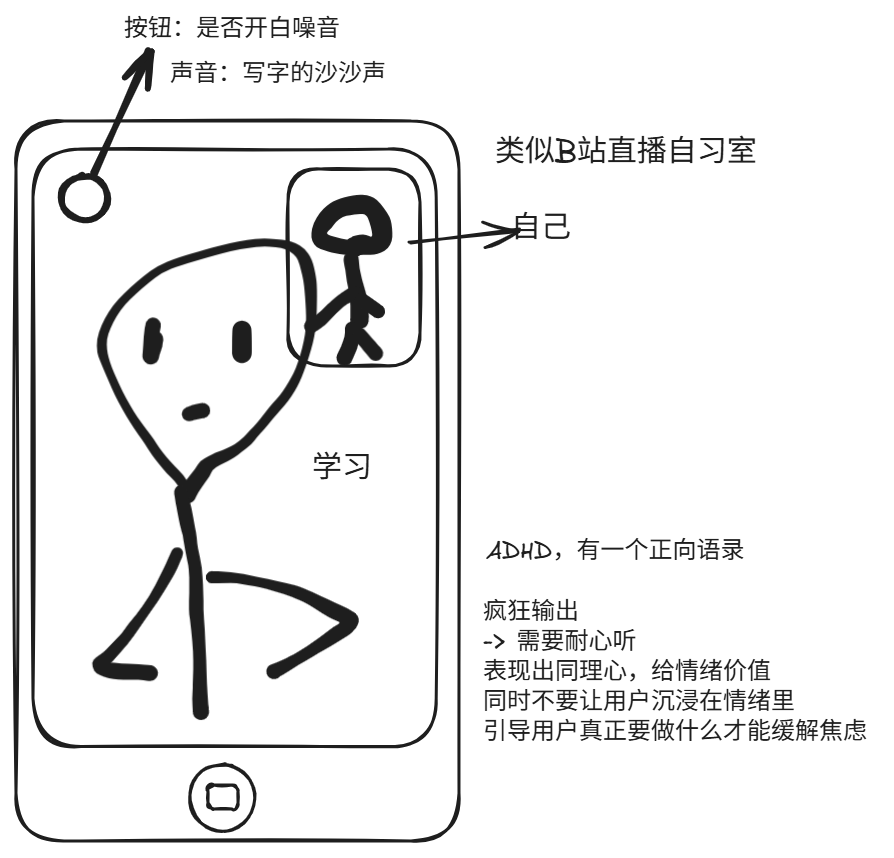} 
    \caption{Serving as a study companion to help an individual stay focused (P7).}
    \Description{A participant-created whiteboard sketch showing how generative AI could simulate a study companion to support sustained attention during individual study.}
    \label{fig:appendix_attention}
\end{figure*}

\begin{figure*}[htbp]
    \centering
    \includegraphics[width=0.28\textwidth]{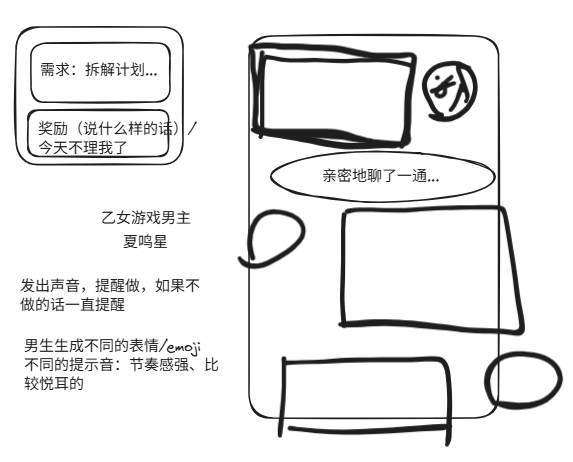} 
    \caption{Emotional support from interactive, multimodal conversations with one’s favorite fictional character (P10).}
    \Description{A participant-created whiteboard sketch showing how generative AI could take on the role of a user’s favorite fictional character to provide interactive, multimodal emotional support as a form of reward.}
    \label{fig:appendix_emotion}
\end{figure*}

\end{document}